\documentclass[showpacs,aps,prd,reprint,superscriptaddress,nofootinbib,longbibliography]{revtex4-2}
\usepackage[colorlinks=true, pdfstartview=FitV,
bookmarks=true, bookmarksnumbered=true, breaklinks]{hyperref}
\usepackage[dvipdfmx]{graphicx}
\usepackage{amsmath,amssymb,bm,color,longtable,mathrsfs,slashed,comment,tikz}
\usepackage{braket}
\usepackage{subfigure}

\definecolor{darkviolet}{rgb}{0.58, 0.0, 0.83}
\definecolor{electricultramarine}{rgb}{0.25, 0.0, 1.0}
\definecolor{brightpink}{rgb}{1.0, 0.0, 0.5}
\hypersetup{linkcolor=brightpink, citecolor=darkviolet, urlcolor=electricultramarine}

\definecolor{lime}{HTML}{A6CE39}
\DeclareRobustCommand{\orcidicon}{
	\hspace{-3mm}
	\begin{tikzpicture}
	\draw[lime, fill=lime] (0,0) 
	circle [radius=0.16] 
	node[white] {{\fontfamily{qag}\selectfont \tiny ID}};
	\draw[white, fill=white] (-0.0625,0.095) 
	circle [radius=0.007];
	\end{tikzpicture}
	\hspace{-3mm}
}

\foreach \x in {A, ..., Z}{\expandafter\xdef\csname orcid\x\endcsname{\noexpand\href{https://orcid.org/\csname orcidauthor\x\endcsname}
			{\noexpand\orcidicon}}
}

\begin{document}

%\title{Mechanical dependence of the nucleon on the scale anomaly in the Skyrme model} 
\title{Gravitational form factors of the nucleon in the Skyrme model based on scale-invariant chiral perturbation theory}

\author{Mitsuru~Tanaka\orcidA{}}
\email[]{tanaka@hken.phys.nagoya-u.ac.jp}
\affiliation{Department of Physics, Nagoya University, Nagoya 464-8602, Japan}

\author{Daisuke Fujii\orcidB{}}
\email[]{daisuke@rcnp.osaka-u.ac.jp}
\affiliation{Advanced Science Research Center, Japan Atomic Energy Agency (JAEA), Tokai, 319-1195, Japan}
\affiliation{Research Center for Nuclear Physics, Osaka University, Ibaraki 567-0048, Japan}

\author{Mamiya~Kawaguchi\orcidC{}}
\email[]{mamiya@aust.edu.cn}
\affiliation{Center for Fundamental Physics, School of Mechanics and Physics, Anhui University of Science and Technology, Huainan, 232001, People’s Republic of China}

\begin{abstract}
We investigate the role of the QCD scale anomaly in the gravitational form factors of the nucleon--particularly the $D(t)$ form factor-- as well as the associated stress distribution and internal forces, using a Skyrme model based on the scale-invariant chiral perturbation theory.
A distinctive feature of this model is the inclusion of both the pion and the scalar meson, which respectively capture the effects of the current quark mass and gluonic quantum contributions to the scale anomaly.
By varying the mass of the scalar meson, we evaluate the sensitivity of the gluonic scale anomaly to the nucleon properties. 
We find that the gluonic scale anomaly plays a crucial role in satisfying the stability conditions of the nucleon and provides an internal confining force. 
Moreover, we also evaluate the momentum-transfer dependence of $D(t)$, which closely reproduces the lattice QCD results. With an appropriate choice of the anomalous dimension associated with the quark mass, its forward-limit value (i.e., the D-term) also matches the lattice data well.

%We investigate the role of the QCD scale anomaly in generating confining forces inside the nucleon, using a Skyrme model based on the scale-invariant chiral perturbation theory. In this model, a scalar meson field is introduced to represent the explicit breaking of scale symmetry, and its coupling to the dilatation current enables a consistent treatment of the scale anomaly via the partially conserved dilatation current (PCDC) relation. We analyze the sensitivity of the pressure distribution to the parameters associated with the scale anomaly, such as the pion decay constant, the scalar meson mass, and anomalous dimension. We also evaluate the force acting on quarks, showing that the gluonic scale anomaly plays the dominant role in confining them inside the nucleon. Finally, we explore how the momentum-transfer dependence of $D(t)$ varies with the model parameters and demonstrate that a specific parameter set reproduces the lattice QCD results with remarkable accuracy. 
\end{abstract}

\maketitle

\section{Introduction}\label{ch:introduction}

The mechanisms that confine quarks and gluons within hadrons still remain one of the most fundamental problems in QCD. A promising route to understanding confinement lies in exploring QCD's non-perturbative phenomena—most notably the quark-antiquark (chiral) condensate, the gluon condensate, and the various patterns of symmetry breaking. 
Recently, the internal stress distribution within hadrons has attracted growing interest for investigating the connection between confinement and non-perturbative QCD phenomena~\cite{Fujii:2024rqd,Fujii:2025aip,Fujii:2025tpk}.
%Recently, we proposed that the internal stress distribution within hadrons offers a valuable probe into the connection between confinement and non-perturbative QCD phenomena~\cite{Fujii:2025aip}. 

%These
The internal stress distributions—specifically, the pressure profile in 2018~\cite{Burkert:2018bqq} and the shear force distribution in 2021~\cite{Burkert:2021ith,Duran:2022xag}—was extracted from experiment for the first time.
%extracted
Furthermore, the upcoming Electron-Ion Collider (EIC) experiment is highly anticipated to provide more detailed measurements.
%Furthermore, there is growing anticipation for more detailed measurements from the upcoming Electron-Ion Collider (EIC) experiment. 
%Against this backdrop
Motivated by these experimental developments, theoretical investigations have rapidly progressed~\cite{Polyakov:2018exb,Shanahan:2018nnv,Lorce:2018egm,Anikin:2019kwi,Avelino:2019esh,Yanagihara:2019foh,Hatta:2019lxo,Freese:2019eww,Azizi:2019ytx,Mamo:2019mka,Neubelt:2019sou,Alharazin:2020yjv,Varma:2020crx,Kim:2020nug,Chakrabarti:2020kdc,Yanagihara:2020tvs,Kim:2020lrs,Tong:2021ctu,Freese:2021czn,Panteleeva:2021iip,Hatta:2021can,Mamo:2021krl,Freese:2021qtb,Gegelia:2021wnj,Kim:2021jjf,Owa:2021hnj,Pefkou:2021fni,Lorce:2021xku,Ji:2021mfb,More:2021stk,Mamo:2022eui,Lorce:2022cle,Fujita:2022jus,Choudhary:2022den,Kim:2022wkc,Alharazin:2022wjj,Won:2022cyy,Tanaka:2022wzy,Ito:2023oby,Lorce:2023zzg,Amor-Quiroz:2023rke,Guo:2023pqw,Won:2023ial,Guo:2023qgu,Czarnecki:2023yqd,Won:2023zmf,Hackett:2023rif,Hatta:2023fqc,Liu:2023cse,Cao:2024zlf,Liu:2024rdm,Yao:2024ixu,Goharipour:2025lep,Dehghan:2025ncw,Goharipour:2025yxm,Broniowski:2025ctl,Ghim:2025gqo,Hatta:2025vhs,Hatta:2025ryj,Dehghan:2025eov,Guo:2025jiz,Liu:2025vfe,Sugimoto:2025btn,Nair:2025sfr,Cao:2025dkv,Coriano:2024wrz,Coriano:2025lge,Coriano:2025lge,Stegeman:2025sca}.
The stress distribution can be obtained from the form factors that characterize the matrix elements of the hadronic energy momentum tensor (EMT), known as the gravitational form factors (GFFs) (for reviews, see Refs.~\cite{Polyakov:2018zvc,Burkert:2023wzr}).
These distributions would be connected to
%represent
the forces that confine quarks and gluons within a hadron , which
%and 
is expected to be closely related to the QCD confinement mechanism. 
Therefore, 
investigating the non-perturbative QCD properties connected to the internal stress distributions would be valuable for gaining a new insight into the QCD confinement mechanism.
%by investigating how non-perturbative QCD phenomena influence the internal pressure, one may gain valuable insights into the confinement mechanism in QCD. 
%The stress distribution reveals the balance between repulsive and attractive forces that ensure the mechanical stability of the nucleon.
%Consequently, the GFFs offer valuable insights into the underlying mechanisms of quark confinement.

From this perspective, a notable non-perturbative QCD quantity would be the scale anomaly, which is represented by the trace part of the EMT.
%In this paper, in continuation of our earlier work~\cite{Fujii:2025aip}, we focus on the breaking of scale symmetry, one of non-perturbative QCD phenomena. 
While QCD is classically scale invariant in the chiral limit, this symmetry is broken by quantum effects and then becomes anomalous. 
This scale anomaly is actually reflected in the definite contribution to the finite nucleon mass.
%The fact that nucleons possess a finite mass directly reflects this breaking of scale invariance.
Indeed, in Ref.~\cite{Ji:1994av}, Ji demonstrated that the scale anomaly accounts for a non-negligible contribution to the nucleon mass.
Given this fact that the nucleon mass is associated with the time component of the EMT, its other components are also expected to be affected by the scale anomaly. However, the influence of the scale anomaly on spatial components such as the pressure and the shear force has not been intensively investigated.
%Since the nucleon mass arises from the time component of the EMT, it is natural to expect that the spatial components, which determine the internal stress, are also affected by the scale anomaly.

To elucidate the role of the scale anomaly in the stress distribution from the low-energy perspective, we employed the Skyrme model based on scale-invariant chiral perturbation theory (sChPT)~\cite{Lanik:1984fc,Ellis:1984jv,Leung:1989hw,Campbell:1990ak,Donoghue:1991qv,Brown:1991kk,Song:1997kx,Lee:2003eg,Park:2003sd,Park:2008zg,Li:2018gng,Crewther:2013vea,Li:2016uzn,Kasai:2016ifi,Hansen:2016fri,Appelquist:2017wcg,Appelquist:2017vyy,Cata:2019edh,Appelquist:2019lgk,Brown:2019ipr,Matsuzaki:2013eva,Zwicky:2023bzk,Zwicky:2023krx,Shifman:2023jqn}
in our earlier work~\cite{Fujii:2025aip}.
In this model framework, a scalar meson field is incorporated into chiral perturbation theory (ChPT) to effectively describe the scale anomaly,
%implement explicit breaking of scale symmetry,
while the Skyrme term is introduced to stabilize the soliton solution,
which is identified with the nucleon (called skyrmion~\cite{Skyrme:1962vh,Adkins:1983ya}).
In more detail, the scalar meson couples to the dilatation current $J^\mu_D$ to account for the anomalous contributions arising from the current quark mass and gluonic quantum corrections, enabling a systematic treatment of the scale anomaly via the partially conserved dilatation current (PCDC) relation.
Within the sChPT framework, the scale anomaly is characterized by the pion mass, the dilaton mass, their decay constants, and the anomalous dimension parameter.
%, and then its strength can be controlled by varying these values. 
By using the skyrmion approach based on sChPT, our earlier work demonstrated that the gluonic scale anomaly gives the dominant contribution to the negative pressure inside the nucleon and plays a leading role in the D-term, defined as a forward limit of the GFF $D(t)$.
%Then, in our model, the strength of the scale anomaly is controlled by parameters such as the pion decay constant, the scalar meson mass, and the anomalous dimension. 

%Our previous work~\cite{Fujii:2025aip}, utilizing this model, demonstrated that the gluonic scale anomaly gives the dominant contribution to the negative pressure inside the nucleon. 
As an extension of our earlier work~\cite{Fujii:2025aip}, in this paper, we develop the analysis by further investigating the following details:
%analyze the following aspects in detail.
\begin{enumerate}
    \item %We investigate the sensitivity of the stress distribution to variations in parameters associated with the scale anomaly. This analysis reveals how the internal stress distribution of hadrons changes as the strength of the scale anomaly is varied.
    By varying the values of meson mass, decay constant, and the anomalous dimension parameter, we control the strength of the scale anomaly and investigate its correlation with the stress distribution inside the nucleon.
    Such insights to be gained in this study are particularly important for understanding how the pressure distribution %inside hadrons
    may change
    %be modified 
    under external conditions, where the strength of the scale anomaly is expected to vary, %change,
    such as finite temperature or finite density. 
    \item 
    We investigate the mechanical stability of the nucleon by conducting a more detailed analysis of its internal stress distribution, the D-term, and the balance of internal forces. Through this analysis, we also explore the connection between the (negative) pressure and the stability condition. We find that the gluonic scale anomaly makes a negative contribution to the internal pressure in the tangential direction, which is essential for satisfying the stability condition in terms of the D-term. Furthermore, we analyze how the gluonic scale anomaly is related to the internal force inside the nucleon. 
    %\blue{ Following the recent proposal in Ref.~\cite{Ji:2025gsq}, we define the force acting on quarks through the EMT, which is responsible for confining the quarks inside the nucleon. To deepen our understanding of the force associated with the internal distributions, we elucidate the role of the scale anomaly. We find that the gluonic scale anomaly dominantly gives rise to the force that confines quarks inside the nucleon.
    %We elucidate the role of the scale anomaly in generating the force that confines quarks and gluons inside hadrons. Following the definition proposed in Ref.\cite{Ji:2025gsq}, we define the force acting on quarks and demonstrate that the scale anomaly—particularly its gluonic component—gives rise to the force that confines quarks inside hadrons. 
    %This clarifies that the negative pressure dominantly generated by the gluonic anomaly, as discussed in Ref.~\cite{Fujii:2025aip,Ji:2025gsq}, is directly related to the confining force acting on quarks.} %that confines quarks within hadrons. 
    \item 
    Building on our previous result for the D-term, we show the momentum transfer dependence of the $D(t)$ form factor in our model and compare it with recent lattice QCD results, demonstrating that the Skyrme model reproduces the momentum transfer dependence of $D(t)$ observed in the lattice QCD simulations.
    %quantitative agreement.
    This reproduction %agreement 
    not only supports the quantitative reliability of our analysis but also provides a robust estimate of the $D$-term, whose precise value has yet to be firmly established.
\end{enumerate}

The remainder of this paper is organized as follows.
Section~\ref{ch:GFFs} provides a review of the GFFs for spin-$\frac{1}{2}$ hadrons, as well as their relation to the internal energy density and stress distributions.
We also introduce the method of decomposing these distributions based on the %trace
scale anomaly, laying the groundwork for evaluating the anomalous contributions.
In addition, we present the nucleon's stability conditions and define the internal force.
In Section~\ref{ch:skyrme model}, we present the Skyrme model 
based on sChPT,
%extended by a scalar meson field, 
which enables the effective incorporation of scale anomaly
%scale symmetry breaking
via the PCDC relation.
Section~\ref{ch:results} presents our numerical results for the energy density and the stress distribution, along with a verification of the stability conditions.
%varying meson masses, decay constants, and the anomalous dimension parameter.
%and analyzes their dependence on parameters characterizing the scale anomaly.
We also discuss the contribution of the scale anomaly to the internal force.
%\blue{We also demonstrate that the scale anomaly generates a confining force acting on quarks.}
%, in agreement with the recent findings of Refs.\cite{Fujii:2025aip,Ji:2025gsq}.
Furthermore, we show the $D(t)$ form factor, which reproduces the recent lattice observations.
%that the $D(t)$ form factor obtained in our model is in good agreement with recent lattice QCD results, providing a quantitatively reliable description of the data.
Finally, Section \ref{ch:summary} summarizes our main results and discusses their broader implications.

\section{Gravitational form factors for spin 1/2 particles
%hadrons
}\label{ch:GFFs}

In this section, we review the definition of
%define
the GFFs of the nucleon—a spin-$1/2$ particle—and present the corresponding expressions for the energy density and stress distributions.
%that can be extracted from them. 
To elucidate the role of the scale anomaly, we decompose the EMT into its trace and traceless parts. 
This general decomposition is founded on Ji’s~\cite{Ji:1994av} mass decomposition, which is formulated in a model-independent manner. 
In addition, we introduce the stability conditions of the nucleon derived from the energy density and stress distributions, and define the internal force inside the nucleon based on the static EMT.
%Even under this treatment, the so-called von Laue condition, which serves as the stability criterion for the pressure inside the nucleon,is manifestly satisfied in the matrix elements of the nucleon's EMT.
%\blue{Furthermore, the force acting on quarks inside the nucleon is introduced through the stress distributions.}

%Through the trace–traceless decomposition of the EMT, the energy density and pressure admit a model-independent separation, and it is generally shown that the spatial integrals of each contribution are proportional to the nucleon mass. 
%These general relations provide the foundation for Ji’s mass decomposition and guarantee the so-called von Laue condition, which is the stability criterion for the pressure.  

\subsection{Matrix elements of the energy-momentum tensor}\label{sec:matrix elements}
The matrix elements of the EMT for spin-$\frac{1}{2}$ particles, under Lorentz, CP and CPT symmetries,
can be generally expressed as
\begin{align}
    &\Braket{p',s'|\Theta_{\mu\nu}(x)|p,s} \notag \\
    &=\bar{u}'\Big[A(t)\frac{P_\mu P_\nu}{M_N}+J(t)\frac{i\left(P_\mu\sigma_{\nu\rho}+P_\nu\sigma_{\mu\rho}\right)\Delta^\rho}{2M_N} \notag \\
    &\hspace{10mm}+D(t)\frac{\Delta_\mu\Delta_\nu-g_{\mu\nu}\Delta^2}{4M_N}\Big]ue^{i(p'-p)x}, \label{GFFs}
\end{align}
where $\sigma_{\mu\nu}$ is defined as $\sigma_{\mu\nu}=\frac{i}{2}[\gamma^\mu,\gamma^\nu]$ with the Dirac matrix $\gamma_\mu$; the momentum $P^\mu$ and $\Delta^\mu$ are defined by $P^\mu=(p^\mu+p'^\mu)/2$ and $\Delta^\mu=p'^\mu-p^\mu$, and $t=\Delta^2$;
$M_N$ is a nucleon mass; 
$s,s'=\pm1/2$ represents the spin state of a hadron;
$u(p,s)$ is the Dirac spinor. 
Here, we employ the Minkowski metric $g_{\mu\nu}={\rm diag}(1,-1,-1,-1)$ and normalization condition as $\bar u(p')u(p)=2E$ 
with the energy $E=\sqrt{M_N^2+\vec{\Delta}^2/4}$
and 
$\braket{p'|p}=2p^0(2\pi)^3\delta^{(3)}(\vec{p}'-\vec{p})$.

In the forward limit, $t=0$, the GFFs take the following values
\begin{align}
    &A(0) =1, \ \ \ J(0) =\frac{1}{2},
\end{align}
these constraints reflect the Poincaré symmetry. %\blue{and indicate that the mass and spin of the nucleon are conserved.}
%fully described at the hadronic level. 
We emphasize that, in contrast, the D-term, defined as $D=D(0)$, is subject to no general constraints.
%there are no general constraints on the D-term.
This difference arises from the fact that mass and angular momentum correspond to the conserved charges (generators) of space-time translations and rotations in the Poincar\'{e} group, whereas the D-term, encoding internal stress distributions, does not. 
%It should be noted that, while the individual quark and gluon contributions to the GFFs are generally renormalization scale dependent, the total form factors $A(t)$, $J(t)$, and $D(t)$ are renormalization scale invariant~\cite{}. 

\subsection{Energy density and stress distributions}
To investigate the spatial structure of the EMT following the physical interpretation proposed in Ref.~\cite{Polyakov:2002yz}, we adopt the Breit frame, where the time component of the momentum transfer vanishes ($\Delta^0 = 0$).
 In this frame, the average momentum and momentum transfer are given by $P^{\mu} = (E, 0, 0, 0)$ and $\Delta^{\mu} = (0, \vec{\Delta})$, respectively, with $t = -|\vec{\Delta}|^2$.
We define the static EMT as the Fourier transform of the matrix elements of the EMT with respect to the spatial momentum transfer $\vec{\Delta}$:
\begin{align}
\Theta_{\mu\nu}^{\mathrm{static}}(\vec{r},\vec{s})=\int\frac{d^3\vec{\Delta}}{(2\pi)^3}e^{-i\vec{r}\cdot\vec{\Delta}}\frac{\Braket{p',s'|\Theta_{\mu\nu}(x=0)|p,s}}{\bar u(p')u(p)}, \label{staticEMT}
\end{align}
%The static EMT is time-independent since $\Delta^0$ dissappears in the Breit frame. 
where, in the Breit frame, the matrix element is written as follows;
\begin{align}
    &\braket{p',s'|\Theta_{00}(0)|p,s} \notag \\
    &=2M_NE\left[A(t)-\frac{t}{4M_N^2}\left(A(t)-2J(t)+D(t)\right)\right]\delta_{ss^\prime} \label{Breit_T00} \\
    &\braket{p',s'|\Theta_{0i}(0)|p,s}=2M_NE\left[J(t)\frac{(-i\vec{\Delta}\times\vec{\sigma}_{s^\prime s})_i}{2M_N}\right] \\
    &\braket{p',s'|\Theta_{ij}(0)|p,s}=2M_NE\left[D(t)\frac{\Delta_i\Delta_j-\delta_{ij}\vec{\Delta}^2}{4M_N^2}\right]\delta_{ss^\prime}, \label{Breit_Tij}
\end{align}
where $\vec{\sigma}_{s^\prime s}=\chi^\dagger_{s^\prime}\vec{\sigma}\chi_s$ with the nucleon Pauli spinors $\chi_s$ in the normalized as $\chi^\dagger_{s^\prime}\chi_s=\delta_{ss^\prime}$ and the Pauli matrix $\vec{\sigma}$. 

In the following, we restrict the analysis to the case $s = s^\prime$ and neglect spin distributions while focusing on the stress distributions within the nucleon. 
For a spherically symmetric system, the energy density $\epsilon(r)$, pressure $p(r)$, and shear force $s(r)$ are defined in terms of the static EMT as follows:
\begin{align}
    \epsilon(r) =& \Theta^{\mathrm{static}}_{00}(r), \label{def of e} \\
    p(r) =& \frac{1}{3}\delta^{ij}\Theta_{ij}^{\mathrm{static}}(r), \label{def of p} \\
    s(r) =& \frac{3}{2} \left( \frac{r^{i}r^{j}}{r^{2}} - \frac{1}{3}\delta^{ij} \right) \Theta_{ij}^{\mathrm{static}}(r). \label{def of s}
\end{align}

From Eq.~\eqref{Breit_Tij}, the GFF $D(t)$, which encodes information about the internal forces and mechanical properties of the nucleon, can be expressed as
\begin{align}
    D(t) = -6 M_{N} \int d^{3}x\, \left( r^{i} r^{j} - \frac{1}{3} \delta^{ij} r^{2} \right) \frac{j_{2}(\Delta r)}{(\Delta r)^{2}}\, \Theta^{\mathrm{static}}_{ij}(r), \label{D(t)}
\end{align}
where $j_{2}(\Delta r)$ is the spherical Bessel function of order two, and $\Delta = |\vec{\Delta}|$ denotes the magnitude of the three-momentum transfer.
The nucleon mass and the D-term can also be related to the energy density and the pressure or shear force distributions as follows:
\begin{align}
    & M_{N} = \int d^{3}x\, \epsilon(r), \label{MN}\\
    & D = M_{N} \int d^{3}x\, r^{2} p(r) = -\frac{4}{15} M_{N} \int d^{3}x\, r^{2} s(r). \label{D-term}
\end{align}

So far, the stress tensor has been expressed in terms of a three-dimensional orthogonal Cartesian coordinate. However, assuming a spherical nucleon shape may be not well suited to this coordinate. Hence, one can rewrite the stress tensor to better reflect spherical symmetry.
By diagonalizing the stress tensor along its eigenvector directions, all off-diagonal (shear) components becomes zero, %vanish,
leaving only the eigenvalues as the principal (normal) stresses.
This diagonalization corresponds to expressing the spatial components of EMT in the spherical coordinate system as
$\Theta_{\alpha\beta}=\mathrm{diag}(p_r,p_\theta,p_\phi)$ with $\alpha,\beta=r,\phi,\theta$. 
%In the spherically symmetric case considered here, where $p_\theta = p_\phi$, this diagonalization allows for \blue{a clear and concise representation of the stress distribution in a two-dimensional planeas illustrated in subsection~\ref{2D_visualization}.},
Note again that 
$p_r$, $p_\theta$ and $p_\phi$ are the eigenvalues of $\Theta$ and their corresponding eigenvectors are denoted by $\vec e_r$, $\vec e_\theta$ and $\vec e_\phi$, respectively.
%\red{Note that these pressures act on an infinitesimal surface element $d\vec S = dS_{r} \vec{e}_{r} + dS_{\theta} \vec{e}_{\theta} + dS_{\phi} \vec{e}_{\phi}$.}
In the case where the nucleon has a spherical shape, we have $p_\theta = p_\phi$.
%Applying \blue{this approach} in spherical coordinates, the stress tensor \blue{yields eigenvalues corresponding to} the radial and tangential pressures acting on an infinitesimal surface element $dS = dS_{r} \vec{e}_{r} + dS_{\theta} \vec{e}_{\theta} + dS_{\phi} \vec{e}_{\phi}$.
Accordingly, the pressures in the radial and tangential directions are given by
\begin{align}
& p_{r}(r) = \frac{2}{3} s(r) + p(r), \label{radial force} \\
& p_{\theta,\phi}(r) = -\frac{1}{3} s(r) + p(r). \label{tangential force}
\end{align}
%where $p_r$, $p_{\theta}$, and $p_{\phi}$ are the pressures acting on the surface elements $dS_r$, $dS_{\theta}$, and $dS_{\phi}$, respectively.
Based on these expressions in Eqs.~\eqref{radial force} and \eqref{tangential force},
we will illustrate the spatial distribution of the pressures projected onto the transverse plane of the nucleon,
%in the nucleon's cross section 
in subsection~\ref{2D_visualization}.
%The quark contribution to the normal and tangential forces as defined in Sec. VI.C, are displayed in a two-dimensional plot in Fig. 14. This figure shows the 3D distributions inside the proton in a slice going through the “equatorial plane”.

\subsection{The decomposition based on the scale anomaly }\label{decomposition}

%Then, we decompose the energy density and pressure distributions according to the trace-traceless decomposition of the EMT, where the trace part encodes the breaking of scale symmetry 
The trace part of the EMT encodes the scale anomaly. %in the underlying theory. 
At the classical level, QCD is scale-invariant in the chiral limit. 
However, once 
the current quark mass and 
the quantum correction are taken into account, the conservation law of the dilatation current becomes anomalous~\cite{Adler:1976zt,Collins:1976yq}:
%the theory is regularized and renormalized at the quantum level, the trace of the EMT acquires an extra term—the scale anomaly:
\begin{align}
    &\Theta^{\mu}_{\; \mu}=
    \partial_\mu J^\mu_D \notag \\
    &=(1+\gamma_m)\sum_fm_f\bar q_fq_f+\frac{\beta(g_s)}{2g_s}{\rm Tr}\left(G_{\mu\nu}G^{\mu\nu}\right), \label{scaleanomaly}
\end{align}
where $J^\mu_D$ is the dilatation current, $\gamma_m$ the quark–mass anomalous dimension, $\beta(g_s)$ the QCD beta function, $G_{\mu\nu}$ the $\mathrm{SU}(N_c)$ field strength, and $m_f$, $q_f$ $(\bar q_f)$ the current quark masses and quark (antiquark) fields with a flavor index $f$.
%The scale anomaly in Eq.~\eqref{scaleanomaly} is renormalization-group invariant, and thus scale symmetry in QCD is broken by quantum effects. 

%\blue{From the above}, 
In order to express the scale anomaly explicitly in energy density and pressure distributions,
we decompose the EMT into the trace and the traceless parts,
\begin{align}
    &\Theta^{\rm static}_{\mu\nu}(\vec{r})=\bar{\Theta}^{\rm static}_{\mu\nu}+\hat{\Theta}^{\rm static}_{\mu\nu} \label{staticEMTdecomp} \\
    &\bar{\Theta}^{\rm static}_{\mu\nu}(\vec{r})=\int\frac{d^3\vec{\Delta}}{(2\pi)^3}e^{-i\vec{r}\cdot\vec{\Delta}}\frac{\braket{p'|\bar{\Theta}_{\mu\nu}(0)|p}}{\bar u(p')u(p)}\\
    &\hat{\Theta}^{\rm static}_{\mu\nu}(\vec{r})=\int\frac{d^3\vec{\Delta}}{(2\pi)^3}e^{-i\vec{r}\cdot\vec{\Delta}}\frac{\braket{p'|\hat{\Theta}_{\mu\nu}(0)|p}}{\bar u(p')u(p)}.
\end{align}
Here, $\bar{\Theta}_{\mu\nu} \equiv \Theta_{\mu\nu}-\frac{1}{4}g_{\mu\nu}{\Theta^\rho}_\rho$ 
%is
represents the traceless part of the EMT, which can be interpreted as the dynamical energy of quarks and gluons, %\red{and can further be decomposed into quark and gluon contributions,
%\begin{equation}
%\bar{\Theta}^{\rm static}_{\mu\nu} =\bar{\Theta}^{{\rm static,} q}_{\mu\nu} +\bar{\Theta}^{{\rm static,} g}_{\mu\nu},
%\end{equation}}
while $\hat{\Theta}_{\mu\nu} \equiv \frac{1}{4}g_{\mu\nu}{\Theta^{\rho}}_\rho$ 
%is 
represents the trace part, which is interpreted as the energy arising from the scale anomaly. %We emphasize that this decomposition is both renormalization-scale and gauge-invariant. 
In accordance with the definition~\eqref{def of e}-\eqref{def of s}, we similarly decompose the energy density $\epsilon=\bar{\epsilon}+\hat{\epsilon}$, pressure $p=\bar{p}+\hat{p}$, and shear force $s=\bar{s}$:
\begin{equation}
\begin{split}
    &\bar{\epsilon}(r) = \bar{\Theta}^{\mathrm{static}}_{00}(r), \quad \hat{\epsilon}(r) = \hat{\Theta}^{\mathrm{static}}_{00}(r), \\
    &\bar{p}(r) = \frac{1}{3}\delta^{ij} \bar{\Theta}^{\mathrm{static}}_{ij}(r), \quad \hat{p}(r) = \frac{1}{3}\delta^{ij} \hat{\Theta}^{\mathrm{static}}_{ij}(r), \\
    &\bar{s}(r) = \frac{3}{2} \left( \frac{r^i r^j}{r^2} - \frac{1}{3}\delta^{ij} \right) \bar{\Theta}^{\mathrm{static}}_{ij}(r),
    \label{decomposed}
\end{split}
\end{equation}
where the trace part of the shear force is absent due to the tensorial structure of the EMT.

\subsection{
Model-independent relations associated with the nucleon’s stability
%Model independent relations
}

In this subsection,  we present model-independent relations associated with the nucleon’s stability, which are expressed through the energy density, stress distribution, and their decomposed components.
%$s$ vanishes from the tensorial structure. 

%As shown below, the energy density, pressure, shear force, and their decomposed components satisfy model-independent relations. 

The static EMT conservation law, $\partial^i \Theta^{\rm static}_{ij} = 0$, 
yields the differential equation,
\begin{align}
    \frac{2}{3}s^{\prime}(r) + \frac{2}{r}s(r) + p^{\prime}(r) = 0, \label{differential equation}
\end{align}
where the prime denotes a derivative with respect to $r$. This relation connects the pressure $p(r)$ and shear force $s(r)$, demonstrating that they are not independent. 
Consequently, although $s(r)$ does not directly appear in the diagonal part of the EMT, it is related to $p(r)$ through this differential equation, implying that the shear force is indirectly connected to the scale anomaly.
%contribute to the EMT trace (and thus the scale anomaly), it is indirectly influenced by the trace part via its link to $p(r)$.  

Moreover, the static EMT conservation also implies the von Laue condition~\cite{Laue:1911lrk}:  
\begin{align}
    \int_{0}^{\infty} dr\, r^{2} p(r) = 0. \label{von laue condition}
\end{align}
%so \blue{$p(r)$ must change sign at least once} to ensure overall mechanical stability. Local stability further requires the normal force
This implies that, to ensure mechanical stability, the pressure must exhibit both positive and negative regions in its radial dependence, and their contributions must cancel out upon integration over space. In addition to the information on the pressure, the following constraint on its local behavior is required in any region inside the nucleon (see Ref.~\cite{Perevalova:2016dln}),
\begin{align}
    p_r(r)=\frac{2}{3} s(r) + p(r) > 0. \label{local stability condition}
\end{align}
This local constraint also ensures that the nucleon does not collapse.
%everywhere inside the nucleon (see Ref.~\cite{Perevalova:2016dln}), preventing collapse. Finally, from
Using this constraint together with Eqs.~\eqref{D-term} and \eqref{local stability condition}, it follows that the D-term takes the negative value,
%is negative,  
\begin{align}
    D < 0. 
    \label{D-term condition}
\end{align}
The negativity of the $D$-term reflects the nucleon’s mechanical stability.
%a necessary condition for the nucleon’s mechanical stability.

%The von Laue condition leads to an even more intriguing insight regarding the interplay between the repulsive and confining (attractive) pressures.~\cite{Lorce:2017xzd,Lorce:2021xku}
%\blue{Decomposing the pressure into its components shows how the von Laue condition balances these contributions to yield a net zero. } First, the spatial integral of the anomalous contribution to the energy density, $\hat{\epsilon}(r)$, is given by
Next, we present the relations obtained from the decomposed components in Eq.~\eqref{decomposed}.
Considering the spatial integral of the anomalous part $\hat\epsilon(r)=\Theta^{{\rm static}\,\mu}_\mu/4$,
one can obtain the anomalous mass $\hat M_N$ originating from the scale anomaly as
\begin{align}
    \hat M_N=4\pi\int_0^\infty drr^2\hat{\epsilon}(r)=\frac{1}{4}M_N. \label{MN/4}
\end{align}
This result implies that the scale anomaly contributes $25\%$ of the nucleon mass.
The remaining $75\%$ originates from the dynamical part,
\begin{align}
    \bar M_N=4\pi\int_0^\infty drr^2\bar{\epsilon}(r)=\frac{3}{4}M_N.
    \label{3MN/4}
\end{align}
%\blue{Given that  $4\pi\int_0^\infty drr^2 (\bar \epsilon + \hat \epsilon) =M_N$, one finds that exactly $75\%$ of the nucleon mass is carried by the dynamical part $\bar \epsilon$ and the remaining $25\%$ by the anomalous part $\hat \epsilon$.}
%\blue{On the other hand}, using Eq.~\eqref{GFFs} with $p = p'$, $s = s'$, and $A(0) = 1$, 
Furthermore, using $\Theta^{{\rm static}\,\mu}_\mu = \hat\epsilon(r)
-3\hat p(r)$,
the spatial integral of %the anomalous contribution to the pressure,
the anomalous part
$\hat{p}(r)$, reads~\cite{Lorce:2021xku,Lorce:2017xzd}
\begin{align}
    4\pi\int_0^\infty drr^2\hat{p}(r) = -\frac{1}{4}M_N.
    \label{hatp}
\end{align}
Applying the von Laue condition to the dynamical part $\bar p(r)$ leads to
\begin{align}
    4\pi\int_0^\infty dr\, r^2\, \bar p(r) = \frac{1}{4}M_N. \label{barp}
\end{align}

\subsection{
Internal force density inside the nucleon
%The force acting on quarks
}\label{force acting on quarks}

%In this work, the pressure we have discussed so far corresponds to the force acting on a hypothetical surface. 
In Ref.~\cite{Burkert:2018bqq}, the negative pressure inside the nucleon, defined in three-dimensional Cartesian coordinate, was referred to as
%was called 
“confining pressure”,  
yet it is unclear whether the negative pressure can definitively be regarded as the confining pressure that binds the quarks and gluons inside the nucleon. 
A natural approach to addressing this ambiguity would be to define a force density based on the static EMT:
\begin{equation}
  \mathcal{F}^j(\vec{r}) \;\equiv\; \partial^i {\Theta}_{ij}^{\mathrm{static}}(\vec{r}).
\end{equation}
%yet there is no obvious reason why a negative pressure acting on an hypothetical surface should imply confining pressure.  
For a spherically symmetric nucleon, this force density is expressed as follows
\begin{equation}
\begin{split}
\vec {\mathcal{F}}(r) &= \mathcal{F}^r(r) \vec e_r,\\
\mathcal{F}^r(r)
&= p_r^\prime(r)  
+\frac{2}{r}(p_r(r) -p_{\theta,\phi}(r)).
\end{split}
\end{equation}
Although the force density is directed along the radial direction, it consists of the contributions from both the radial and tangential pressures. 
Furthermore, this radial force can be decomposed into the dynamical and anomalous parts, which must cancel each other out as required by the static EMT conservation law,
\begin{equation}
\mathcal{F}^r(r) =
\bar{\mathcal{F}}^{r}(r)
+
%\bar{\mathcal{F}}^{r,q}+\bar{\mathcal{F}}^{r,g}
\hat{\mathcal{F}}^{r}(r) =0.
\label{decomposed_force}
\end{equation}
This indicates the force balance inside the nucleon, which is equivalent to Eq.~\eqref{differential equation} and represents one of the stability conditions.

\section{Skyrme model based on the scale-invariant chiral perturbation theory}\label{ch:skyrme model}
In this section, we present the analytical expressions for the nucleon’s internal distributions—energy density, pressure, and shear force—
by following the definitions given in Sec.~\ref{ch:GFFs} and using the Skyrme model based on the sChPT.
In this model framework, the lightest scalar meson field is incorporated through the low-energy theorem associated with the dilatation current, which serves as a central ingredient in decomposing the nucleon’s internal distributions in terms of the scale anomaly. 
%In this section, we extend the Skyrme model by including a scalar meson field to implement scale symmetry breaking effectively. Using the PCDC relation, we introduce the scalar field as a central ingredient. We then derive explicit formulas for the nucleon’s internal distributions—energy density, pressure, and shear force—based on the definitions given in Sec.~\ref{ch:GFFs}. 

\subsection{Model Lagrangian and PCDC relation}

To reflect the scale anomaly in the Skyrme model, we incorporate the scalar field into the ChPT Lagrangian 
%~\cite{Gasser:1983yg,Gasser:1984gg}
through the low-energy theorem. This theorem, known as the PCDC relation, connects the divergence of the dilatation current to the overlap amplitude between the broken current associated with the scale symmetry and the lightest isosinglet scalar meson $\phi$:
%the explicit breaking of QCD scale symmetry by the anomaly, we incorporate its effect into the Skyrme model using the PCDC relation. The PCDC relation connects the divergence of the dilatation current to the overlap amplitude between the broken current and the lightest isosinglet scalar meson: 
\begin{eqnarray}
\Braket{0|\partial_\mu J_D^\mu(x)|\phi(p)}
&=&
- f_\phi m_\phi^2 
e^{-ip\cdot x},
\label{PCDC}
\end{eqnarray}
where %$\phi$ is the scalar meson field, and 
$f_{\phi}$ and $m_\phi$ are the decay constant and mass of the scalar meson.
%To properly reproduce the scale properties of QCD, we extend the conventional Skyrme model by introducing a scalar meson field~\cite{}:
An effective model that satisfies this PCDC relation is given by the following Lagrangian~\cite{Lanik:1984fc,Ellis:1984jv,Leung:1989hw,Campbell:1990ak,Donoghue:1991qv,Brown:1991kk,Song:1997kx,Lee:2003eg,Park:2003sd,Park:2008zg,Li:2018gng,Crewther:2013vea,Li:2016uzn,Kasai:2016ifi,Hansen:2016fri,Appelquist:2017wcg,Appelquist:2017vyy,Cata:2019edh,Appelquist:2019lgk,Brown:2019ipr,Matsuzaki:2013eva,Zwicky:2023bzk,Zwicky:2023krx,Shifman:2023jqn}:
\begin{align}
\mathcal{L}=&\frac{f_{\pi}^{2}}{4} \left( \frac{\chi}{f_{\phi}} \right)^{2} \mathrm{Tr} \left( \partial_{\mu}U^{\dagger}\partial^{\mu}U \right) + \frac{1}{32e^{2}} \mathrm{Tr}\left( [ U^{\dagger}\partial_{\mu}U, U^{\dagger}\partial_{\nu}U ] \right) \notag \\
&+ \frac{f_{\pi}^{2} m_{\pi}^{2}}{4} \left( \frac{\chi}{f_{\phi}} \right)^{3-\gamma_{m}} \mathrm{Tr} \left( U+U^{\dagger} \right) \notag \\
&+ \frac{1}{2} \partial_{\mu} \chi \partial^{\mu} \chi - \frac{1}{4} m_{\phi0}^{2} f_{\phi}^{2}  \left(\frac{\chi}{f_{\phi}}\right)^{4} \bigg[ \mathrm{ln} \left(\frac{\chi}{f_{\phi}}\right) - \frac{1}{4} \bigg], \label{schpt lagragian}
\end{align}
where
$f_{\pi}$ and $m_{\pi}$ are the pion decay constant and mass, respectively, $e$ is the Skyrme parameter, $\gamma_m$ represents the anomalous dimension parameter within the effective model approach, and $m_{\phi0}$ corresponds to the scalar meson mass in the chiral limit.
%\blue{$U$ denotes the $\mathrm{SU}(2)$ chiral field defined by $U=\exp(i\vec \pi\cdot \vec \tau/f_\pi)$ with the pion field $\pi^a$ and the pauli matrix ($a=1,2,3$).}
%The field $\chi$ is the conformal compensator with scale dimension one, nonlinearly parametrized as $\chi = f_\phi e^{\phi/f_\phi}$.
The Lagrangian in Eq.~\eqref{schpt lagragian} is formulated as an extension of the ChPT Lagrangian typically used in the conventional Skyrme model.
The fundamental building block of ChPT is the $SU(2)$ chiral field in the nonlinear representation, defined as $U=\exp(i\vec \pi\cdot \vec \tau/f_\pi)$, where $\vec \pi$ denote the pion fields and $\vec \tau$ are the Pauli matrices in the isospin space.
The chiral field is scale invariant and carries the scale dimension of zero. The scalar field $\chi$, introduced as the extension, is the conformal compensator and is  nonlinearly parametrized as $\chi = f_\phi e^{\phi/f_\phi}$.
This $\chi$ field has the scale dimension of 1. By considering the scale transformation based on the corresponding scale dimensions, one can easily find the scale anomaly in the effective model,

\begin{equation}
\begin{split}
\partial_\mu j^\mu_D =&
-(1+\gamma_m)\frac{m_\pi^2 f_\pi^2}{4}
\left(\frac{\chi}{f_\phi}\right)^{3-\gamma_m}
{\rm tr}\left[U+U^\dagger\right]
 \\
&
-\frac{m_{\phi0}^2 f_\phi^2}{4}\left(\frac{\chi}{f_\phi}\right)^4.
\end{split}
\label{scale_anomaly}
\end{equation}
This result clearly satisfies the PCDC relation in Eq.~\eqref{PCDC}. 

To gain further insight into the effective model result for the scale anomaly, we consider taking the vacuum expectation value of the relevant operators.
Suppose that the vacuum expectation values of $U$ and $\chi$ are, respectively, given by
\begin{equation}
\begin{split}
\langle 0| U |0\rangle = 1,
\quad
\langle 0| \chi |0\rangle = f_\phi.
\end{split}
\end{equation}
From this, the vacuum expectation value of the scale anomaly is evaluated as
%\blue{Let us} clarify how the QCD scale anomaly is mimicked in our model.The divergence of the dilatation current is expressed as
\begin{align}
\Braket{0|\partial^\mu J^D_\mu(x)|0}=  - \left(1+\gamma_m\right)f_\pi^2m_\pi^2 -\frac{f_\phi^2m_{\phi0}^2}{4}. \label{sanomaly_vac}
\end{align}
Since the pion mass term is interpreted as representing the contribution from the current quark masses of QCD via Gell-Mann--Oakes--Renner relation, $f_\pi^2m_\pi^2=-m_f\braket{0|\bar qq|0}$, %~\eqref{gor relation}, 
we find that the first term in Eq.~(\ref{sanomaly_vac}) 
corresponds to
%can be identified with
the quark mass contribution to the scale anomaly in Eq.~\eqref{scaleanomaly}.
In contrast, the remaining second term in Eq.~(\ref{sanomaly_vac}) can be interpreted as corresponding to the gluonic contribution in Eq.~\eqref{scaleanomaly}.
In other words, 
the current quark mass part of the scale anomaly is described by the pion mass and its decay constant,
while the gluonic anomalous part is mimicked by the scalar meson mass and its decay constant. Hence, within the effective model, we can control the strength of each scale-anomalous part by varying the values of meson masses and decay constants.
%within our framework, $\gamma_m$ mimics the anomalous dimension associated with the current quark mass term, while the scalar meson mass and its decay constant effectively emulate the gluonic quantum corrections that appear in the scale anomaly of the underlying QCD. 
%By introducing the scalar field, \blue{we are now able to evaluate how the scale anomaly induced by the quark mass and gluons contributes to each physical quantity within the framework of the Skyrme model.} 
We emphasize that %this separation  
this separate treatment of these anomalous contributions could not be achieved within the framework of the conventional Skyrme model.

The contribution from each component of the scale anomaly is also reflected in the scalar meson mass.
By taking into account the matching between the effective model and QCD considered above, the scalar meson mass derived from the effective model can be expressed as
%We also comment on the relationship between the physical mass of the lightest scalar meson, $m_\phi$, and the parameter $m_{\phi0}$ appearing in the Lagrangian~\eqref{schpt lagragian}. \blue{The physical mass $m_\phi$ is related to $m_{\phi0}$ via}
\begin{align}
    &m_\phi^2 = m_{\phi0}^2 - (3-\gamma_m)^2 \frac{f_\pi^2 m_\pi^2}{f_\phi^2} \notag \\
    &= - \frac{4}{f_\phi^2} \langle 0 | \frac{\beta(g_s)}{2g_s} {\rm Tr}\left(G_{\mu\nu}G^{\mu\nu}\right) | 0 \rangle
    + (3-\gamma_m)^2 \frac{m_f \langle 0 | \bar{q} q | 0 \rangle}{f_\phi^2}. \label{physical scalar meson mass}
\end{align}
This shows that, in the chiral limit, the scalar meson mass is governed by the gluonic scale anomaly.
When including the current quark mass, the quark condensate contributes negatively to the scalar meson mass, resulting in its reduction.
%This relation shows that, in the chiral limit, $m_{\phi0}$ corresponds to the physical scalar meson mass and is determined solely by the gluonic quantum corrections. \blue{When the current quark masses are nonzero, the quark condensate also contributes to the scalar meson mass.}

\subsection{Classical solutions} \label{classical solutions}

To describe the nucleon based on the sChPT, we employ the skyrmion approach~\cite{Skyrme:1962vh,Adkins:1983ya}, in which the baryon is represented as a soliton solution known as the skyrmion. This skyrmion has a finite size and carries a nonzero baryon number. Hence, it can be identified as the nucleon with the baryon number one.
%\blue{In this subsection, we obtain the classical soliton solution of the Lagrangian \eqref{schpt lagragian}. As described in the next subsection, the static EMT is then constructed by inserting this classical solution into the EMT derived from this Lagrangian. }

To obtain the soliton solution, we impose the hedgehog ansatz for the chiral field $U$ and spherical symmetric configuration for the conformal compensator $\chi$ as
%symmetry for the scalar meson field as:
\begin{align}
    & U(\vec x) = \mathrm{exp} \left( i \vec{\tau} \cdot \vec{\hat x} F(r) \right) \label{hedgehog} \\ 
    & \chi(\vec x) =f_{\phi}C(r). \label{scalar meson ansatz}
\end{align}
where %\blue{$\vec{\tau}$} is the Pauli matrix in the isospin space, and 
$F(r)$ and $C(r)$ dimensionless functions. 
Substituting the ansatz into the sChPT Lagrangian and performing the spatial integration, we obtain the static energy functional as follows:
\begin{align}
E[F,C]&=4 \pi \int_{0}^{\infty} dr r^{2} \bigg[  \frac{f_{\pi}^{2}}{2} C^{2} \left( F^{\prime 2} + 2 \frac{\mathrm{sin}^{2}F}{r^{2}} \right) \notag \\
&+  \frac{\mathrm{sin}^{2}F}{2 e^{2} r^{2}} \left( \frac{\mathrm{sin}^{2}F}{r^{2}} + 2 F^{\prime 2} \right) - f_{\pi}^{2} m_{\pi}^{2} C^{3-\gamma_{m}} \mathrm{cos}F \notag \\
&+ \frac{f_{\phi}^{2}}{2} \bigg\{ C^{\prime 2} + \frac{m_{\phi 0}^{2}}{2} \left( C^{4} (\ln C-\frac{1}{4}) \right) \bigg\} - \epsilon^{\mathrm{sub}} \bigg], \label{soliton mass_scalar}
\end{align}
where 
$\epsilon^{\mathrm{sub}}$ is introduced to subtract constant contributions from the energy density.
%$\epsilon^{\mathrm{sub}}$ is the subtraction constant that is necessary to remove \blue{the divergence of the soliton mass}.
We see that 
setting $C(r)=1$, the static energy
%the scalar meson decouples at $C(r)=1$ and 
$E$ reduces to the conventional skyrmion energy. 

By minimizing the energy functional $E[F,C]$, the equations of motion (EoMs) for $F(r)$ and $C(r)$ are given by
\begin{align}
 &\left( r^{2} C^{2} + 2 \mathrm{sin}^{2}F \right) F^{\prime \prime}+ 2 \left( r C^{2} + r^{2} C C^{\prime} \right) F^{\prime} \notag \\
 &+ 2 F^{\prime 2} \mathrm{sin}F \mathrm{cos}F - 2 C^{2} \mathrm{sin}F \mathrm{cos}F- 2 \frac{ \mathrm{sin}^{3}F \mathrm{cos}F}{r^{2}} \notag \\
 &\hspace{28mm}- \left( \frac{m_{\pi}}{e f_{\pi}} \right)^{2} r^{2} C^{3-\gamma_{m}} \mathrm{sin}F = 0, \label{eom_chiral field} \\
 &C^{\prime \prime} + \frac{2}{r} C^{\prime} - \left( \frac{f_{\pi}}{f_{\phi}} \right)^{2} C \left( F^{\prime 2} + 2 \frac{\mathrm{sin}^{2}F}{r^{2}} \right) \notag \\
 &+ (3-\gamma_{m}) \left( \frac{m_{\pi}}{e f_{\phi}} \right)^{2} C^2 \mathrm{cos}F - \left( \frac{m_{\phi 0}}{e f_{\pi}} \right)^{2} C^{3} \ln C =0. \label{eom_scalar field}
\end{align}
To construct a soliton with baryon number one, we impose the boundary condition on the profile function $F(r)$ as follows:
\begin{align}
    F(0)=\pi, \quad F(r \rightarrow \infty)=0. \label{boundary condition of f}
\end{align}
The boundary conditions for $C(r)$ are chosen so that the soliton solution 
is regular at $r=0$ and its energy
finite:
%is finite and the ansatz is regular at $r=0$,
\begin{align}
    \frac{dC}{dr} \bigg\lvert_{r=0} =0, \quad C(r \rightarrow \infty)=C_\infty, \label{boundary condition of C}
\end{align}
where $C_\infty $ is %a stationary point 
determined by the stationary condition in the vacuum,
%the follwoing equation:
\begin{align}
    0 =& \frac{dV(F,C)}{dC} \bigg|_{r \rightarrow \infty} \notag \\ 
    =& (3-\gamma_{m}) f_{\pi}^{2} m_{\pi}^{2} - f_{\phi}^{2} m_{\phi 0}^{2} C_{\infty}^{4} \mathrm{ln} C_{\infty}. \label{stationary point}
\end{align}

In the following, we briefly provide the practical techniques used in the numerical calculation.
To accurately evaluate the GFFs (Eq.~\eqref{D(t)}) and to numerically satisfy global relations—such as the von Laue condition (Eq.~\eqref{von laue condition}) and other relations (Eqs.~\eqref{MN/4}–\eqref{barp})—it is necessary to integrate over $r$ up to infinity. However, 
in the numerical calculations, it is not possible to solve  
Eq.~\eqref{eom_chiral field} and \eqref{eom_scalar field}
up to spatial infinity and the computation must be truncated at a finite value of $r$.
In addition, even with truncation, it is still difficult to capture the behavior of the profile functions $F(r)$ and $C(r)$ in the asymptotic region far from $r=0$.
To analyze their asymptotic behaviors, it is useful to employ the Pad\'{e}-like approximate solutions~\cite{Ponciano:2004cs}, given as follows
%the numerical solutions of Eq.~\eqref{eom_chiral field} and \eqref{eom_scalar field} are only available up to a finite $r$. 
%It is therefore essential to construct analytic approximations of the profile functions $F(r)$ and $C(r)$. 
%Following Ref.~\cite{Ponciano:2004cs}, we employ the Pad\'{e}-like approximate solutions for $F(r)$ and $C(r)$ as follows:
\begin{align}
    F(r) &= \frac{\pi + \sum_{k=1}^{n} a_{k}r^{k}}{1+\sum_{k=1}^{n} b_{k}r^{k} + b_{n+1}r^{n+1} \exp(m_{\pi}r)}, \label{F asymptotic form} \\
    C(r) &= C_\infty  - \frac{\sum_{k=0}^{n}c_{k}r^{k}}{1+\sum_{k=1}^{n} d_{k}r^{k} + d_{n+1}r^{n+1} \exp(m_{\phi0}r)}, \label{C asymptotic form}
\end{align}
where $a_k$, $b_k$, $c_k$, and $d_k$ are determined by fitting the above ansatz to the numerical solutions for $F(r)$ and $C(r)$. 
These approximate analytical solutions exactly satisfy the boundary conditions in Eq.~\eqref{boundary condition of f} and \eqref{boundary condition of C}.
%, at $r=0$ and reproduce the correct asymptotic behaviors $\exp(-m_\pi r)/r$ and $C_\infty [1-\exp(-m_{\phi0}r)/r]$ as $r\to\infty$, which are consistent with those derived from the EoMs, Eqs.\eqref{eom_chiral field} and \eqref{eom_scalar field}. 
We find that $n=3$ is already sufficient to reproduce all global quantities with high accuracy.
%\blue{the full numerical solutions} with high accuracy for all global relations. 

\subsection{The static EMT from the Skyrme model}

%\blue{In the Skyrme model, the nucleon emerges not as an elementary field but as a quantized soliton. Consequently, the EMT matrix element~\eqref{GFFs} must be reformulated as a quantum-mechanical matrix element between eigenstates of \blue{the soliton's collective coordinates.} }

%\blue{The static classical solutions of the soliton, $U_{\rm cl}$ and $\chi_{\rm cl}$, are promoted to dynamical fields by introducing the collective coordinates for translation and rotation, $(\vec{X}(t), A(t))$, as}

As we have seen in the previous subsection, a baryon is described as a soliton solution of the $U$ and $\chi$ fields. These solutions do not carry spin and isospin quantum numbers, which are referred to as the classical solutions, $U_{\rm cl}(\vec x)$ and $\chi_{\rm cl}(\vec x)$. This may raise concerns about whether the static EMT can be appropriately described using classical solutions. Nevertheless, within the large $N_c$ limit, the static EMT can still be expressed at the classical level, and the GFF $D(t)$ can also be evaluated accordingly~\cite{Carson:1991fu,GarciaMartin-Caro:2023klo}.   
In this subsection, we begin with the quantization procedure for the skyrmion and demonstrate that the static EMT can be reduced to a form that is sufficiently described at the classical level.

To assign spin and isospin to the skyrmion, one performs collective quantization of the classical solitons by introducing translation and rotation:  
\begin{align}
    &U(t,\, \vec{x}) = A(t)\, U_{\rm cl}[R(B(t))(\vec{x} - \vec{X}(t))]\, A^\dagger(t), \\
    &\chi(t,\, \vec{x}) = \chi_{\rm cl}(R(B(t))(\vec{x} - \vec{X}(t))),
\end{align}
where the translatoinal, isospin and rotational zero mode is parametrized by the vector $\vec{X}$ and $SU(2)$ matrices $A(t)$ and $B(t)$, respectively with $R_{ij}(B(t))={\rm Tr}[\tau^iB\tau^jB^\dagger]/2\in SO(3)$.
Note that here, $t$ denotes the time component.
By treating the collective coordinates as the canonical variables, the quantum numbers are assigned to the skyrmion. 
Within the Skyrme model, the matrix element of $\Theta_{ij}$ is then expressed in the Breit frame as follows,
%\blue{Using the collective coordinates of the soliton, the matrix element of $\Theta_{ij}$ for the skyrmion is expressed in the Breit frame as follows,}}
%\blue{In the Breit frame, the quantum-mechanical matrix elements of the spatial component of the EMT between momentum eigenstates for the soliton are then evaluated as}
\begin{align}
    &\big<\vec{q}/2\big|\Theta_{ij}[U(t,\vec{x}),\chi(t,\vec{x})]\big|-\vec{q}/2\big> \notag \\
    &=\exp[-i\vec{q}\cdot\vec{x}]R(B)^T_{ia}R(B)^T_{jb}\int d{x^\prime}^3\exp[i\vec{q}\cdot R(B)^T\vec{x^\prime}] \notag \\
    &\hspace{40mm}\times\Theta_{ab}[U(t,\vec{x}),\chi(t,\vec{x})]. \label{QM_Matrix}
\end{align}

Since the contribution of collective rotations generally appears at subleading order in the large-$N_c$ expansion, we approximate the EMT with collective rotations as
\begin{align}
    \Theta_{ij}[U(t,\vec{x}),\chi(t,\vec{x})] \simeq \Theta^{\rm cl}_{ij}[U_{\rm cl}(\vec{x}),\chi_{\rm cl}(\vec{x})].
\end{align}

Under this approximation, the GFF $D(t)$ obtained from the matrix elements in Eq.~\eqref{QM_Matrix} reduce to the expression
\begin{align}
    D(t) \simeq -6 M_{N} \int d^{3}x\, \left( r^{i} r^{j} - \frac{1}{3} \delta^{ij} r^{2} \right) \frac{j_{2}(\Delta r)}{(\Delta r)^{2}}\, \Theta^{\mathrm{cl}}_{ij}.
\end{align}
where the nonrelativistic limit $E \rightarrow M_N$ is taken,
%~\cite{Carson:1991fu,Cebulla:2007ei}
and comparing with the general expression for the GFF $D(t)$ in Eq.~\eqref{D(t)}, the static EMT tensor within the skyrmion approach reads
%式(10)と見比べてと入れる？
%in the nonrelativistic limit $E \rightarrow M_N$~\cite{Carson:1991fu,Cebulla:2007ei}. By comparing with Eq.~\eqref{D(t)}, the static EMT~\eqref{staticEMT} is approximated as
\begin{align}
    \Theta^{\rm static}_{ij}(r)\simeq\Theta^{\rm cl}_{ij}[U_{\rm cl}(\vec{x}),\chi_{\rm cl}(\vec{x})].
\end{align}
The time component of the static EMT is also expressed as
$\Theta^{\rm static}_{00}(r)\simeq\Theta^{\rm cl}_{00}[U_{\rm cl}(\vec{x}),\chi_{\rm cl}(\vec{x})]$ within the leading order approximation in the large-$N_c$ expansion.
%A similar argument also applies to the 00-component of the static EMT. Therefore, in this work, we proceed by identifying $\Theta^{\rm static}_{\mu\nu}(r)$ with $\Theta^{\rm cl}_{\blue{\mu\nu}}[U_{\rm cl}(\vec{x})]=\Theta^{\rm cl}_{ij}[F(r),C(r)]$.

In general, the EMT is obtained as $T_{\mu\nu} = -\frac{2}{\sqrt{-g}} \frac{\delta (\mathcal{L} \sqrt{-g})}{\delta g^{\mu\nu}}$. However, this EMT in sChPT 
does not reproduce the scale anomaly in Eq.~\eqref{scale_anomaly}. To reconcile this discrepancy, 
an improved term should be added to the %canonical
EMT, which is given by $\theta_{\mu\nu} = -\left( \partial_\mu \partial_\nu - g_{\mu\nu} \partial_\rho \partial^\rho \right) \chi^2/6$.
Accordingly, the improved EMT in sChPT is evaluated as

%\blue{To obtain $\Theta^{\rm cl}_{ij}[F(r),C(r)]$, we define the EMT from the Lagrangian~\eqref{schpt lagragian}. The improved EMT is constructed from the canonical EMT, $T_{\mu\nu} = -\frac{2}{\sqrt{-g}} \frac{\delta (\mathcal{L} \sqrt{-g})}{\delta g^{\mu\nu}}$, and the improved term, $\theta_{\mu\nu} = -\left( \partial_\mu \partial_\nu - g_{\mu\nu} \partial_\rho \partial^\rho \right) \chi^2/6$, as}
\begin{widetext}
\begin{align}
    &\Theta_{\mu\nu} = T_{\mu\nu} + \theta_{\mu\nu} = \bar{\Theta}_{\mu\nu} + \hat{\Theta}_{\mu\nu} %- \epsilon^{\rm sub}_{\mu \nu}, 
    \\   &\bar{\Theta}_{\mu\nu}=\frac{f_\pi^2}{2}\left(\frac{\chi}{f_\phi}\right)^2{\rm Tr}\left(\partial_\mu U^\dagger\partial_\nu U\right)+\frac{1}{8e^2}{\rm Tr}\left(\left[U^\dagger\partial_\mu U,U^\dagger\partial_\rho U\right]\left[U^\dagger\partial_\nu U,U^\dagger\partial^\rho U\right]\right)+\frac{2}{3}\partial_\mu\chi\partial_\nu\chi-\frac{1}{3}\chi\partial_\mu\partial_\nu\chi \notag \\
    & \hspace{12mm}- \frac{1}{4} g_{\mu\nu}\left\{\frac{f_\pi^2}{2}\left(\frac{\chi}{f_\phi}\right)^2{\rm Tr}\left(\partial_\rho U^\dagger\partial^\rho U\right)+\frac{1}{8e^2}{\rm Tr}\left[U^\dagger\partial_\rho U,U^\dagger\partial_\sigma U\right]^2+\frac{2}{3}\partial_\rho\chi\partial^\rho\chi-\frac{1}{3}\chi\partial_\rho\partial^\rho\chi\right\}, \label{barEMTsChPT} \\
    &
\hat{\Theta}_{\mu\nu}=\hat{\Theta}_{\mu\nu}^{q}+
\hat{\Theta}_{\mu\nu}^g,
\\
&
\hat{\Theta}_{\mu\nu}^{q}=
-\frac{1}{4}g_{\mu\nu} (1+\gamma_{m}) \frac{f_\pi^2m_\pi^2}{4}\left(\frac{\chi}{f_\phi}\right)^{3-\gamma_{m}}{\rm Tr}\left(U+U^{\dagger}\right),\\
&
\hat{\Theta}_{\mu\nu}^{g}=
-\frac{1}{4}g_{\mu\nu}\frac{m_{\phi 0}^{2}f_\phi^2}{4}\left(\frac{\chi}{f_\phi}\right)^4.
%\blue{\hat{\Theta}_{\mu\nu}=-\frac{1}{4}g_{\mu\nu}\left\{ (1+\gamma_{m}) \frac{f_\pi^2m_\pi^2}{\red{4}%2}\left(\frac{\chi}{f_\phi}\right)^{3-\gamma_{m}}{\rm Tr}\left(U+U^{\dagger}\right)+\frac{m_{\phi 0}^{2}f_\phi^2}{4}\left(\frac{\red{\chi}}{f_\chi}\right)^4\right\}}. 
\label{hatEMTsChPT}
\end{align}
\end{widetext}
%where $\epsilon^{\rm sub}_{\mu \nu}$ denotes the subtraction term introduced to ensure that the EMT density \red{is normalized to} vanish at \red{the vacuum.}
%spatial infinity. 
By substituting the classical solutions into the improved EMT, we obtain the static EMT in the skyrmion approach $\Theta^{\rm cl}_{\mu\nu}[F(r),C(r)]$. 
The decomposed components of the energy density and stress distribution
are given by
%From the components of the improved EMT, we obtain the following expressions for the energy density, pressure, and shear force: 
\begin{widetext}
\begin{align}
    &\bar{\epsilon}(r) =3\bar p(r) = \frac{f_\pi^2}{4}C^2\left(F'^2+2\frac{\sin^2F}{r^2}\right)+\frac{\sin^2F}{2e^2r^2}\left(
2F'^2+\frac{\sin^2F}{r^2}\right)+\frac{f_\phi^2}{6}C'^2{-\frac{f_\phi^2}{6}\frac{CC'}{r}}-\frac{f_\phi^2}{12}CC'',
\label{energy density1}
%\notag 
\\ 
    &\hat{\epsilon}(r)=-\hat p(r)= \hat{\epsilon}^q(r) + \hat{\epsilon}^g(r)=-\hat{p}^q(r) - \hat{p}^g(r),
     \label{energy density2} 
    \\
    &\hat{\epsilon}^q(r)=-\hat p^q(r) = -\frac{1}{4}(1+\gamma_{m})f_\pi^2m_\pi^2C^{3-\gamma_{m}}\cos F
    -\hat{\epsilon}^{q,\rm{sub}},
    \label{energy density3}
    \\   
    &\hat{\epsilon}^g(r) =-\hat p^g(r)= -\frac{1}{4}\left[\frac{1}{4}m_{\phi 0}^{2}f_\phi^2C^4\right]
    -\hat\epsilon^{g,\rm{sub}},
    \label{energy density4}\\%\blue{\hat{\epsilon}(r) = -\frac{1}{4}\left[(1+\gamma_{m})f_\pi^2m_\pi^2C^{3-\gamma_{m}}\cos F(r)+\frac{1}{4}m_{\phi 0}^{2}f_\phi^2C^4\right]}
    %&\bar{p}(r) =  \frac{F_\pi^2}{12}C^2\left(F'^2+2\frac{\sin^2F}{r^2}\right)+\frac{\sin^2F}{6e^2 r^2}\left(2F'^2+\frac{\sin^2F}{r^2}\right)+\frac{f_\chi^2}{18}C'^2{-\frac{f_\chi^2}{18}\frac{CC'}{r}}-\frac{f_\chi^2}{36}CC'' \notag \\    
    %&\blue{\hat{p}(r) = \frac{1}{4}\left[(1+\gamma_{m})f_\pi^2m_\pi^2C^{3-\gamma_{m}}\cos F(r)+\frac{1}{4}m_{\phi0}^{2}f_\phi^2C^4\right] }
%\label{pressure} \\
    &
    \bar s(r)=\left(f_\pi^2C^2+\frac{\sin^2F}{e^2r^2}\right)\left(F'^2-\frac{\sin^2F}{r^2}\right)+\frac{f_\phi^2}{3}\left(2C'^2+\frac{CC'}{r}-CC''\right), \label{shear force}
\end{align}
\end{widetext}
where $\hat\epsilon^{q(g),{\rm sub}}$ denotes the subtraction term introduced to ensure that the EMT density is normalized to vanish at spatial infinity.
%\blue{where we again note that the subtraction term is required to ensure that these distributions vanish at spatial infinity.}
These decomposed components show that the trace part of $\hat\epsilon$ ($\hat p$) is further decomposed into the current quark mass part of the scale anomaly $\hat\epsilon^q$ ($\hat p^q$) 
and the gluonic anomalous part $\hat\epsilon^g$ ($\hat p^g$). 
This further decomposition is thanks to the distinctive features of sChPT, where the scale anomaly can be separated into contributions from the pion mass and the scalar meson mass, as was discussed around Eq.~\eqref{sanomaly_vac}.

In the skyrmion approach, 
the nucleon mass $M_N$ is obtained by integrating the energy density over space, following the general expression in Eq.~\eqref{MN}.
%\blue{\begin{equation}
%\begin{split}
%M_N &= \bar M_N +
%\hat M_N,\\
%\bar M_N&= 4\pi\int_0^\infty drr^2 \bar %\epsilon,\\ 
%\hat M_N&= 4\pi\int_0^\infty drr^2 \hat %\epsilon.
%\end{split}
%\end{equation}}
%the nucleon mass can also be evaluated from the scale anomaly as
%$M_N= 4\pi\int_0^\infty drr^2 \Theta^{\rm{cl}\,\mu}_\mu$, if von Laue condition in Eq.~\eqref{von laue condition} is satisfied. 
%With $\hat \epsilon =-\hat p$ in Eq.~\eqref{energy density2} , as estimated from the above equations, the nucleon mass can be rewritten as $M_N= 4\hat M_N$.This implies that the anomalous contribution accounts for $25\%$ of the nucleon mass, consistent with the general discussion in Eq.~\eqref{MN/4}.
The decomposed energies obtained in the skyrmion approach clearly indicate that the anomalous mass $\hat M_N$ indeed accounts for $25\%$ of the nucleon mass. 
Interestingly, the anomalous mass consists of two parts: the current quark mass part and the gluonic anomalous part,
\begin{equation}
\begin{split}
\hat M_N&=
\hat M_N^{q} + \hat M_N^{g}
,\\
 \hat M_N^{q(g)}&= 4\pi\int_0^\infty drr^2 \hat \epsilon^{q(g)}.
\end{split}
\end{equation}
The detailed amounts of the anomalous contributions will be evaluated in the next section.

\section{Results}\label{ch:results}
In this section, we numerically elucidate the influence of the scale anomaly on the energy density, the pressure, the force density and the GFF $D(t)$. 
Since our aim is to evaluate the contribution of the scale anomaly, we fix the pion mass to its physical value and vary the scalar meson mass, considering three different values: $m_{\phi0} = 550,\,720,\,1000~\mathrm{MeV}$ (the strength of the gluonic scale anomaly can be controlled by the scalar meson mass.).
As for the decay constant of the scalar meson, it is fixed at $f_\phi=240~\mathrm{MeV}$, which provides a reasonable description of the skyrmion properties within sChPT with $m_{\phi0}=720~\mathrm{MeV}$~\cite{Lee:2003eg}.

Additionally, since the anomalous dimension also affects the strength of the scale anomaly, we also vary its value.
In Ref.~\cite{Crewther:2013vea}, the value of $\gamma_m$ in the effective model is discussed to lie within the range $-1<\gamma_m\le 2$. Following this range, we consider $\gamma_m = -0.9,\;0,\;2$.

In the skyrmion approach, the parameters $f_\pi$and $e$  are highly sensitive to the nucleon mass and other properties. Therefore, we adopt two parameter sets.
One set, $f_\pi=68~\mathrm{MeV}$ and $e=5.45$, is chosen to reproduce the nucleon and $\Delta$ masses and obtain a realistic soliton size~\cite{Adkins:1983ya},  though the pion decay constant deviates from the physical value. 
The other set, $f_\pi = 93~\mathrm{MeV}$ and $e=4.75$, is fixed to the physical value of the pion decay constant, with the value of $e$ determined by referring to the axial coupling $g_A$~\cite{Lee:2003eg,Brown:1984sx}.

%\subsection{Model parameters}

%\blue{The model contains six parameters: $m_\pi$, $f_\pi$, $e$, $f_\phi$, $m_{\phi0}$, and $\gamma_m$. We adopt the empirical pion mass $m_\pi = 138~\mathrm{MeV}$ and explore the remaining parameters as follows.}

%To reproduce the nucleon and $\Delta$ masses and obtain a realistic soliton size, we follow Ref.~\cite{Adkins:1983ya} and adopt $f_\pi=68~\mathrm{MeV}$ and $e=5.45$. The physical value may not yield optimal agreement with hadron observables, since $f_\pi$ and $e$ serve as effective energy and length scales for the soliton. However, for comparison, we fix the alternative set $f_\pi = 93~\mathrm{MeV}$ and $e=4.75$ from the axial coupling $g_A$, as adopted in Refs.~\cite{Lee:2003eg,Brown:1984sx}.

%For the lightest scalar meson, we adopt $m_{\phi0}=720~\mathrm{MeV}$ and $f_\phi=240~\mathrm{MeV}$,  following Ref.~\cite{Lee:2003eg}. These values are chosen in order to reasonably reproduce the skyrmion properties within sChPT and to effectively account for gluonic quantum corrections. To investigate how the stress distribution responds to the scalar-meson mass that controls the strength of the scale anomaly, we examined three values:
%\begin{align}
%    m_{\phi0} = 550,\,720,\,1000~\mathrm{MeV}.
%\end{align}
%Finally, to explore the effect of the quark-mass anomalous dimension, we consider the values $\gamma_m = -0.9,\;0,\;2$ within the range $-1<\gamma_m\le 2$ discussed in Ref.~\cite{Crewther:2013vea}.

\subsection{Energy density and mass decomposition}

\begin{figure}
    \includegraphics[scale=0.4]{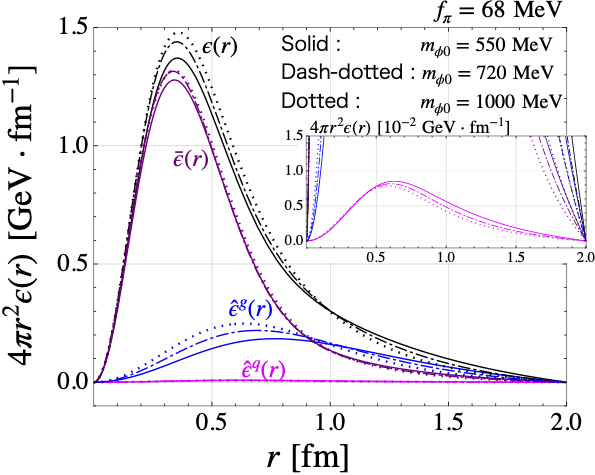}
    \caption{
    %Energy density $\epsilon(r)$ inside the nucleon for $f_\pi = 68~\mathrm{MeV}$ and its decomposition into dynamical [$\bar{\epsilon}(r)$], quark mass [$\hat{\epsilon}^q(r)$], and gluonic anomaly [$\hat{\epsilon}^g(r)$] contributions. Solid, dash-dotted, and dotted lines correspond to $m_{\phi 0} = 550,\ 720,\ 1000~\mathrm{MeV}$, respectively.
    Spatial distribution of the energy density inside nucleon for the case of $f_\pi = 68~\mathrm{MeV}$ and $\gamma_m=0$, and its decomposition.
    The dynamical part $\bar{\epsilon}(r)$, the current quark mass part of the scale anomaly  $\hat{\epsilon}^q(r)$, and that of the gluonic anomaly part $\hat{\epsilon}^g(r)$ are depicted in purple, magenta, and blue, respectively. Their sum $\epsilon(r)$ is represented by the black line. The solid, dash-dotted, and dotted lines indicate the results for $m_{\phi 0} = 550~\mathrm{MeV}$, $720~\mathrm{MeV}$, and $1000~\mathrm{MeV}$, respectively. An inset shows a close-up view of $\hat{\epsilon}^q(r)$.
    }
    \label{FIG1_energy_density_68}
\end{figure}

\begin{figure}
    \includegraphics[scale=0.4]{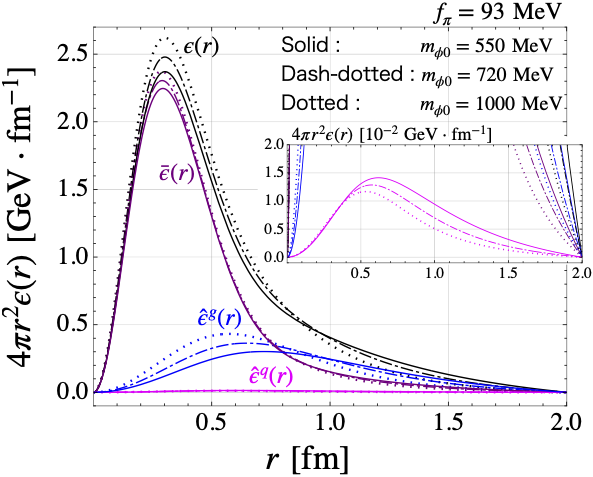}
    \caption{Same as Fig.~\ref{FIG1_energy_density_68}, but for $f_\pi = 93~\mathrm{MeV}$ and $\gamma_m=0$. %The energy density becomes slightly more localised at the core.
    }
    \label{FIG2_energy_density_93}
\end{figure}

\begin{figure}
    \includegraphics[scale=0.4]{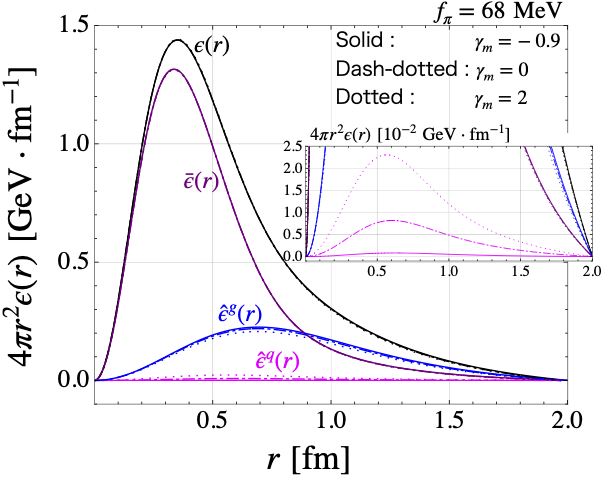}
    \caption{
    Impact of $\gamma_m$ on the energy density for $f_\pi = 68~\mathrm{MeV}$. The scalar meson mass is fixed at $m_{\phi0}=720~\mathrm{MeV}$.
    %The decomposition of energy density for $f_\pi = 68~\mathrm{MeV}$ with various values of the anomalous dimension $\gamma_m = -0.9,\ 0,\ 2$. %The quark mass contribution scales with $1 + \gamma_m$, while the gluonic anomaly term adjusts accordingly.
    }
    \label{FIG3_energy_density_ad_68}
\end{figure}

%We begin by presenting
In this subsection, we present the spatial distribution of 
the energy density  and its decomposition based on Eqs.~(\ref{energy density1}, \ref{energy density2}, \ref{energy density3}, \ref{energy density4}).
%for the case $\gamma_m = 0$. 
%Figure~\ref{FIG1_energy_density_68} shows

First, we display the result for $f_\pi = 68~\mathrm{MeV}$ and $\gamma_m = 0$  in Fig.~\ref{FIG1_energy_density_68}. 
This figure shows that the energy density is dominated by the dynamical one, while the contributions from the scale anomaly are subleading.
We also investigate the effect of the scalar meson mass on the energy density, noting that changes in this mass effectively modulate the strength of the gluonic scale anomaly.
In the region near the origin ($r=0$), increasing the scalar meson mass leads to a larger energy density. However, in the asymptotic region far from $r=0$, a larger scalar meson mass causes the energy density to fall off more rapidly toward zero.
This trend stems from the behavior of the dynamical energy density $\bar\epsilon$ and the gluonic anomalous part $\hat\epsilon^g$. Although the current quark mass part of the scale anomaly $\hat \epsilon^q$ behaves oppositely, its magnitude is negligible.
%\blue{The dynamical energy contribution $\bar{\epsilon}(r)$, the quark mass contribution $\hat{\epsilon}^q(r)$, and the gluonic anomaly contribution $\hat{\epsilon}^g(r)$ are depicted in purple, magenta, and blue, respectively, while the black line represents their sum. The solid, dash-dotted, and dotted lines correspond to the results for $m_{\phi 0} = 550~\mathrm{MeV}$, $720~\mathrm{MeV}$, and $1000~\mathrm{MeV}$, respectively. An inset shows a close-up view of $\hat{\epsilon}^q(r)$.} 

%\blue{The behavior of the energy density under variation of the parameter $m_{\phi 0}$, which controls the strength of the gluonic anomaly, is as follows. In the central region of the nucleon (small-$r$), each contribution to the energy density becomes larger as $m_{\phi 0}$ increases. In contrast, near the nucleon surface (large-$r$), each contribution falls off more rapidly for larger $m_{\phi 0}$, in accordance with the asymptotic behavior of $C(r)$. In particular, the gluonic anomaly contribution becomes more spatially extended as $m_{\phi 0}$ decreases. This feature will play a more significant role in the discussion of the pressure distribution.}

Using the evaluated energy density, we estimate the nucleon mass. In the current  setup with $f_\pi = 68~\mathrm{MeV}$ and $\gamma_m = 0$,
the values of $M_N$ corresponding to each scalar meson mas are listed in Table~\ref{tab1}. 
The values of $M_N$ for all scalar meson masses are close to the empirical value.
Regarding the anomalous part $\hat M_N$, which should account for $25\%$ of $M_N$, the gluonic anomalous mass $\hat M_N^g$ is dominant, comprising about $24\%$ of $M_N$.

%\blue{The nucleon mass is obtained by integrating the energy density. Table~\ref{tab1} summarizes the total mass, $\int d^3x\epsilon(r)$, and the gluonic anomaly contribution, $\int d^3x\hat{\epsilon}^g(r)$. In this case, the total mass is close to the empirical value.}

We now turn to the alternative setup with $f_\pi = 93~\mathrm{MeV}$ and $\gamma_m=0$,  and present the result in Fig.~\ref{FIG2_energy_density_93}.
Qualitatively, the behavior is similar to that observed in the previous setup with  $f_\pi = 68~\mathrm{MeV}$ and $\gamma_m=0$, shown in Fig.~\ref{FIG1_energy_density_68}.
In contrast, the spatial distribution of the energy density shows enhancement near the origin in the current setup, resulting in a larger nucleon mass. Nevertheless, 
the gluonic anomalous mass $\hat M_N^g$ remains nearly unchanged, accounting for approximately $24\%$ of $M_N$, as shown in Table~\ref{tab1}.

%\blue{Figure~\ref{FIG2_energy_density_93} shows the energy density for $f_\pi = 93~\mathrm{MeV}$. While the qualitative behavior remains similar, the energy density becomes slightly more localized at the center, leading to a higher nucleon mass (see Table~\ref{tab1}).}

We also consider the impact of $\gamma_m$ on the spatial distribution of the energy density.
Figure~\ref{FIG3_energy_density_ad_68} shows the results for varying  $\gamma_m = -0.9,\ 0,\ 2$, while $f_\pi$ and $m_\phi$ are fixed at $f_\pi = 68~\mathrm{MeV}$ and $m_{\phi0} =720~\mathrm{MeV}$. Although varying $\gamma_m$ does not significantly affect the overall energy density, an increase in 
$\gamma_m$ slightly enhances $\hat \epsilon^q(r)$ while reducing $\hat \epsilon^g(r)$. This is because the current quark mass part of the scale anomaly $\hat \epsilon^q(r)$ scales with the anomalous dimension $\gamma_m$, while
the gluonic anomalous part $\hat \epsilon^g(r)$ correspondingly diminishes to preserve the total anomalous mass proportion of $25\%$.

Incidentally, using the obtained energy density, we also numerically evaluate the mass radius, which is defined as
\begin{equation}
 \sqrt{\braket{r^2}^{\rm mass}}=\left( \frac{\int d^{3}x r^{2} \epsilon(r)}{\int d^{3}x \epsilon(r)} \right)^{1/2}.
\end{equation}
For the case of $f_\pi=68~{\rm MeV}$ with $m_{\phi0}=720~{\rm MeV}$ and $\gamma_m=0$, 
the mass radius is estimated to be $\sqrt{\braket{r^2}^{\rm mass}}=0.74 \ {\rm fm}$. This evaluation is qualitatively consistent with lattice results~\cite{Hackett:2023rif}.
Results for other parameter sets are listed in Table~\ref{tab1}.

%\blue{Figure~\ref{FIG3_energy_density_ad_68} shows the results for $\gamma_m = -0.9,\ 0,\ 2$ with $f_\pi = 68~\mathrm{MeV}$. The overall distribution remains largely unchanged. As expected from Eq.~\eqref{energy density}, the magnitude of $\hat{\epsilon}^q(r)$ is rescaled by $1 + \gamma_m$, while $\hat{\epsilon}^g(r)$ adjusts in the opposite direction to preserve the constraint~\eqref{MN/4}.}

\begin{table*}[t!]
\centering
\caption{Summary of calculated nucleon observables for various parameter sets $(f_\pi, m_{\phi0})$ and anomalous dimension $\gamma_m$. Listed are the nucleon mass $M_N$, the gluonic anomaly contribution to the mass (in percent), the D-term value $D(0)$, the mass radius $\sqrt{\braket{r^2}^{\mathrm{mass}}}$, and the mechanical radius $\sqrt{\braket{r^2}^{\mathrm{mech}}}$.}
\begin{tabular}{lcccccc}
\hline\hline
$(f_\pi, \ m_{\phi 0})/{\rm MeV}$ & Anomalous dim. $\gamma_m$ & $M_N/{\rm MeV}$ & Gluonic anomaly (\%) & D-term: $D=D(0)$ & $\sqrt{\braket{r^2}^{\rm mass}}/{\rm fm}$ & $\sqrt{\braket{r^2}^{\rm mech}}/{\rm fm}$ \\
\hline
$(68, \ 550)$ & 0 & 923.6 & 24.1\% & -5.69 & 0.81 & 0.82 \\
$(68, \ 720)$ & 0 & 927.6 & 24.2\% & -4.12 & 0.74 & 0.74 \\
$(68, \ 1000)$ & 0 & 932.5 & 24.2\% & -3.03 & 0.68 & 0.72 \\
\hline
$(93, \ 550)$ & 0 & 1397.4 & 24.0\% & -12.14 & 0.75 & 0.76 \\
$(93, \ 720)$ & 0 & 1409.1 & 24.2\% & -8.64 & 0.69 & 0.66 \\
$(93, \ 1000)$ & 0 & 1425.4 & 24.3\% & -5.69 & 0.61 & 0.62 \\
\hline
$(68, \ 720)$ & -0.9 & 928.8 & 24.9\% & -4.19 & 0.75 & 0.75 \\
$(68, \ 720)$ & 2 & 925.3 & 22.8\% & -3.98 & 0.76 & 0.71 \\
\hline\hline
\end{tabular}
  \label{tab1}
\end{table*}

\subsection{Pressure and its decomposition}
\label{Pressure_subsec}

\begin{figure}
\begin{center}
    \includegraphics[scale=0.4]{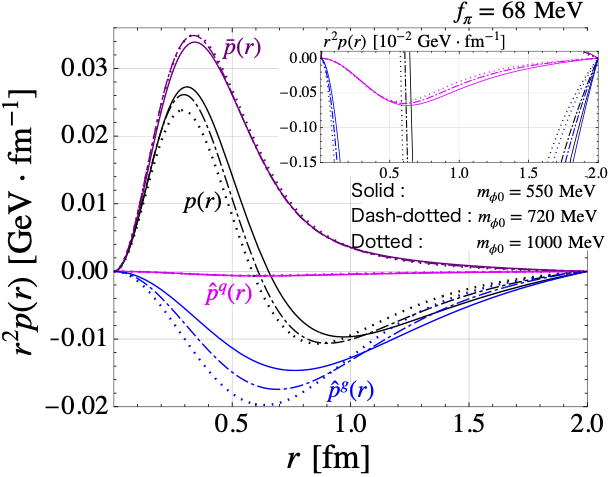}
    %\subfigure{ (a) }
    %\includegraphics[scale=0.28]{FIG4_force_on_quark_68.png}
    %\subfigure{ (b) }
\end{center}
    \caption{
    Same as Fig.~\ref{FIG1_energy_density_68}, but showing the pressure instead of the energy density.
    %Spatial distribution of the pressure inside the nucleon for $f_\pi = 68~\mathrm{MeV}$ and $\gamma_m = 0$, and its decomposition.The dynamical part $\bar{p}(r)$, the anomalous part of the quark contribution $\hat{p}^q(r)$, and that of the gluonic contribution $\hat{p}^g(r)$ are depicted in purple, magenta, and blue, respectibely.Their sum $p(r)$ is represented by the black line. An inset shows an close-up view of $\hat{p}^q(r)$.
    %\blue{(a) Pressure distribution $p(r)$ inside the nucleon for $f_\pi = 68~\mathrm{MeV}$ and $\gamma_m = 0$, and its decomposition into $\bar{p}(r)$, $\hat{p}^q(r)$, and $\hat{p}^g(r)$. (b) Radial distribution of the force on quarks arising from the scale anomaly. A reversal in the force direction reflects the development of an attractive pressure that confines quarks inside the nucleon.}
    }
    \label{FIG4_pressure_68}
\end{figure}

\begin{figure}
\begin{center}
    \includegraphics[scale=0.4]{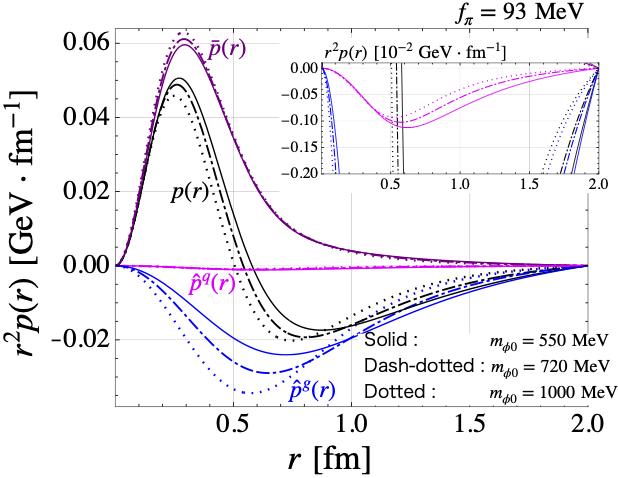}
    %\subfigure{ (a) }
    %\includegraphics[scale=0.28]{FIG5_force_on_quark_93.png}
    %\subfigure{ (b) }
\end{center}
    \caption{
    Spatial distribution of the pressure in the alternative setup with $f_\pi = 93~\mathrm{MeV}$ and $\gamma_m=0$.
    %Same as Fig.~\ref{FIG4_pressure_68}, but for $f_\pi = 93~\mathrm{MeV}$ \red{and $\gamma_m=0$}. %The pressure is slightly more concentrated toward the core.
    }
    \label{FIG5_pressure_93}
\end{figure}

\begin{figure}
    \includegraphics[scale=0.4]{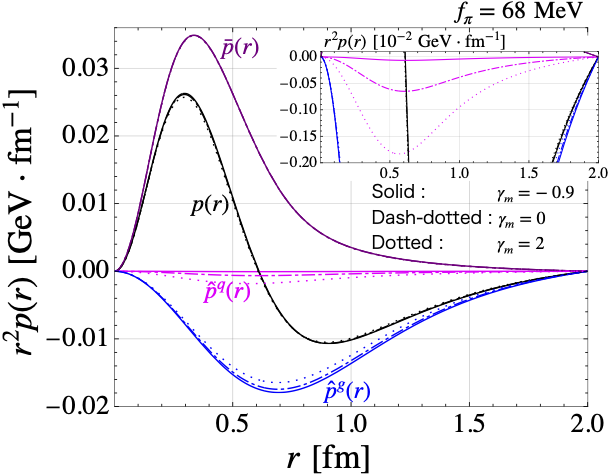}
    \caption{
    Same as in Fig.~\ref{FIG3_energy_density_ad_68}, but showing the pressure instead of the energy density.
    %Pressure decomposition for $f_\pi = 68~\mathrm{MeV}$ and $\gamma_m = -0.9,\ 0,\ 2$. While the total pressure is unchanged, $\hat{p}^q(r)$ and $\hat{p}^g(r)$ vary consistently with Eq.~(\ref{hatp}).
    }
    \label{FIG6_pressure_ad_68}
\end{figure}

Having analyzed the spatial distribution of the energy density in the previous subsection, we now focus on the pressure and its decomposition.

Figure~\ref{FIG4_pressure_68} (a) presents the spatial distribution of the pressure distribution for $f_\pi = 68~\mathrm{MeV}$ and $\gamma_m = 0$, %\blue{decomposed into the dynamical contribution $\bar{p}(r)$ (purple), the quark mass contribution $\hat{p}^q(r)$ (magenta), and the gluonic anomaly contribution $\hat{p}^g(r)$ (blue), with the total shown in black. An inset shows an close-up view of $\hat{p}^q(r)$.}
The dynamical part is positive throughout space and is localized near the origin ($r=0$).
On the other hand, the anomalous part takes negative values and is dominated by the gluonic anomaly contribution, which retains substantial values even at locations far from $r=0$.
As a result, the total pressure, obtained by summing the dynamical and anomalous contributions, is positive near $r=0$, decreases with distance, and eventually flips to negative values.
At larger distances, it asymptotically approaches zero from below.
Regarding the effect of the scalar meson mass, the total pressure is suppressed in the region near the origin. This behavior is in contrast to that observed in the energy density.
This implies that the anomalous part near $r=0$ is more sensitive to the scalar meson mass than the dynamical part, which can be interpreted as the suppression in pressure induced by the increased strength of the gluonic scale anomaly.
Looking at the  asymptotic region far from the origin, we find that the asymptotic behavior is similar to that of the energy density.

By performing the spatial integral of the obtained pressure, we confirm that the present Skyrme model satisfies the von Laue condition in Eq.~\eqref{von laue condition}. 
Furthermore,
the integrated contributions of the anomalous part $\hat p$ and the dynamical part $\bar p$ also fulfill the relations in Eqs.~\eqref{hatp} and~\eqref{barp}, respectively. 
These results ensure that the Skyrme model based on sChPT reliably describes a stable object.

We then plot the result for the alternative setup with $f_\pi = 93~\mathrm{MeV}$ and $\gamma_m=0$ in Fig.~\ref{FIG5_pressure_93}. As shown earlier for the energy density in the previous subsection, the qualitative behavior of the pressure remains unchanged compared to the previous setup with $f_\pi = 68~\mathrm{MeV}$ and $\gamma_m=0$. Only the overall magnitude of the pressure becomes larger.

%\blue{Figure~\ref{FIG5_pressure_93} (a) shows the result for $f_\pi = 93~\mathrm{MeV}$. Compared to FIG.~\ref{FIG4_pressure_68} (a), the pressure distribution is slightly more localized toward the core region. Figure~\ref{FIG5_pressure_93} (b) shows the spatial distribution of the force on quarks arising from the scale anomaly.} 

The impact of $\gamma_m$ on the pressure is shown in Fig.~\ref{FIG6_pressure_ad_68}.
Consistent with the analysis of the energy density, the total pressure remains largely insensitive to variations in $\gamma_m$.
Regarding the behavior of the anomalous part,   the magnitude of $\hat p^q$ increases with $\gamma_m$, while $\hat p^g$ is correspondingly suppressed to satisfy the relation in Eq.~\eqref{hatp}.  

%\blue{Figure~\ref{FIG6_pressure_ad_68} shows the $\gamma_m$ dependence. The total pressure remains largely insensitive to variations in $\gamma_m$ within the explored range. As expected from Eq.~(\ref{pressure}), the magnitude of $\hat{p}^q(r)$ is rescaled by $1 + \gamma_m$, while $\hat{p}^g(r)$ adjusts in the opposite direction to preserve the constraint~(\ref{hatp}). }

%\blue{Since the essential structure of the pressure remains largely unchanged under variations of $\gamma_m$, we restrict our analysis to the representative case of $\gamma_m = 0$ in the following. An exception is made in the discussion of the D-term, where the $\gamma_m$ dependence has a non-negligible impact and will be examined separately.}

\subsection{The radial and tangential pressure in a 2D plane}
\label{2D_visualization}

\begin{figure}
    \includegraphics[scale=0.3]{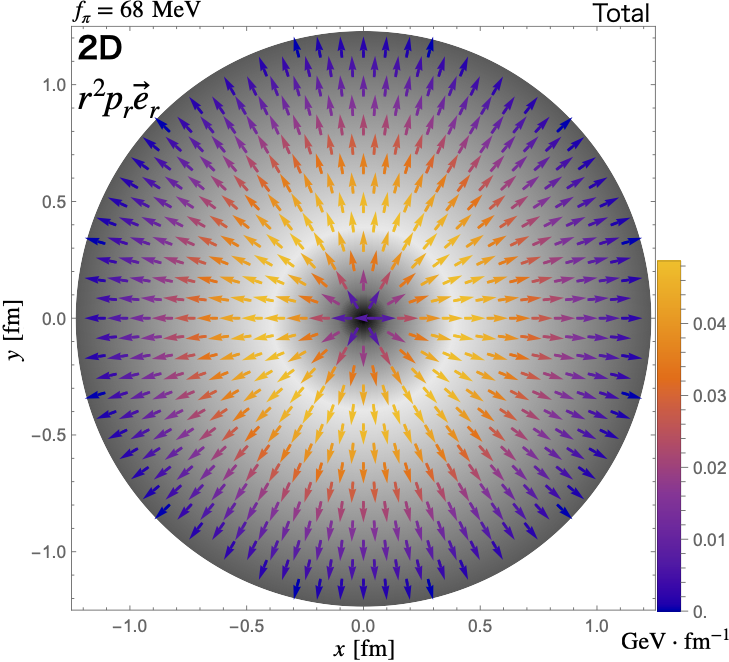}
    \caption{
    Spatial distribution of
    the radial pressure together with its corresponding eigenvector, $p_r \vec{e}_r$, on the traverse x-y plane 
    for the case of $f_\pi=68~{\rm MeV}$ with $m_{\phi0}=720~{\rm MeV}$ and $\gamma_m=0$.
    %The radial pressure $r^2p_r \vec{e}_r$ inside the nucleon in a two dimensional plane. The arrow direction shows the eigenvector $\vec{e}_r$, and the color indicates its magnitude. Stretching (Squeezing) corresponds to positive (negative) eigenvalues.
    }
    \label{FIG7_pr_68}
\end{figure}

\begin{figure}
\begin{center}
    \includegraphics[scale=0.3]{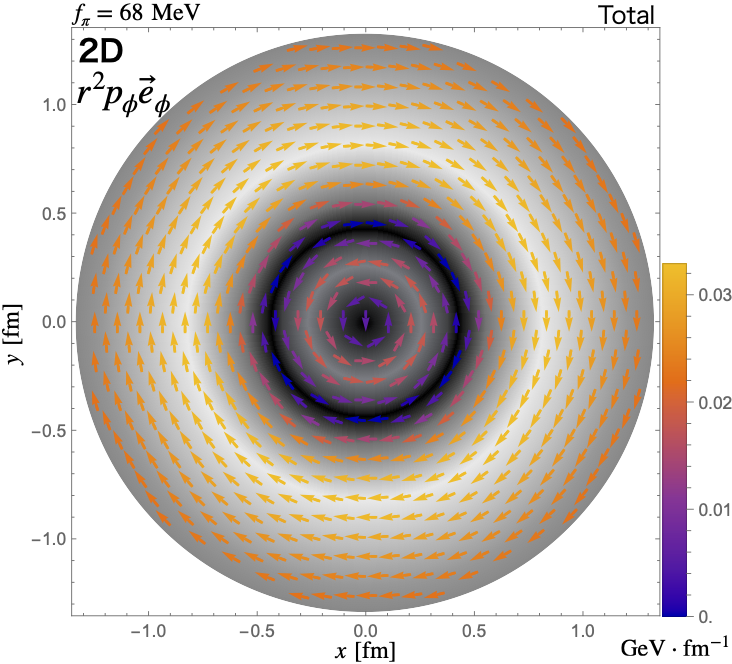}
    \subfigure{ (a) }
    \includegraphics[scale=0.3]{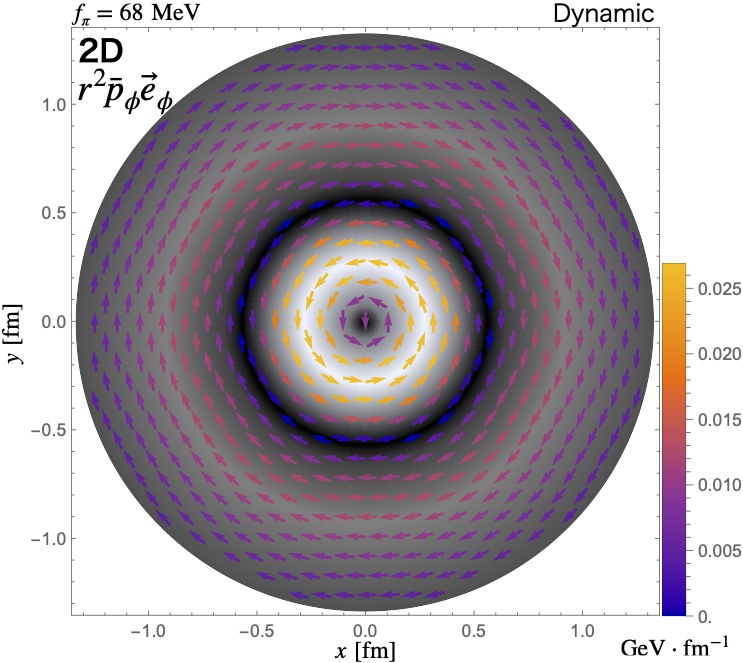}
    \subfigure{ (b) }
    \includegraphics[scale=0.3]{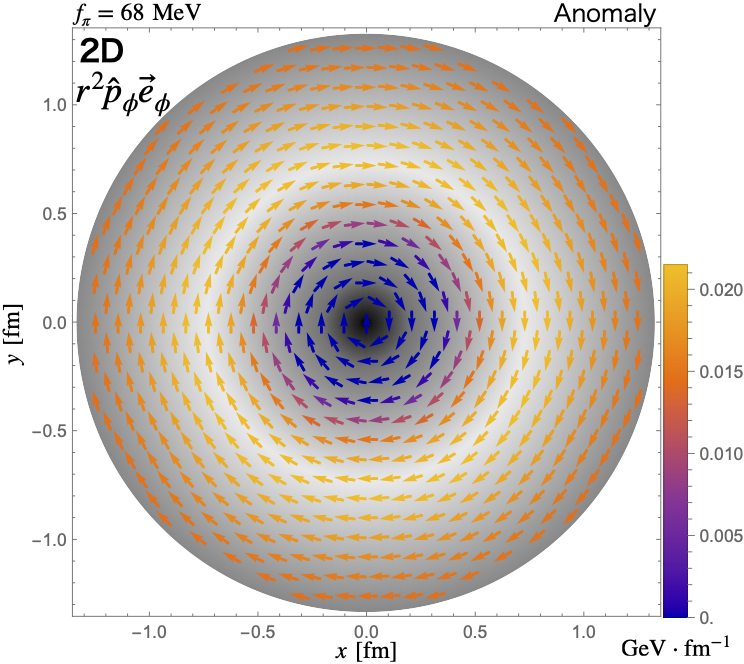}
    \subfigure{ (c) }
\end{center}
    \caption{
    (a) Spatial distribution of the tangential pressure together with its corresponding eigenvector, $p_\phi \vec{e}_\phi$, on the transverse x-y plane for the case of $f_\pi=68~{\rm MeV}$ with $m_{\phi0}=720~{\rm MeV}$ and $\gamma_m=0$. (b) and (c) show its decomposition into the dynamical part ($\bar{p}_\phi$) and the anomalous part ($\hat{p}_\phi$), respectively.
    %(a) Tangential pressure $r^2p_\phi \vec{e}_\phi$ inside the nucleon. The vector direction shows the tangential eigenvector, and the color indicates magnitude. (b) and (c) show the decomposition into dynamical ($\bar{F}_\phi$) and anomaly ($\hat{F}_\phi$) contributions, respectively.
    }
    \label{FIG8_pt_68}
\end{figure}

In this subsection, we present the spatial distribution of the pressure projected onto the transverse x-y plane of the nucleon,
and discuss the stability based on the spherical coordinates in Eqs.~\eqref{radial force} and \eqref{tangential force}. 

First, we show the radial component of the pressure together with its corresponding eigenvector, $p_r\vec e_r$, in Fig.~\ref{FIG7_pr_68}. The arrows in this figure indicate the direction of the eigenvector: when an arrow points outward from the center, it represents a positive radial pressure in that direction; when it points inward, the radial pressure is negative.
In other words, a positive/negative pressure corresponds to a stretching/squeezing effect on nucleon.
The color of each arrow indicates the magnitude of the pressure (its absolute value).
The figure clearly shows that the radial pressure points outward at every spatial point, indicating that the nucleon undergoes the stretching effect along the radial direction throughout the entire region. 
Such behavior is consistent with the stability condition in Eq.~\eqref{local stability condition}.

It is worth noting that using the radial pressure, we numerically evaluate the mechanical radius, defined as 
\begin{align}
    \sqrt{\braket{r^2}^{\rm mech}}=\left( \frac{\int d^{3}x r^{2} p_{r}(r)}{\int d^{3}x p_{r}(r)} \right)^{1/2}.
\end{align}
For the case of $f_\pi=68~{\rm MeV}$ with $m_{\phi0}=720~{\rm MeV}$ and $\gamma_m=0$, the mechanical radius is estimated to be $\sqrt{\braket{r^2}^{\rm mech}}=0.74 \ {\rm fm}$, which is also qualitatively consistent with lattice result~\cite{Hackett:2023rif}.

We also show the tangential component of the pressure together with its corresponding eigenvector, $p_\phi\vec e_{\phi}$, in Fig.~\ref{FIG8_pt_68}. 
Similarly to the radial case, counterclockwise rotation indicates positive 
tangential pressure, while clockwise rotation indicates negative tangential pressure. Furthermore, the color represents its magnitude.
The panel~(a) shows that the tangential pressure is positive (stretching) near the nucleon core, but its sign flips around $r \simeq 0.4~\mathrm{fm}$, beyond which the tangential pressure becomes negative (squeezing). 
As will be shown below, this negative behavior plays a crucial role in ensuring the stability of the nucleon, as it is closely related to the negative D-term discussed in Eq.~\eqref{D-term condition}. The D-term can be rewritten in terms of the radial coordinate as
\begin{align}
    D = M_N \int d^3x\, r^2 \frac{p_r(r)+2p_\phi(r)}{3}.
\end{align}
As shown above, the radial pressure $p_r$ is positive throughout the spatial region. Hence, a sufficiently negative tangential pressure $p_\phi$ is essential to yield the negative D-term, which is required by the stability condition in Eq.~\eqref{D-term condition}.
To further examine the origin of this negative contribution, we consider the decomposition of the tangential pressure in the following.

Panels (b) and (c) in Fig.~\ref{FIG8_pt_68}
show the decomposition of the tangential pressure into the dynamical part ($\bar{p}_\phi$) and the anomalous part ($\hat{p}_\phi$). Both parts contribute negatively in the outer region of the nucleon. However, the anomalous part is dominant, while the dynamical part is negligibly small. As discussed in the previous section, the anomalous part is primarily governed by the gluonic scale anomaly. These findings emphasize the crucial role of the gluonic scale anomaly in generating the negative tangential pressure that yields the negative D-term.

\subsection{Internal confining force}\label{internal confining force}

\begin{figure}
\begin{center}
        \includegraphics[scale=0.28]{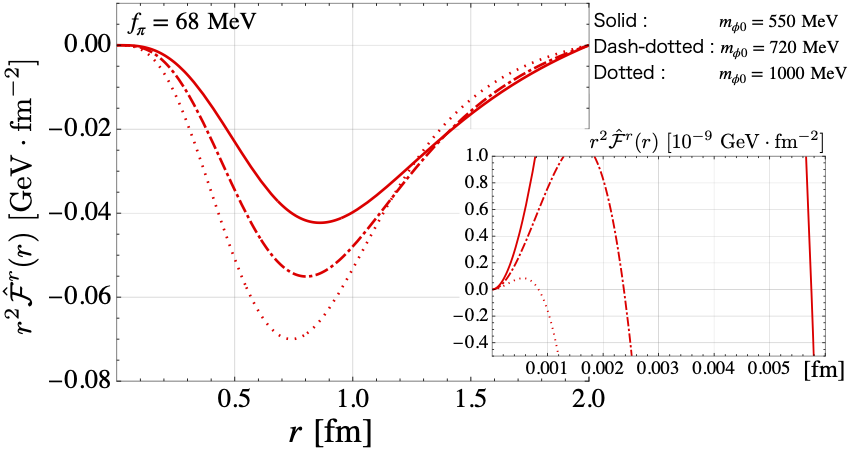}
         \subfigure{ (a) }
        \includegraphics[scale=0.28]{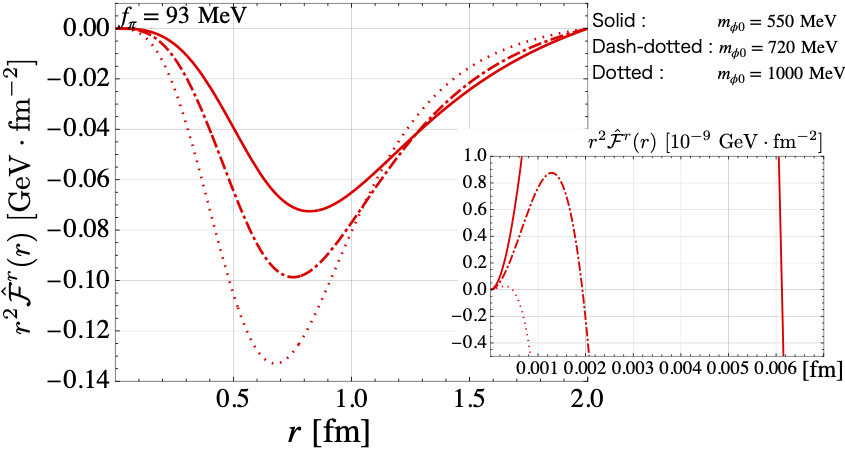}
         \subfigure{ (b) }
    %\subfigure{ (b) }
\end{center}
    \caption{ 
    The anomalous part of the force density for $f_\pi = 68~\mathrm{MeV}$ (a) and $f_\pi = 98~\mathrm{MeV}$ (b), with the anomalous dimension fixed at $\gamma_m=0$. 
    }
    \label{Fig_force}
\end{figure}

In this subsection, we present numerical results for the internal force inside the nucleon based on Eq.~\eqref{decomposed_force}, and discuss its role in the confining mechanism of the nucleon.

 To examine the force, we focus on the anomalous part, $\hat{\cal F}_r$, and present its spatial distribution in Fig.\ref{Fig_force}.
Note that the dynamical force exhibits behavior opposite to the anomalous part in order to ensure the force balance inside the nucleon, as shown in Eq.\eqref{decomposed_force}.
Regardless of the setup, the anomalous force takes small positive values only in the very vicinity of the nucleon core.
However, it becomes negative throughout most of the interior of the nucleon.
Since the negative force is directed inward from the outside, it represents a confining force.
This implies that the scale anomaly—especially the gluonic scale anomaly—plays a crucial role in the confinement mechanism of the nucleon.   

We also investigate the effect of the scalar meson mass on the force. As the scalar meson mass increases, the strength of the anomalous force becomes intense, implying that a larger scale anomaly results in a more pronounced confining effect.

%\blue{By computing the force density acting on quarks, as defined in Eq.(\ref{force_on_quarks}), one finds that the negative pressure arising from the trace contributions $\hat{p}^q(r)$ and $\hat{p}^g(r)$ corresponds to a force that confines quarks within the hadron. Figure~\ref{FIG4_pressure_68} (b) illustrates the spatial distribution of this force induced by the scale anomaly. As illustrated in the inset, a repulsive force acts near the center of the nucleon, pushing quarks outward. Beyond a certain radius—dependent on the scalar meson mass $m_{\phi0}$—the force reverses direction and becomes attractive, indicating the onset of a confining behavior that extends over a wide spatial region. This result also indicates that the scale anomaly plays a crucial role in the confinement of quarks.}

%%%%%%%%%%%%%%
\subsection{$D(t)$ form factor
%Comparison with lattice results
}

\begin{figure}
    \includegraphics[scale=0.4]{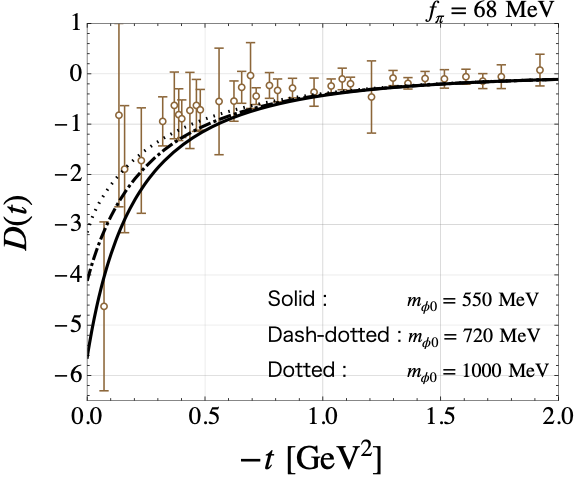}
    \caption{
    Momentum transfer dependence of $D(t)$ form factor for $f_\pi =63\,{\rm MeV}$
    (black curve), compared with the lattice results from~\cite{Hackett:2023rif} (points with error bars).
    %Comparison between our result for the gravitational form factor $D(t)$ (black curve) and lattice QCD data~\cite{Hackett:2023rif} (points with error bars). The best agreement is found for $f_\pi = 68~\mathrm{MeV}, m_\phi = 720~\mathrm{MeV}$.
    }
    \label{FIG10_D_lattice_68}
\end{figure}

Based on the pressure results obtained above, we evaluate the momentum transfer dependence of $D(t)$ form factor and compare it with the recent lattice data~\cite{Hackett:2023rif}, as shown in Fig.~\ref{FIG10_D_lattice_68}. 
The figure indicates that the momentum transfer dependence in the Skyrme model qualitatively agrees with that in the lattice result. In particular, for $f_\pi = 68\mathrm{MeV}$ and $m_\phi = 720~\mathrm{MeV}$, the model successfully reproduces the lattice data.
In this case, the D-term is evaluated as
\begin{align}
    D=D(0)=-4.12.
\end{align}
This result is in qualitative agreement with recent lattice QCD findings obtained via dipole fits, within the quoted systematic uncertainties~\cite{Hackett:2023rif}. However, their analysis using the $z$-expansion method yields a somewhat larger D-term compared to our prediction.
In addition, recent model-independent estimates based on dispersion relations have also extracted the D-term~\cite{Cao:2024zlf}. Compared to these previous observations, our result is somewhat smaller. To reconcile this discrepancy, we take into account the finite value of the anomalous dimension. As shown in Table~\ref{tab1}, in particular for $\gamma_m = 2$, the D-term reaches $D = -3.98$, which is close to the values obtained from both the lattice $z$-expansion analysis and the model-independent estimate\footnote{
In our present setup the model parameters were adjusted to reproduce the lattice results for the $D(t)$ form factor, but this parameter setup causes the $A(t)$ form factor to deviate from the lattice results outside the error bars.
}.

We also comment on the impact of the scalar meson mass on the D-term. Unlike the nucleon mass, the D-term is significantly affected by changes in the scalar meson mass, as shown in Table~\ref{tab1}. As discussed in Sec.~\ref{Pressure_subsec}, this is likely because the anomalous part $\hat p$ is more sensitive to changes in the mass than the dynamical part $\bar p$, which leads to a substantial variation in the D-term .

\section{Summary and outlook}\label{ch:summary}
In this study, we extend our previous analysis further. As a brief reminder, in our earlier work using a Skyrme model based on sChPT, we revealed that the gluonic scale anomaly gives a negative contribution to the pressure in Cartesian coordinates and constitutes the dominant part of the D-term~\cite{Fujii:2025aip}. Building on this previous result, the main results obtained in the present study are as follows:

\begin{itemize}
    \item By varying the scalar meson mass, which controls the strength of the gluonic scale anomaly, we investigated the response of the gluonic scale anomaly in the energy density and the pressure. While the energy density somewhat changes with the scalar meson mass, the nucleon mass remains largely unaffected. In contrast, the anomalous part of the pressure exhibits a strong sensitivity to the scalar meson mass, leading to a significant variation in the D-term. Since the magnitude of the scale anomaly is expected to change in high-temperature and high-density environments, the insights obtained in this work may be useful for studying the properties of nucleons under such extreme conditions.   
    \item
    To better understand the internal structure of the nucleon, we analyzed the pressure distribution in spherical coordinates. We confirmed that the radial pressure satisfies the stability condition of the nucleon. For the tangential pressure, we found that it is positive in the inner region and becomes negative toward the outer region. This negative tangential pressure plays a crucial role in satisfying the stability condition associated with a negative D-term. We then identified that the main contribution to this negative tangential pressure originates from the gluonic scale anomaly. Furthermore, by examining the internal force density, we found that the forces are balanced inside the nucleon, and the confining force is predominantly generated by the gluonic scale anomaly.
    \item
    We also investigated the momentum transfer dependence of the GFF $D(t)$. The result obtained from the Skyrme model successfully reproduces the momentum-transfer dependence observed in recent lattice QCD calculations~\cite{Hackett:2023rif}. Based on the best-fit agreement with the lattice results, the D-term is found to be $D=-4.12$. However, this estimate is somewhat smaller than those reported in some previous studies~\cite{Hackett:2023rif,Cao:2024zlf}. To reconsinel this deviation, we find that this value can be brought closer to earlier results by adjusting the anomalous dimension. In addition, we also make a comparison of the mass radius and the mechanical radius with the lattice QCD results~\cite{Hackett:2023rif}. This agreement with lattice QCD reinforces the quantitative reliability of our model analysis and, at the same time, provides a robust estimate of the D-term, whose precise value has not yet been firmly established.

\end{itemize}

In closing, we comment on the remaining issues and the implications of our analysis below. 

First, in this study, our analysis is based on classical soliton solutions.
In order to yield more realistic results,
one needs to quantize the soliton, which corresponds to the next-to-leading order in the large $N_c$ expansion. However, as pointed out by the previous studies, the quantized soliton would fail to satisfy the stability condition for the nucleon~\cite{Cebulla:2007ei,Kim:2020lrs}. Resolving this issue remains an important direction for future work.

We also offer a few remarks on the current model framework. In the present analysis, the sChPT Lagrangian includes only the contributions from the pion and the scalar meson. Extending the model to incorporate higher excited states, such as vector mesons,  is expected to yield more refined and comprehensive results. Further investigation based on an extended framework would be a valuable direction for future research.

Although this study has addressed the internal force structure, the skyrmion approach cannot describe the forces acting on the quark constituents of the nucleon.
Nevertheless, recent studies have proposed an approach to characterize the forces acting on quarks through the static EMT~\cite{Ji:2025gsq}. In light of this development, it may be possible to infer the forces on quarks even within the framework of the Skyrme model.
Alternatively, one could investigate the quark-level forces more directly by employing models that incorporate explicit quark degrees of freedom, such as the chiral quark soliton model~\cite{Diakonov:1986yh,Diakonov:1987ty,Goeke:2007fp} or the quark-meson coupling model~\cite{Guichon:1987jp,Saito:1994ki}.

There are also ongoing discussions regarding the relation between GFFs and spatial distributions.
Our analysis follows the interpretation of hadrons as a continuum medium proposed in Ref.~\cite{Polyakov:2002yz}, wherein the static EMT in the Breit frame is interpreted as three-dimensional spatial distributions such as pressure and shear forces.
Based on this interpretation, in this study, we evaluated the density distribution in the Breit frame on the equal-time surface $t=0$ in the instant form.
However, it has been pointed out that, for light hadrons such as the nucleon, relativistic effects can lead to ambiguities in the definition of such densities~\cite{Miller:2018ybm,Jaffe:2020ebz}.
To avoid that problem, two-dimensional transverse densities in the light-front form have been widely discussed~\cite{Fujii:2025tpk,Fujii:2025paw}.
In the present work, however, we leave them for future work.
Furthermore, alternative interpretations, distinct from the continuum medium viewpoint of Ref.~\cite{Polyakov:2002yz}, have been proposed~\cite{Ji:2025gsq,Ji:2025qax}, and discussions based on those frameworks could be of interest; however, they lie beyond the scope of the present work.

One of the key features of the present study is the decomposition of the static EMT into dynamical and anomalous parts. Based on this decomposition, we analyzed the pressure distribution and force balance inside the nucleon. This method is universally applicable to any type of hadron. Recently, some of the present authors have applied this decomposition analysis to the pion and found that the scale anomaly plays a significant role even in the pion's internal structure~\cite{Fujii:2025tpk}. Since this decomposition can also be applied to other hadrons, such as excited states of the nucleon and heavy hadrons, this approach may contribute to a deeper understanding of the confinement mechanism in hadronic systems.

\acknowledgments

The author M.T. would like to take this opportunity to thank the financial support from "THERS Make New Standards Program for the Next Generation Researchers" and JST SPRING, Grant Number JPMJSP2125.
This work of D.F. was supported in part by the Japan Society for the Promotion of Science (JSPS) KAKENHI (Grants No. JP24K17054) and the COREnet project of RCNP, Osaka University.
The work of M.K. is supported by RFIS-NSFC under Grant No. W2433019.

%This work was financially supported by .

\appendix

\bibliography{ref}

\begin{thebibliography}{124}%
\makeatletter
\providecommand \@ifxundefined [1]{%
 \@ifx{#1\undefined}
}%
\providecommand \@ifnum [1]{%
 \ifnum #1\expandafter \@firstoftwo
 \else \expandafter \@secondoftwo
 \fi
}%
\providecommand \@ifx [1]{%
 \ifx #1\expandafter \@firstoftwo
 \else \expandafter \@secondoftwo
 \fi
}%
\providecommand \natexlab [1]{#1}%
\providecommand \enquote  [1]{``#1''}%
\providecommand \bibnamefont  [1]{#1}%
\providecommand \bibfnamefont [1]{#1}%
\providecommand \citenamefont [1]{#1}%
\providecommand \href@noop [0]{\@secondoftwo}%
\providecommand \href [0]{\begingroup \@sanitize@url \@href}%
\providecommand \@href[1]{\@@startlink{#1}\@@href}%
\providecommand \@@href[1]{\endgroup#1\@@endlink}%
\providecommand \@sanitize@url [0]{\catcode `\\12\catcode `\$12\catcode `\&12\catcode `\#12\catcode `\^12\catcode `\_12\catcode `\%12\relax}%
\providecommand \@@startlink[1]{}%
\providecommand \@@endlink[0]{}%
\providecommand \url  [0]{\begingroup\@sanitize@url \@url }%
\providecommand \@url [1]{\endgroup\@href {#1}{\urlprefix }}%
\providecommand \urlprefix  [0]{URL }%
\providecommand \Eprint [0]{\href }%
\providecommand \doibase [0]{https://doi.org/}%
\providecommand \selectlanguage [0]{\@gobble}%
\providecommand \bibinfo  [0]{\@secondoftwo}%
\providecommand \bibfield  [0]{\@secondoftwo}%
\providecommand \translation [1]{[#1]}%
\providecommand \BibitemOpen [0]{}%
\providecommand \bibitemStop [0]{}%
\providecommand \bibitemNoStop [0]{.\EOS\space}%
\providecommand \EOS [0]{\spacefactor3000\relax}%
\providecommand \BibitemShut  [1]{\csname bibitem#1\endcsname}%
\let\auto@bib@innerbib\@empty
%</preamble>
\bibitem [{\citenamefont {Fujii}\ \emph {et~al.}(2024)\citenamefont {Fujii}, \citenamefont {Iwanaka},\ and\ \citenamefont {Tanaka}}]{Fujii:2024rqd}%
  \BibitemOpen
  \bibfield  {author} {\bibinfo {author} {\bibfnamefont {D.}~\bibnamefont {Fujii}}, \bibinfo {author} {\bibfnamefont {A.}~\bibnamefont {Iwanaka}},\ and\ \bibinfo {author} {\bibfnamefont {M.}~\bibnamefont {Tanaka}},\ }\bibfield  {title} {\bibinfo {title} {{Gravitational form factors of pion from top-down holographic QCD}},\ }\href {https://doi.org/10.1103/PhysRevD.110.L091501} {\bibfield  {journal} {\bibinfo  {journal} {Phys. Rev. D}\ }\textbf {\bibinfo {volume} {110}},\ \bibinfo {pages} {L091501} (\bibinfo {year} {2024})},\ \Eprint {https://arxiv.org/abs/2407.21113} {arXiv:2407.21113 [hep-ph]} \BibitemShut {NoStop}%
\bibitem [{\citenamefont {Fujii}\ \emph {et~al.}(2025{\natexlab{a}})\citenamefont {Fujii}, \citenamefont {Kawaguchi},\ and\ \citenamefont {Tanaka}}]{Fujii:2025aip}%
  \BibitemOpen
  \bibfield  {author} {\bibinfo {author} {\bibfnamefont {D.}~\bibnamefont {Fujii}}, \bibinfo {author} {\bibfnamefont {M.}~\bibnamefont {Kawaguchi}},\ and\ \bibinfo {author} {\bibfnamefont {M.}~\bibnamefont {Tanaka}},\ }\bibfield  {title} {\bibinfo {title} {{Dominance of gluonic scale anomaly in confining pressure inside nucleon and D-term}},\ }\href {https://doi.org/10.1016/j.physletb.2025.139559} {\bibfield  {journal} {\bibinfo  {journal} {Phys. Lett. B}\ }\textbf {\bibinfo {volume} {866}},\ \bibinfo {pages} {139559} (\bibinfo {year} {2025}{\natexlab{a}})},\ \Eprint {https://arxiv.org/abs/2503.09686} {arXiv:2503.09686 [hep-ph]} \BibitemShut {NoStop}%
\bibitem [{\citenamefont {Fujii}\ \emph {et~al.}(2025{\natexlab{b}})\citenamefont {Fujii}, \citenamefont {Iwanaka},\ and\ \citenamefont {Tanaka}}]{Fujii:2025tpk}%
  \BibitemOpen
  \bibfield  {author} {\bibinfo {author} {\bibfnamefont {D.}~\bibnamefont {Fujii}}, \bibinfo {author} {\bibfnamefont {A.}~\bibnamefont {Iwanaka}},\ and\ \bibinfo {author} {\bibfnamefont {M.}~\bibnamefont {Tanaka}},\ }\bibfield  {title} {\bibinfo {title} {{Dominance of scale anomaly in confining pressure inside pions on light front in the top-down holographic QCD}},\ }\href@noop {} {\  (\bibinfo {year} {2025}{\natexlab{b}})},\ \Eprint {https://arxiv.org/abs/2507.18690} {arXiv:2507.18690 [hep-ph]} \BibitemShut {NoStop}%
\bibitem [{\citenamefont {Burkert}\ \emph {et~al.}(2018)\citenamefont {Burkert}, \citenamefont {Elouadrhiri},\ and\ \citenamefont {Girod}}]{Burkert:2018bqq}%
  \BibitemOpen
  \bibfield  {author} {\bibinfo {author} {\bibfnamefont {V.~D.}\ \bibnamefont {Burkert}}, \bibinfo {author} {\bibfnamefont {L.}~\bibnamefont {Elouadrhiri}},\ and\ \bibinfo {author} {\bibfnamefont {F.~X.}\ \bibnamefont {Girod}},\ }\bibfield  {title} {\bibinfo {title} {{The pressure distribution inside the proton}},\ }\href {https://doi.org/10.1038/s41586-018-0060-z} {\bibfield  {journal} {\bibinfo  {journal} {Nature}\ }\textbf {\bibinfo {volume} {557}},\ \bibinfo {pages} {396} (\bibinfo {year} {2018})}\BibitemShut {NoStop}%
\bibitem [{\citenamefont {Burkert}\ \emph {et~al.}(2021)\citenamefont {Burkert}, \citenamefont {Elouadrhiri},\ and\ \citenamefont {Girod}}]{Burkert:2021ith}%
  \BibitemOpen
  \bibfield  {author} {\bibinfo {author} {\bibfnamefont {V.~D.}\ \bibnamefont {Burkert}}, \bibinfo {author} {\bibfnamefont {L.}~\bibnamefont {Elouadrhiri}},\ and\ \bibinfo {author} {\bibfnamefont {F.~X.}\ \bibnamefont {Girod}},\ }\bibfield  {title} {\bibinfo {title} {{Determination of shear forces inside the proton}},\ }\href@noop {} {\  (\bibinfo {year} {2021})},\ \Eprint {https://arxiv.org/abs/2104.02031} {arXiv:2104.02031 [nucl-ex]} \BibitemShut {NoStop}%
\bibitem [{\citenamefont {Duran}\ \emph {et~al.}(2023)\citenamefont {Duran} \emph {et~al.}}]{Duran:2022xag}%
  \BibitemOpen
  \bibfield  {author} {\bibinfo {author} {\bibfnamefont {B.}~\bibnamefont {Duran}} \emph {et~al.},\ }\bibfield  {title} {\bibinfo {title} {{Determining the gluonic gravitational form factors of the proton}},\ }\href {https://doi.org/10.1038/s41586-023-05730-4} {\bibfield  {journal} {\bibinfo  {journal} {Nature}\ }\textbf {\bibinfo {volume} {615}},\ \bibinfo {pages} {813} (\bibinfo {year} {2023})},\ \Eprint {https://arxiv.org/abs/2207.05212} {arXiv:2207.05212 [nucl-ex]} \BibitemShut {NoStop}%
\bibitem [{\citenamefont {Polyakov}\ and\ \citenamefont {Son}(2018)}]{Polyakov:2018exb}%
  \BibitemOpen
  \bibfield  {author} {\bibinfo {author} {\bibfnamefont {M.~V.}\ \bibnamefont {Polyakov}}\ and\ \bibinfo {author} {\bibfnamefont {H.-D.}\ \bibnamefont {Son}},\ }\bibfield  {title} {\bibinfo {title} {{Nucleon gravitational form factors from instantons: forces between quark and gluon subsystems}},\ }\href {https://doi.org/10.1007/JHEP09(2018)156} {\bibfield  {journal} {\bibinfo  {journal} {JHEP}\ }\textbf {\bibinfo {volume} {09}},\ \bibinfo {pages} {156}},\ \Eprint {https://arxiv.org/abs/1808.00155} {arXiv:1808.00155 [hep-ph]} \BibitemShut {NoStop}%
\bibitem [{\citenamefont {Shanahan}\ and\ \citenamefont {Detmold}(2019)}]{Shanahan:2018nnv}%
  \BibitemOpen
  \bibfield  {author} {\bibinfo {author} {\bibfnamefont {P.~E.}\ \bibnamefont {Shanahan}}\ and\ \bibinfo {author} {\bibfnamefont {W.}~\bibnamefont {Detmold}},\ }\bibfield  {title} {\bibinfo {title} {{Pressure Distribution and Shear Forces inside the Proton}},\ }\href {https://doi.org/10.1103/PhysRevLett.122.072003} {\bibfield  {journal} {\bibinfo  {journal} {Phys. Rev. Lett.}\ }\textbf {\bibinfo {volume} {122}},\ \bibinfo {pages} {072003} (\bibinfo {year} {2019})},\ \Eprint {https://arxiv.org/abs/1810.07589} {arXiv:1810.07589 [nucl-th]} \BibitemShut {NoStop}%
\bibitem [{\citenamefont {Lorc\'e}\ \emph {et~al.}(2019)\citenamefont {Lorc\'e}, \citenamefont {Moutarde},\ and\ \citenamefont {Trawi\'nski}}]{Lorce:2018egm}%
  \BibitemOpen
  \bibfield  {author} {\bibinfo {author} {\bibfnamefont {C.}~\bibnamefont {Lorc\'e}}, \bibinfo {author} {\bibfnamefont {H.}~\bibnamefont {Moutarde}},\ and\ \bibinfo {author} {\bibfnamefont {A.~P.}\ \bibnamefont {Trawi\'nski}},\ }\bibfield  {title} {\bibinfo {title} {{Revisiting the mechanical properties of the nucleon}},\ }\href {https://doi.org/10.1140/epjc/s10052-019-6572-3} {\bibfield  {journal} {\bibinfo  {journal} {Eur. Phys. J. C}\ }\textbf {\bibinfo {volume} {79}},\ \bibinfo {pages} {89} (\bibinfo {year} {2019})},\ \Eprint {https://arxiv.org/abs/1810.09837} {arXiv:1810.09837 [hep-ph]} \BibitemShut {NoStop}%
\bibitem [{\citenamefont {Anikin}(2019)}]{Anikin:2019kwi}%
  \BibitemOpen
  \bibfield  {author} {\bibinfo {author} {\bibfnamefont {I.~V.}\ \bibnamefont {Anikin}},\ }\bibfield  {title} {\bibinfo {title} {{Gravitational form factors within light-cone sum rules at leading order}},\ }\href {https://doi.org/10.1103/PhysRevD.99.094026} {\bibfield  {journal} {\bibinfo  {journal} {Phys. Rev. D}\ }\textbf {\bibinfo {volume} {99}},\ \bibinfo {pages} {094026} (\bibinfo {year} {2019})},\ \Eprint {https://arxiv.org/abs/1902.00094} {arXiv:1902.00094 [hep-ph]} \BibitemShut {NoStop}%
\bibitem [{\citenamefont {Avelino}(2019)}]{Avelino:2019esh}%
  \BibitemOpen
  \bibfield  {author} {\bibinfo {author} {\bibfnamefont {P.~P.}\ \bibnamefont {Avelino}},\ }\bibfield  {title} {\bibinfo {title} {{Probing gravity at sub-femtometer scales through the pressure distribution inside the proton}},\ }\href {https://doi.org/10.1016/j.physletb.2019.05.056} {\bibfield  {journal} {\bibinfo  {journal} {Phys. Lett. B}\ }\textbf {\bibinfo {volume} {795}},\ \bibinfo {pages} {627} (\bibinfo {year} {2019})},\ \Eprint {https://arxiv.org/abs/1902.01318} {arXiv:1902.01318 [gr-qc]} \BibitemShut {NoStop}%
\bibitem [{\citenamefont {Yanagihara}\ and\ \citenamefont {Kitazawa}(2019)}]{Yanagihara:2019foh}%
  \BibitemOpen
  \bibfield  {author} {\bibinfo {author} {\bibfnamefont {R.}~\bibnamefont {Yanagihara}}\ and\ \bibinfo {author} {\bibfnamefont {M.}~\bibnamefont {Kitazawa}},\ }\bibfield  {title} {\bibinfo {title} {{A study of stress-tensor distribution around the flux tube in the Abelian\textendash{}Higgs model}},\ }\href {https://doi.org/10.1093/ptep/ptz093} {\bibfield  {journal} {\bibinfo  {journal} {PTEP}\ }\textbf {\bibinfo {volume} {2019}},\ \bibinfo {pages} {093B02} (\bibinfo {year} {2019})},\ \bibinfo {note} {[Erratum: PTEP 2020, 079201 (2020)]},\ \Eprint {https://arxiv.org/abs/1905.10056} {arXiv:1905.10056 [hep-ph]} \BibitemShut {NoStop}%
\bibitem [{\citenamefont {Hatta}\ \emph {et~al.}(2019)\citenamefont {Hatta}, \citenamefont {Rajan},\ and\ \citenamefont {Yang}}]{Hatta:2019lxo}%
  \BibitemOpen
  \bibfield  {author} {\bibinfo {author} {\bibfnamefont {Y.}~\bibnamefont {Hatta}}, \bibinfo {author} {\bibfnamefont {A.}~\bibnamefont {Rajan}},\ and\ \bibinfo {author} {\bibfnamefont {D.-L.}\ \bibnamefont {Yang}},\ }\bibfield  {title} {\bibinfo {title} {{Near threshold J/\ensuremath{\psi} and \ensuremath{\Upsilon} photoproduction at JLab and RHIC}},\ }\href {https://doi.org/10.1103/PhysRevD.100.014032} {\bibfield  {journal} {\bibinfo  {journal} {Phys. Rev. D}\ }\textbf {\bibinfo {volume} {100}},\ \bibinfo {pages} {014032} (\bibinfo {year} {2019})},\ \Eprint {https://arxiv.org/abs/1906.00894} {arXiv:1906.00894 [hep-ph]} \BibitemShut {NoStop}%
\bibitem [{\citenamefont {Freese}\ and\ \citenamefont {Clo\"et}(2020)}]{Freese:2019eww}%
  \BibitemOpen
  \bibfield  {author} {\bibinfo {author} {\bibfnamefont {A.}~\bibnamefont {Freese}}\ and\ \bibinfo {author} {\bibfnamefont {I.~C.}\ \bibnamefont {Clo\"et}},\ }\bibfield  {title} {\bibinfo {title} {{Impact of dynamical chiral symmetry breaking and dynamical diquark correlations on proton generalized parton distributions}},\ }\href {https://doi.org/10.1103/PhysRevC.101.035203} {\bibfield  {journal} {\bibinfo  {journal} {Phys. Rev. C}\ }\textbf {\bibinfo {volume} {101}},\ \bibinfo {pages} {035203} (\bibinfo {year} {2020})},\ \Eprint {https://arxiv.org/abs/1907.08256} {arXiv:1907.08256 [nucl-th]} \BibitemShut {NoStop}%
\bibitem [{\citenamefont {Azizi}\ and\ \citenamefont {\"Ozdem}(2020)}]{Azizi:2019ytx}%
  \BibitemOpen
  \bibfield  {author} {\bibinfo {author} {\bibfnamefont {K.}~\bibnamefont {Azizi}}\ and\ \bibinfo {author} {\bibfnamefont {U.}~\bibnamefont {\"Ozdem}},\ }\bibfield  {title} {\bibinfo {title} {{Nucleon\textquoteright{}s energy\textendash{}momentum tensor form factors in light-cone QCD}},\ }\href {https://doi.org/10.1140/epjc/s10052-020-7676-5} {\bibfield  {journal} {\bibinfo  {journal} {Eur. Phys. J. C}\ }\textbf {\bibinfo {volume} {80}},\ \bibinfo {pages} {104} (\bibinfo {year} {2020})},\ \Eprint {https://arxiv.org/abs/1908.06143} {arXiv:1908.06143 [hep-ph]} \BibitemShut {NoStop}%
\bibitem [{\citenamefont {Mamo}\ and\ \citenamefont {Zahed}(2020)}]{Mamo:2019mka}%
  \BibitemOpen
  \bibfield  {author} {\bibinfo {author} {\bibfnamefont {K.~A.}\ \bibnamefont {Mamo}}\ and\ \bibinfo {author} {\bibfnamefont {I.}~\bibnamefont {Zahed}},\ }\bibfield  {title} {\bibinfo {title} {{Diffractive photoproduction of $J/\psi$ and $\Upsilon$ using holographic QCD: gravitational form factors and GPD of gluons in the proton}},\ }\href {https://doi.org/10.1103/PhysRevD.101.086003} {\bibfield  {journal} {\bibinfo  {journal} {Phys. Rev. D}\ }\textbf {\bibinfo {volume} {101}},\ \bibinfo {pages} {086003} (\bibinfo {year} {2020})},\ \Eprint {https://arxiv.org/abs/1910.04707} {arXiv:1910.04707 [hep-ph]} \BibitemShut {NoStop}%
\bibitem [{\citenamefont {Neubelt}\ \emph {et~al.}(2020)\citenamefont {Neubelt}, \citenamefont {Sampino}, \citenamefont {Hudson}, \citenamefont {Tezgin},\ and\ \citenamefont {Schweitzer}}]{Neubelt:2019sou}%
  \BibitemOpen
  \bibfield  {author} {\bibinfo {author} {\bibfnamefont {M.~J.}\ \bibnamefont {Neubelt}}, \bibinfo {author} {\bibfnamefont {A.}~\bibnamefont {Sampino}}, \bibinfo {author} {\bibfnamefont {J.}~\bibnamefont {Hudson}}, \bibinfo {author} {\bibfnamefont {K.}~\bibnamefont {Tezgin}},\ and\ \bibinfo {author} {\bibfnamefont {P.}~\bibnamefont {Schweitzer}},\ }\bibfield  {title} {\bibinfo {title} {{Energy momentum tensor and the D-term in the bag model}},\ }\href {https://doi.org/10.1103/PhysRevD.101.034013} {\bibfield  {journal} {\bibinfo  {journal} {Phys. Rev. D}\ }\textbf {\bibinfo {volume} {101}},\ \bibinfo {pages} {034013} (\bibinfo {year} {2020})},\ \Eprint {https://arxiv.org/abs/1911.08906} {arXiv:1911.08906 [hep-ph]} \BibitemShut {NoStop}%
\bibitem [{\citenamefont {Alharazin}\ \emph {et~al.}(2020)\citenamefont {Alharazin}, \citenamefont {Djukanovic}, \citenamefont {Gegelia},\ and\ \citenamefont {Polyakov}}]{Alharazin:2020yjv}%
  \BibitemOpen
  \bibfield  {author} {\bibinfo {author} {\bibfnamefont {H.}~\bibnamefont {Alharazin}}, \bibinfo {author} {\bibfnamefont {D.}~\bibnamefont {Djukanovic}}, \bibinfo {author} {\bibfnamefont {J.}~\bibnamefont {Gegelia}},\ and\ \bibinfo {author} {\bibfnamefont {M.~V.}\ \bibnamefont {Polyakov}},\ }\bibfield  {title} {\bibinfo {title} {{Chiral theory of nucleons and pions in the presence of an external gravitational field}},\ }\href {https://doi.org/10.1103/PhysRevD.102.076023} {\bibfield  {journal} {\bibinfo  {journal} {Phys. Rev. D}\ }\textbf {\bibinfo {volume} {102}},\ \bibinfo {pages} {076023} (\bibinfo {year} {2020})},\ \Eprint {https://arxiv.org/abs/2006.05890} {arXiv:2006.05890 [hep-ph]} \BibitemShut {NoStop}%
\bibitem [{\citenamefont {Varma}\ and\ \citenamefont {Schweitzer}(2020)}]{Varma:2020crx}%
  \BibitemOpen
  \bibfield  {author} {\bibinfo {author} {\bibfnamefont {M.}~\bibnamefont {Varma}}\ and\ \bibinfo {author} {\bibfnamefont {P.}~\bibnamefont {Schweitzer}},\ }\bibfield  {title} {\bibinfo {title} {{Effects of long-range forces on the D-term and the energy-momentum structure}},\ }\href {https://doi.org/10.1103/PhysRevD.102.014047} {\bibfield  {journal} {\bibinfo  {journal} {Phys. Rev. D}\ }\textbf {\bibinfo {volume} {102}},\ \bibinfo {pages} {014047} (\bibinfo {year} {2020})},\ \Eprint {https://arxiv.org/abs/2006.06602} {arXiv:2006.06602 [hep-ph]} \BibitemShut {NoStop}%
\bibitem [{\citenamefont {Kim}\ \emph {et~al.}(2021)\citenamefont {Kim}, \citenamefont {Kim}, \citenamefont {Polyakov},\ and\ \citenamefont {Son}}]{Kim:2020nug}%
  \BibitemOpen
  \bibfield  {author} {\bibinfo {author} {\bibfnamefont {J.-Y.}\ \bibnamefont {Kim}}, \bibinfo {author} {\bibfnamefont {H.-C.}\ \bibnamefont {Kim}}, \bibinfo {author} {\bibfnamefont {M.~V.}\ \bibnamefont {Polyakov}},\ and\ \bibinfo {author} {\bibfnamefont {H.-D.}\ \bibnamefont {Son}},\ }\bibfield  {title} {\bibinfo {title} {{Strong force fields and stabilities of the nucleon and singly heavy baryon $\Sigma_c$}},\ }\href {https://doi.org/10.1103/PhysRevD.103.014015} {\bibfield  {journal} {\bibinfo  {journal} {Phys. Rev. D}\ }\textbf {\bibinfo {volume} {103}},\ \bibinfo {pages} {014015} (\bibinfo {year} {2021})},\ \Eprint {https://arxiv.org/abs/2008.06652} {arXiv:2008.06652 [hep-ph]} \BibitemShut {NoStop}%
\bibitem [{\citenamefont {Chakrabarti}\ \emph {et~al.}(2020)\citenamefont {Chakrabarti}, \citenamefont {Mondal}, \citenamefont {Mukherjee}, \citenamefont {Nair},\ and\ \citenamefont {Zhao}}]{Chakrabarti:2020kdc}%
  \BibitemOpen
  \bibfield  {author} {\bibinfo {author} {\bibfnamefont {D.}~\bibnamefont {Chakrabarti}}, \bibinfo {author} {\bibfnamefont {C.}~\bibnamefont {Mondal}}, \bibinfo {author} {\bibfnamefont {A.}~\bibnamefont {Mukherjee}}, \bibinfo {author} {\bibfnamefont {S.}~\bibnamefont {Nair}},\ and\ \bibinfo {author} {\bibfnamefont {X.}~\bibnamefont {Zhao}},\ }\bibfield  {title} {\bibinfo {title} {{Gravitational form factors and mechanical properties of proton in a light-front quark-diquark model}},\ }\href {https://doi.org/10.1103/PhysRevD.102.113011} {\bibfield  {journal} {\bibinfo  {journal} {Phys. Rev. D}\ }\textbf {\bibinfo {volume} {102}},\ \bibinfo {pages} {113011} (\bibinfo {year} {2020})},\ \Eprint {https://arxiv.org/abs/2010.04215} {arXiv:2010.04215 [hep-ph]} \BibitemShut {NoStop}%
\bibitem [{\citenamefont {Yanagihara}\ \emph {et~al.}(2020)\citenamefont {Yanagihara}, \citenamefont {Kitazawa}, \citenamefont {Asakawa},\ and\ \citenamefont {Hatsuda}}]{Yanagihara:2020tvs}%
  \BibitemOpen
  \bibfield  {author} {\bibinfo {author} {\bibfnamefont {R.}~\bibnamefont {Yanagihara}}, \bibinfo {author} {\bibfnamefont {M.}~\bibnamefont {Kitazawa}}, \bibinfo {author} {\bibfnamefont {M.}~\bibnamefont {Asakawa}},\ and\ \bibinfo {author} {\bibfnamefont {T.}~\bibnamefont {Hatsuda}},\ }\bibfield  {title} {\bibinfo {title} {{Distribution of Energy-Momentum Tensor around a Static Quark in the Deconfined Phase of SU(3) Yang-Mills Theory}},\ }\href {https://doi.org/10.1103/PhysRevD.102.114522} {\bibfield  {journal} {\bibinfo  {journal} {Phys. Rev. D}\ }\textbf {\bibinfo {volume} {102}},\ \bibinfo {pages} {114522} (\bibinfo {year} {2020})},\ \Eprint {https://arxiv.org/abs/2010.13465} {arXiv:2010.13465 [hep-lat]} \BibitemShut {NoStop}%
\bibitem [{\citenamefont {Kim}\ and\ \citenamefont {Sun}(2021)}]{Kim:2020lrs}%
  \BibitemOpen
  \bibfield  {author} {\bibinfo {author} {\bibfnamefont {J.-Y.}\ \bibnamefont {Kim}}\ and\ \bibinfo {author} {\bibfnamefont {B.-D.}\ \bibnamefont {Sun}},\ }\bibfield  {title} {\bibinfo {title} {{Gravitational form factors of a baryon with spin-3/2}},\ }\href {https://doi.org/10.1140/epjc/s10052-021-08852-z} {\bibfield  {journal} {\bibinfo  {journal} {Eur. Phys. J. C}\ }\textbf {\bibinfo {volume} {81}},\ \bibinfo {pages} {85} (\bibinfo {year} {2021})},\ \Eprint {https://arxiv.org/abs/2011.00292} {arXiv:2011.00292 [hep-ph]} \BibitemShut {NoStop}%
\bibitem [{\citenamefont {Tong}\ \emph {et~al.}(2021)\citenamefont {Tong}, \citenamefont {Ma},\ and\ \citenamefont {Yuan}}]{Tong:2021ctu}%
  \BibitemOpen
  \bibfield  {author} {\bibinfo {author} {\bibfnamefont {X.-B.}\ \bibnamefont {Tong}}, \bibinfo {author} {\bibfnamefont {J.-P.}\ \bibnamefont {Ma}},\ and\ \bibinfo {author} {\bibfnamefont {F.}~\bibnamefont {Yuan}},\ }\bibfield  {title} {\bibinfo {title} {{Gluon gravitational form factors at large momentum transfer}},\ }\href {https://doi.org/10.1016/j.physletb.2021.136751} {\bibfield  {journal} {\bibinfo  {journal} {Phys. Lett. B}\ }\textbf {\bibinfo {volume} {823}},\ \bibinfo {pages} {136751} (\bibinfo {year} {2021})},\ \Eprint {https://arxiv.org/abs/2101.02395} {arXiv:2101.02395 [hep-ph]} \BibitemShut {NoStop}%
\bibitem [{\citenamefont {Freese}\ and\ \citenamefont {Miller}(2021{\natexlab{a}})}]{Freese:2021czn}%
  \BibitemOpen
  \bibfield  {author} {\bibinfo {author} {\bibfnamefont {A.}~\bibnamefont {Freese}}\ and\ \bibinfo {author} {\bibfnamefont {G.~A.}\ \bibnamefont {Miller}},\ }\bibfield  {title} {\bibinfo {title} {{Forces within hadrons on the light front}},\ }\href {https://doi.org/10.1103/PhysRevD.103.094023} {\bibfield  {journal} {\bibinfo  {journal} {Phys. Rev. D}\ }\textbf {\bibinfo {volume} {103}},\ \bibinfo {pages} {094023} (\bibinfo {year} {2021}{\natexlab{a}})},\ \Eprint {https://arxiv.org/abs/2102.01683} {arXiv:2102.01683 [hep-ph]} \BibitemShut {NoStop}%
\bibitem [{\citenamefont {Panteleeva}\ and\ \citenamefont {Polyakov}(2021)}]{Panteleeva:2021iip}%
  \BibitemOpen
  \bibfield  {author} {\bibinfo {author} {\bibfnamefont {J.~Y.}\ \bibnamefont {Panteleeva}}\ and\ \bibinfo {author} {\bibfnamefont {M.~V.}\ \bibnamefont {Polyakov}},\ }\bibfield  {title} {\bibinfo {title} {{Forces inside the nucleon on the light front from 3D Breit frame force distributions: Abel tomography case}},\ }\href {https://doi.org/10.1103/PhysRevD.104.014008} {\bibfield  {journal} {\bibinfo  {journal} {Phys. Rev. D}\ }\textbf {\bibinfo {volume} {104}},\ \bibinfo {pages} {014008} (\bibinfo {year} {2021})},\ \Eprint {https://arxiv.org/abs/2102.10902} {arXiv:2102.10902 [hep-ph]} \BibitemShut {NoStop}%
\bibitem [{\citenamefont {Hatta}\ and\ \citenamefont {Strikman}(2021)}]{Hatta:2021can}%
  \BibitemOpen
  \bibfield  {author} {\bibinfo {author} {\bibfnamefont {Y.}~\bibnamefont {Hatta}}\ and\ \bibinfo {author} {\bibfnamefont {M.}~\bibnamefont {Strikman}},\ }\bibfield  {title} {\bibinfo {title} {{$\phi$-meson lepto-production near threshold and the strangeness $D$-term}},\ }\href {https://doi.org/10.1016/j.physletb.2021.136295} {\bibfield  {journal} {\bibinfo  {journal} {Phys. Lett. B}\ }\textbf {\bibinfo {volume} {817}},\ \bibinfo {pages} {136295} (\bibinfo {year} {2021})},\ \Eprint {https://arxiv.org/abs/2102.12631} {arXiv:2102.12631 [hep-ph]} \BibitemShut {NoStop}%
\bibitem [{\citenamefont {Mamo}\ and\ \citenamefont {Zahed}(2021)}]{Mamo:2021krl}%
  \BibitemOpen
  \bibfield  {author} {\bibinfo {author} {\bibfnamefont {K.~A.}\ \bibnamefont {Mamo}}\ and\ \bibinfo {author} {\bibfnamefont {I.}~\bibnamefont {Zahed}},\ }\bibfield  {title} {\bibinfo {title} {{Nucleon mass radii and distribution: Holographic QCD, Lattice QCD and GlueX data}},\ }\href {https://doi.org/10.1103/PhysRevD.103.094010} {\bibfield  {journal} {\bibinfo  {journal} {Phys. Rev. D}\ }\textbf {\bibinfo {volume} {103}},\ \bibinfo {pages} {094010} (\bibinfo {year} {2021})},\ \Eprint {https://arxiv.org/abs/2103.03186} {arXiv:2103.03186 [hep-ph]} \BibitemShut {NoStop}%
\bibitem [{\citenamefont {Freese}\ and\ \citenamefont {Miller}(2021{\natexlab{b}})}]{Freese:2021qtb}%
  \BibitemOpen
  \bibfield  {author} {\bibinfo {author} {\bibfnamefont {A.}~\bibnamefont {Freese}}\ and\ \bibinfo {author} {\bibfnamefont {G.~A.}\ \bibnamefont {Miller}},\ }\bibfield  {title} {\bibinfo {title} {{Genuine empirical pressure within the proton}},\ }\href {https://doi.org/10.1103/PhysRevD.104.014024} {\bibfield  {journal} {\bibinfo  {journal} {Phys. Rev. D}\ }\textbf {\bibinfo {volume} {104}},\ \bibinfo {pages} {014024} (\bibinfo {year} {2021}{\natexlab{b}})},\ \Eprint {https://arxiv.org/abs/2104.03213} {arXiv:2104.03213 [hep-ph]} \BibitemShut {NoStop}%
\bibitem [{\citenamefont {Gegelia}\ and\ \citenamefont {Polyakov}(2021)}]{Gegelia:2021wnj}%
  \BibitemOpen
  \bibfield  {author} {\bibinfo {author} {\bibfnamefont {J.}~\bibnamefont {Gegelia}}\ and\ \bibinfo {author} {\bibfnamefont {M.~V.}\ \bibnamefont {Polyakov}},\ }\bibfield  {title} {\bibinfo {title} {{A bound on the nucleon Druck-term from chiral EFT in curved space-time and mechanical stability conditions}},\ }\href {https://doi.org/10.1016/j.physletb.2021.136572} {\bibfield  {journal} {\bibinfo  {journal} {Phys. Lett. B}\ }\textbf {\bibinfo {volume} {820}},\ \bibinfo {pages} {136572} (\bibinfo {year} {2021})},\ \Eprint {https://arxiv.org/abs/2104.13954} {arXiv:2104.13954 [hep-ph]} \BibitemShut {NoStop}%
\bibitem [{\citenamefont {Kim}\ and\ \citenamefont {Kim}(2021)}]{Kim:2021jjf}%
  \BibitemOpen
  \bibfield  {author} {\bibinfo {author} {\bibfnamefont {J.-Y.}\ \bibnamefont {Kim}}\ and\ \bibinfo {author} {\bibfnamefont {H.-C.}\ \bibnamefont {Kim}},\ }\bibfield  {title} {\bibinfo {title} {{Energy-momentum tensor of the nucleon on the light front: Abel tomography case}},\ }\href {https://doi.org/10.1103/PhysRevD.104.074019} {\bibfield  {journal} {\bibinfo  {journal} {Phys. Rev. D}\ }\textbf {\bibinfo {volume} {104}},\ \bibinfo {pages} {074019} (\bibinfo {year} {2021})},\ \Eprint {https://arxiv.org/abs/2105.10279} {arXiv:2105.10279 [hep-ph]} \BibitemShut {NoStop}%
\bibitem [{\citenamefont {Owa}\ \emph {et~al.}(2022)\citenamefont {Owa}, \citenamefont {Thomas},\ and\ \citenamefont {Wang}}]{Owa:2021hnj}%
  \BibitemOpen
  \bibfield  {author} {\bibinfo {author} {\bibfnamefont {S.}~\bibnamefont {Owa}}, \bibinfo {author} {\bibfnamefont {A.~W.}\ \bibnamefont {Thomas}},\ and\ \bibinfo {author} {\bibfnamefont {X.~G.}\ \bibnamefont {Wang}},\ }\bibfield  {title} {\bibinfo {title} {{Effect of the pion field on the distributions of pressure and shear in the proton}},\ }\href {https://doi.org/10.1016/j.physletb.2022.137136} {\bibfield  {journal} {\bibinfo  {journal} {Phys. Lett. B}\ }\textbf {\bibinfo {volume} {829}},\ \bibinfo {pages} {137136} (\bibinfo {year} {2022})},\ \Eprint {https://arxiv.org/abs/2106.00929} {arXiv:2106.00929 [hep-ph]} \BibitemShut {NoStop}%
\bibitem [{\citenamefont {Pefkou}\ \emph {et~al.}(2022)\citenamefont {Pefkou}, \citenamefont {Hackett},\ and\ \citenamefont {Shanahan}}]{Pefkou:2021fni}%
  \BibitemOpen
  \bibfield  {author} {\bibinfo {author} {\bibfnamefont {D.~A.}\ \bibnamefont {Pefkou}}, \bibinfo {author} {\bibfnamefont {D.~C.}\ \bibnamefont {Hackett}},\ and\ \bibinfo {author} {\bibfnamefont {P.~E.}\ \bibnamefont {Shanahan}},\ }\bibfield  {title} {\bibinfo {title} {{Gluon gravitational structure of hadrons of different spin}},\ }\href {https://doi.org/10.1103/PhysRevD.105.054509} {\bibfield  {journal} {\bibinfo  {journal} {Phys. Rev. D}\ }\textbf {\bibinfo {volume} {105}},\ \bibinfo {pages} {054509} (\bibinfo {year} {2022})},\ \Eprint {https://arxiv.org/abs/2107.10368} {arXiv:2107.10368 [hep-lat]} \BibitemShut {NoStop}%
\bibitem [{\citenamefont {Lorc\'e}\ \emph {et~al.}(2021)\citenamefont {Lorc\'e}, \citenamefont {Metz}, \citenamefont {Pasquini},\ and\ \citenamefont {Rodini}}]{Lorce:2021xku}%
  \BibitemOpen
  \bibfield  {author} {\bibinfo {author} {\bibfnamefont {C.}~\bibnamefont {Lorc\'e}}, \bibinfo {author} {\bibfnamefont {A.}~\bibnamefont {Metz}}, \bibinfo {author} {\bibfnamefont {B.}~\bibnamefont {Pasquini}},\ and\ \bibinfo {author} {\bibfnamefont {S.}~\bibnamefont {Rodini}},\ }\bibfield  {title} {\bibinfo {title} {{Energy-momentum tensor in QCD: nucleon mass decomposition and mechanical equilibrium}},\ }\href {https://doi.org/10.1007/JHEP11(2021)121} {\bibfield  {journal} {\bibinfo  {journal} {JHEP}\ }\textbf {\bibinfo {volume} {11}},\ \bibinfo {pages} {121}},\ \Eprint {https://arxiv.org/abs/2109.11785} {arXiv:2109.11785 [hep-ph]} \BibitemShut {NoStop}%
\bibitem [{\citenamefont {Ji}\ and\ \citenamefont {Liu}(2022)}]{Ji:2021mfb}%
  \BibitemOpen
  \bibfield  {author} {\bibinfo {author} {\bibfnamefont {X.}~\bibnamefont {Ji}}\ and\ \bibinfo {author} {\bibfnamefont {Y.}~\bibnamefont {Liu}},\ }\bibfield  {title} {\bibinfo {title} {{Momentum-Current Gravitational Multipoles of Hadrons}},\ }\href {https://doi.org/10.1103/PhysRevD.106.034028} {\bibfield  {journal} {\bibinfo  {journal} {Phys. Rev. D}\ }\textbf {\bibinfo {volume} {106}},\ \bibinfo {pages} {034028} (\bibinfo {year} {2022})},\ \Eprint {https://arxiv.org/abs/2110.14781} {arXiv:2110.14781 [hep-ph]} \BibitemShut {NoStop}%
\bibitem [{\citenamefont {More}\ \emph {et~al.}(2022)\citenamefont {More}, \citenamefont {Mukherjee}, \citenamefont {Nair},\ and\ \citenamefont {Saha}}]{More:2021stk}%
  \BibitemOpen
  \bibfield  {author} {\bibinfo {author} {\bibfnamefont {J.}~\bibnamefont {More}}, \bibinfo {author} {\bibfnamefont {A.}~\bibnamefont {Mukherjee}}, \bibinfo {author} {\bibfnamefont {S.}~\bibnamefont {Nair}},\ and\ \bibinfo {author} {\bibfnamefont {S.}~\bibnamefont {Saha}},\ }\bibfield  {title} {\bibinfo {title} {{Gravitational form factors and mechanical properties of a quark at one loop in light-front Hamiltonian QCD}},\ }\href {https://doi.org/10.1103/PhysRevD.105.056017} {\bibfield  {journal} {\bibinfo  {journal} {Phys. Rev. D}\ }\textbf {\bibinfo {volume} {105}},\ \bibinfo {pages} {056017} (\bibinfo {year} {2022})},\ \Eprint {https://arxiv.org/abs/2112.06550} {arXiv:2112.06550 [hep-ph]} \BibitemShut {NoStop}%
\bibitem [{\citenamefont {Mamo}\ and\ \citenamefont {Zahed}(2022)}]{Mamo:2022eui}%
  \BibitemOpen
  \bibfield  {author} {\bibinfo {author} {\bibfnamefont {K.~A.}\ \bibnamefont {Mamo}}\ and\ \bibinfo {author} {\bibfnamefont {I.}~\bibnamefont {Zahed}},\ }\bibfield  {title} {\bibinfo {title} {{J/\ensuremath{\psi} near threshold in holographic QCD: A and D gravitational form factors}},\ }\href {https://doi.org/10.1103/PhysRevD.106.086004} {\bibfield  {journal} {\bibinfo  {journal} {Phys. Rev. D}\ }\textbf {\bibinfo {volume} {106}},\ \bibinfo {pages} {086004} (\bibinfo {year} {2022})},\ \Eprint {https://arxiv.org/abs/2204.08857} {arXiv:2204.08857 [hep-ph]} \BibitemShut {NoStop}%
\bibitem [{\citenamefont {Lorc\'e}\ \emph {et~al.}(2022)\citenamefont {Lorc\'e}, \citenamefont {Schweitzer},\ and\ \citenamefont {Tezgin}}]{Lorce:2022cle}%
  \BibitemOpen
  \bibfield  {author} {\bibinfo {author} {\bibfnamefont {C.}~\bibnamefont {Lorc\'e}}, \bibinfo {author} {\bibfnamefont {P.}~\bibnamefont {Schweitzer}},\ and\ \bibinfo {author} {\bibfnamefont {K.}~\bibnamefont {Tezgin}},\ }\bibfield  {title} {\bibinfo {title} {{2D energy-momentum tensor distributions of nucleon in a large-Nc quark model from ultrarelativistic to nonrelativistic limit}},\ }\href {https://doi.org/10.1103/PhysRevD.106.014012} {\bibfield  {journal} {\bibinfo  {journal} {Phys. Rev. D}\ }\textbf {\bibinfo {volume} {106}},\ \bibinfo {pages} {014012} (\bibinfo {year} {2022})},\ \Eprint {https://arxiv.org/abs/2202.01192} {arXiv:2202.01192 [hep-ph]} \BibitemShut {NoStop}%
\bibitem [{\citenamefont {Fujita}\ \emph {et~al.}(2022)\citenamefont {Fujita}, \citenamefont {Hatta}, \citenamefont {Sugimoto},\ and\ \citenamefont {Ueda}}]{Fujita:2022jus}%
  \BibitemOpen
  \bibfield  {author} {\bibinfo {author} {\bibfnamefont {M.}~\bibnamefont {Fujita}}, \bibinfo {author} {\bibfnamefont {Y.}~\bibnamefont {Hatta}}, \bibinfo {author} {\bibfnamefont {S.}~\bibnamefont {Sugimoto}},\ and\ \bibinfo {author} {\bibfnamefont {T.}~\bibnamefont {Ueda}},\ }\bibfield  {title} {\bibinfo {title} {{Nucleon D-term in holographic quantum chromodynamics}},\ }\href {https://doi.org/10.1093/ptep/ptac110} {\bibfield  {journal} {\bibinfo  {journal} {PTEP}\ }\textbf {\bibinfo {volume} {2022}},\ \bibinfo {pages} {093B06} (\bibinfo {year} {2022})},\ \Eprint {https://arxiv.org/abs/2206.06578} {arXiv:2206.06578 [hep-th]} \BibitemShut {NoStop}%
\bibitem [{\citenamefont {Choudhary}\ \emph {et~al.}(2022)\citenamefont {Choudhary}, \citenamefont {Gurjar}, \citenamefont {Chakrabarti},\ and\ \citenamefont {Mukherjee}}]{Choudhary:2022den}%
  \BibitemOpen
  \bibfield  {author} {\bibinfo {author} {\bibfnamefont {P.}~\bibnamefont {Choudhary}}, \bibinfo {author} {\bibfnamefont {B.}~\bibnamefont {Gurjar}}, \bibinfo {author} {\bibfnamefont {D.}~\bibnamefont {Chakrabarti}},\ and\ \bibinfo {author} {\bibfnamefont {A.}~\bibnamefont {Mukherjee}},\ }\bibfield  {title} {\bibinfo {title} {{Gravitational form factors and mechanical properties of the proton: Connections between distributions in 2D and 3D}},\ }\href {https://doi.org/10.1103/PhysRevD.106.076004} {\bibfield  {journal} {\bibinfo  {journal} {Phys. Rev. D}\ }\textbf {\bibinfo {volume} {106}},\ \bibinfo {pages} {076004} (\bibinfo {year} {2022})},\ \Eprint {https://arxiv.org/abs/2206.12206} {arXiv:2206.12206 [hep-ph]} \BibitemShut {NoStop}%
\bibitem [{\citenamefont {Kim}\ \emph {et~al.}(2023)\citenamefont {Kim}, \citenamefont {Sun}, \citenamefont {Fu},\ and\ \citenamefont {Kim}}]{Kim:2022wkc}%
  \BibitemOpen
  \bibfield  {author} {\bibinfo {author} {\bibfnamefont {J.-Y.}\ \bibnamefont {Kim}}, \bibinfo {author} {\bibfnamefont {B.-D.}\ \bibnamefont {Sun}}, \bibinfo {author} {\bibfnamefont {D.}~\bibnamefont {Fu}},\ and\ \bibinfo {author} {\bibfnamefont {H.-C.}\ \bibnamefont {Kim}},\ }\bibfield  {title} {\bibinfo {title} {{Mechanical structure of a spin-1 particle}},\ }\href {https://doi.org/10.1103/PhysRevD.107.054007} {\bibfield  {journal} {\bibinfo  {journal} {Phys. Rev. D}\ }\textbf {\bibinfo {volume} {107}},\ \bibinfo {pages} {054007} (\bibinfo {year} {2023})},\ \Eprint {https://arxiv.org/abs/2208.01240} {arXiv:2208.01240 [hep-ph]} \BibitemShut {NoStop}%
\bibitem [{\citenamefont {Alharazin}\ \emph {et~al.}(2022)\citenamefont {Alharazin}, \citenamefont {Epelbaum}, \citenamefont {Gegelia}, \citenamefont {Mei\ss{}ner},\ and\ \citenamefont {Sun}}]{Alharazin:2022wjj}%
  \BibitemOpen
  \bibfield  {author} {\bibinfo {author} {\bibfnamefont {H.}~\bibnamefont {Alharazin}}, \bibinfo {author} {\bibfnamefont {E.}~\bibnamefont {Epelbaum}}, \bibinfo {author} {\bibfnamefont {J.}~\bibnamefont {Gegelia}}, \bibinfo {author} {\bibfnamefont {U.~G.}\ \bibnamefont {Mei\ss{}ner}},\ and\ \bibinfo {author} {\bibfnamefont {B.~D.}\ \bibnamefont {Sun}},\ }\bibfield  {title} {\bibinfo {title} {{Gravitational form factors of the delta resonance in chiral EFT}},\ }\href {https://doi.org/10.1140/epjc/s10052-022-10882-0} {\bibfield  {journal} {\bibinfo  {journal} {Eur. Phys. J. C}\ }\textbf {\bibinfo {volume} {82}},\ \bibinfo {pages} {907} (\bibinfo {year} {2022})},\ \Eprint {https://arxiv.org/abs/2209.01233} {arXiv:2209.01233 [hep-ph]} \BibitemShut {NoStop}%
\bibitem [{\citenamefont {Won}\ \emph {et~al.}(2022)\citenamefont {Won}, \citenamefont {Kim},\ and\ \citenamefont {Kim}}]{Won:2022cyy}%
  \BibitemOpen
  \bibfield  {author} {\bibinfo {author} {\bibfnamefont {H.-Y.}\ \bibnamefont {Won}}, \bibinfo {author} {\bibfnamefont {J.-Y.}\ \bibnamefont {Kim}},\ and\ \bibinfo {author} {\bibfnamefont {H.-C.}\ \bibnamefont {Kim}},\ }\bibfield  {title} {\bibinfo {title} {{Gravitational form factors of the baryon octet with flavor SU(3) symmetry breaking}},\ }\href {https://doi.org/10.1103/PhysRevD.106.114009} {\bibfield  {journal} {\bibinfo  {journal} {Phys. Rev. D}\ }\textbf {\bibinfo {volume} {106}},\ \bibinfo {pages} {114009} (\bibinfo {year} {2022})},\ \Eprint {https://arxiv.org/abs/2210.03320} {arXiv:2210.03320 [hep-ph]} \BibitemShut {NoStop}%
\bibitem [{\citenamefont {Tanaka}(2023)}]{Tanaka:2022wzy}%
  \BibitemOpen
  \bibfield  {author} {\bibinfo {author} {\bibfnamefont {K.}~\bibnamefont {Tanaka}},\ }\bibfield  {title} {\bibinfo {title} {{Twist-four gravitational form factor at NNLO QCD from trace anomaly constraints}},\ }\href {https://doi.org/10.1007/JHEP03(2023)013} {\bibfield  {journal} {\bibinfo  {journal} {JHEP}\ }\textbf {\bibinfo {volume} {03}},\ \bibinfo {pages} {013}},\ \Eprint {https://arxiv.org/abs/2212.09417} {arXiv:2212.09417 [hep-ph]} \BibitemShut {NoStop}%
\bibitem [{\citenamefont {Ito}\ and\ \citenamefont {Kitazawa}(2023)}]{Ito:2023oby}%
  \BibitemOpen
  \bibfield  {author} {\bibinfo {author} {\bibfnamefont {H.}~\bibnamefont {Ito}}\ and\ \bibinfo {author} {\bibfnamefont {M.}~\bibnamefont {Kitazawa}},\ }\bibfield  {title} {\bibinfo {title} {{Gravitational form factors of a kink in 1 + 1 dimensional \ensuremath{\phi}$^{4}$ model}},\ }\href {https://doi.org/10.1007/JHEP08(2023)033} {\bibfield  {journal} {\bibinfo  {journal} {JHEP}\ }\textbf {\bibinfo {volume} {08}},\ \bibinfo {pages} {033}},\ \Eprint {https://arxiv.org/abs/2302.08762} {arXiv:2302.08762 [hep-th]} \BibitemShut {NoStop}%
\bibitem [{\citenamefont {Lorc\'e}\ and\ \citenamefont {Song}(2023)}]{Lorce:2023zzg}%
  \BibitemOpen
  \bibfield  {author} {\bibinfo {author} {\bibfnamefont {C.}~\bibnamefont {Lorc\'e}}\ and\ \bibinfo {author} {\bibfnamefont {Q.-T.}\ \bibnamefont {Song}},\ }\bibfield  {title} {\bibinfo {title} {{Gravitational transverse-momentum distributions}},\ }\href {https://doi.org/10.1016/j.physletb.2023.138016} {\bibfield  {journal} {\bibinfo  {journal} {Phys. Lett. B}\ }\textbf {\bibinfo {volume} {843}},\ \bibinfo {pages} {138016} (\bibinfo {year} {2023})},\ \Eprint {https://arxiv.org/abs/2303.11538} {arXiv:2303.11538 [hep-ph]} \BibitemShut {NoStop}%
\bibitem [{\citenamefont {Amor-Quiroz}\ \emph {et~al.}(2023)\citenamefont {Amor-Quiroz}, \citenamefont {Focillon}, \citenamefont {Lorc\'e},\ and\ \citenamefont {Rodini}}]{Amor-Quiroz:2023rke}%
  \BibitemOpen
  \bibfield  {author} {\bibinfo {author} {\bibfnamefont {A.}~\bibnamefont {Amor-Quiroz}}, \bibinfo {author} {\bibfnamefont {W.}~\bibnamefont {Focillon}}, \bibinfo {author} {\bibfnamefont {C.}~\bibnamefont {Lorc\'e}},\ and\ \bibinfo {author} {\bibfnamefont {S.}~\bibnamefont {Rodini}},\ }\bibfield  {title} {\bibinfo {title} {{Energy\textendash{}momentum tensor in the scalar diquark model}},\ }\href {https://doi.org/10.1140/epjc/s10052-023-12190-7} {\bibfield  {journal} {\bibinfo  {journal} {Eur. Phys. J. C}\ }\textbf {\bibinfo {volume} {83}},\ \bibinfo {pages} {1012} (\bibinfo {year} {2023})},\ \Eprint {https://arxiv.org/abs/2304.10339} {arXiv:2304.10339 [hep-ph]} \BibitemShut {NoStop}%
\bibitem [{\citenamefont {Guo}\ \emph {et~al.}(2023)\citenamefont {Guo}, \citenamefont {Ji}, \citenamefont {Liu},\ and\ \citenamefont {Yang}}]{Guo:2023pqw}%
  \BibitemOpen
  \bibfield  {author} {\bibinfo {author} {\bibfnamefont {Y.}~\bibnamefont {Guo}}, \bibinfo {author} {\bibfnamefont {X.}~\bibnamefont {Ji}}, \bibinfo {author} {\bibfnamefont {Y.}~\bibnamefont {Liu}},\ and\ \bibinfo {author} {\bibfnamefont {J.}~\bibnamefont {Yang}},\ }\bibfield  {title} {\bibinfo {title} {{Updated analysis of near-threshold heavy quarkonium production for probe of proton\textquoteright{}s gluonic gravitational form factors}},\ }\href {https://doi.org/10.1103/PhysRevD.108.034003} {\bibfield  {journal} {\bibinfo  {journal} {Phys. Rev. D}\ }\textbf {\bibinfo {volume} {108}},\ \bibinfo {pages} {034003} (\bibinfo {year} {2023})},\ \Eprint {https://arxiv.org/abs/2305.06992} {arXiv:2305.06992 [hep-ph]} \BibitemShut {NoStop}%
\bibitem [{\citenamefont {Won}\ \emph {et~al.}(2023)\citenamefont {Won}, \citenamefont {Kim},\ and\ \citenamefont {Kim}}]{Won:2023ial}%
  \BibitemOpen
  \bibfield  {author} {\bibinfo {author} {\bibfnamefont {H.-Y.}\ \bibnamefont {Won}}, \bibinfo {author} {\bibfnamefont {H.-C.}\ \bibnamefont {Kim}},\ and\ \bibinfo {author} {\bibfnamefont {J.-Y.}\ \bibnamefont {Kim}},\ }\bibfield  {title} {\bibinfo {title} {{Role of strange quarks in the D-term and cosmological constant term of the proton}},\ }\href {https://doi.org/10.1103/PhysRevD.108.094018} {\bibfield  {journal} {\bibinfo  {journal} {Phys. Rev. D}\ }\textbf {\bibinfo {volume} {108}},\ \bibinfo {pages} {094018} (\bibinfo {year} {2023})},\ \Eprint {https://arxiv.org/abs/2307.00740} {arXiv:2307.00740 [hep-ph]} \BibitemShut {NoStop}%
\bibitem [{\citenamefont {Guo}\ \emph {et~al.}(2024)\citenamefont {Guo}, \citenamefont {Ji},\ and\ \citenamefont {Yuan}}]{Guo:2023qgu}%
  \BibitemOpen
  \bibfield  {author} {\bibinfo {author} {\bibfnamefont {Y.}~\bibnamefont {Guo}}, \bibinfo {author} {\bibfnamefont {X.}~\bibnamefont {Ji}},\ and\ \bibinfo {author} {\bibfnamefont {F.}~\bibnamefont {Yuan}},\ }\bibfield  {title} {\bibinfo {title} {{Proton\textquoteright{}s gluon GPDs at large skewness and gravitational form factors from near threshold heavy quarkonium photoproduction}},\ }\href {https://doi.org/10.1103/PhysRevD.109.014014} {\bibfield  {journal} {\bibinfo  {journal} {Phys. Rev. D}\ }\textbf {\bibinfo {volume} {109}},\ \bibinfo {pages} {014014} (\bibinfo {year} {2024})},\ \Eprint {https://arxiv.org/abs/2308.13006} {arXiv:2308.13006 [hep-ph]} \BibitemShut {NoStop}%
\bibitem [{\citenamefont {Czarnecki}\ \emph {et~al.}(2023)\citenamefont {Czarnecki}, \citenamefont {Liu},\ and\ \citenamefont {Reza}}]{Czarnecki:2023yqd}%
  \BibitemOpen
  \bibfield  {author} {\bibinfo {author} {\bibfnamefont {A.}~\bibnamefont {Czarnecki}}, \bibinfo {author} {\bibfnamefont {Y.}~\bibnamefont {Liu}},\ and\ \bibinfo {author} {\bibfnamefont {S.~N.}\ \bibnamefont {Reza}},\ }\bibfield  {title} {\bibinfo {title} {{Energy-momentum Tensor of a Hydrogen Atom: Stability, $D$-term, and the Lamb Shift}},\ }\href {https://doi.org/10.5506/APhysPolBSupp.16.7-A19} {\bibfield  {journal} {\bibinfo  {journal} {Acta Phys. Polon. Supp.}\ }\textbf {\bibinfo {volume} {16}},\ \bibinfo {pages} {7} (\bibinfo {year} {2023})},\ \Eprint {https://arxiv.org/abs/2309.10994} {arXiv:2309.10994 [hep-ph]} \BibitemShut {NoStop}%
\bibitem [{\citenamefont {Won}\ \emph {et~al.}(2024)\citenamefont {Won}, \citenamefont {Kim},\ and\ \citenamefont {Kim}}]{Won:2023zmf}%
  \BibitemOpen
  \bibfield  {author} {\bibinfo {author} {\bibfnamefont {H.-Y.}\ \bibnamefont {Won}}, \bibinfo {author} {\bibfnamefont {H.-C.}\ \bibnamefont {Kim}},\ and\ \bibinfo {author} {\bibfnamefont {J.-Y.}\ \bibnamefont {Kim}},\ }\bibfield  {title} {\bibinfo {title} {{Mechanical structure of the nucleon and the baryon octet: twist-2 case}},\ }\href {https://doi.org/10.1007/JHEP05(2024)173} {\bibfield  {journal} {\bibinfo  {journal} {JHEP}\ }\textbf {\bibinfo {volume} {05}},\ \bibinfo {pages} {173}},\ \Eprint {https://arxiv.org/abs/2310.04670} {arXiv:2310.04670 [hep-ph]} \BibitemShut {NoStop}%
\bibitem [{\citenamefont {Hackett}\ \emph {et~al.}(2024)\citenamefont {Hackett}, \citenamefont {Pefkou},\ and\ \citenamefont {Shanahan}}]{Hackett:2023rif}%
  \BibitemOpen
  \bibfield  {author} {\bibinfo {author} {\bibfnamefont {D.~C.}\ \bibnamefont {Hackett}}, \bibinfo {author} {\bibfnamefont {D.~A.}\ \bibnamefont {Pefkou}},\ and\ \bibinfo {author} {\bibfnamefont {P.~E.}\ \bibnamefont {Shanahan}},\ }\bibfield  {title} {\bibinfo {title} {{Gravitational Form Factors of the Proton from Lattice QCD}},\ }\href {https://doi.org/10.1103/PhysRevLett.132.251904} {\bibfield  {journal} {\bibinfo  {journal} {Phys. Rev. Lett.}\ }\textbf {\bibinfo {volume} {132}},\ \bibinfo {pages} {251904} (\bibinfo {year} {2024})},\ \Eprint {https://arxiv.org/abs/2310.08484} {arXiv:2310.08484 [hep-lat]} \BibitemShut {NoStop}%
\bibitem [{\citenamefont {Hatta}(2024)}]{Hatta:2023fqc}%
  \BibitemOpen
  \bibfield  {author} {\bibinfo {author} {\bibfnamefont {Y.}~\bibnamefont {Hatta}},\ }\bibfield  {title} {\bibinfo {title} {{Accessing the gravitational form factors of the nucleon and nuclei through a massive graviton}},\ }\href {https://doi.org/10.1103/PhysRevD.109.L051502} {\bibfield  {journal} {\bibinfo  {journal} {Phys. Rev. D}\ }\textbf {\bibinfo {volume} {109}},\ \bibinfo {pages} {L051502} (\bibinfo {year} {2024})},\ \Eprint {https://arxiv.org/abs/2311.14470} {arXiv:2311.14470 [hep-ph]} \BibitemShut {NoStop}%
\bibitem [{\citenamefont {Liu}(2024)}]{Liu:2023cse}%
  \BibitemOpen
  \bibfield  {author} {\bibinfo {author} {\bibfnamefont {K.-F.}\ \bibnamefont {Liu}},\ }\bibfield  {title} {\bibinfo {title} {{Hadrons, superconductor vortices, and cosmological constant}},\ }\href {https://doi.org/10.1016/j.physletb.2023.138418} {\bibfield  {journal} {\bibinfo  {journal} {Phys. Lett. B}\ }\textbf {\bibinfo {volume} {849}},\ \bibinfo {pages} {138418} (\bibinfo {year} {2024})},\ \Eprint {https://arxiv.org/abs/2302.11600} {arXiv:2302.11600 [hep-ph]} \BibitemShut {NoStop}%
\bibitem [{\citenamefont {Cao}\ \emph {et~al.}(2024)\citenamefont {Cao}, \citenamefont {Guo}, \citenamefont {Li},\ and\ \citenamefont {Yao}}]{Cao:2024zlf}%
  \BibitemOpen
  \bibfield  {author} {\bibinfo {author} {\bibfnamefont {X.-H.}\ \bibnamefont {Cao}}, \bibinfo {author} {\bibfnamefont {F.-K.}\ \bibnamefont {Guo}}, \bibinfo {author} {\bibfnamefont {Q.-Z.}\ \bibnamefont {Li}},\ and\ \bibinfo {author} {\bibfnamefont {D.-L.}\ \bibnamefont {Yao}},\ }\bibfield  {title} {\bibinfo {title} {{Precise Determination of Nucleon Gravitational Form Factors}},\ }\href@noop {} {\  (\bibinfo {year} {2024})},\ \Eprint {https://arxiv.org/abs/2411.13398} {arXiv:2411.13398 [hep-ph]} \BibitemShut {NoStop}%
\bibitem [{\citenamefont {Liu}\ \emph {et~al.}(2024)\citenamefont {Liu}, \citenamefont {Shuryak},\ and\ \citenamefont {Zahed}}]{Liu:2024rdm}%
  \BibitemOpen
  \bibfield  {author} {\bibinfo {author} {\bibfnamefont {W.-Y.}\ \bibnamefont {Liu}}, \bibinfo {author} {\bibfnamefont {E.}~\bibnamefont {Shuryak}},\ and\ \bibinfo {author} {\bibfnamefont {I.}~\bibnamefont {Zahed}},\ }\bibfield  {title} {\bibinfo {title} {{Glue in hadrons at medium resolution and the QCD instanton vacuum}},\ }\href {https://doi.org/10.1103/PhysRevD.110.054005} {\bibfield  {journal} {\bibinfo  {journal} {Phys. Rev. D}\ }\textbf {\bibinfo {volume} {110}},\ \bibinfo {pages} {054005} (\bibinfo {year} {2024})},\ \Eprint {https://arxiv.org/abs/2404.03047} {arXiv:2404.03047 [hep-ph]} \BibitemShut {NoStop}%
\bibitem [{\citenamefont {Yao}\ \emph {et~al.}(2024)\citenamefont {Yao}, \citenamefont {Xu}, \citenamefont {Binosi}, \citenamefont {Cui}, \citenamefont {Ding}, \citenamefont {Raya}, \citenamefont {Roberts}, \citenamefont {Rodr\'\i{}guez-Quintero},\ and\ \citenamefont {Schmidt}}]{Yao:2024ixu}%
  \BibitemOpen
  \bibfield  {author} {\bibinfo {author} {\bibfnamefont {Z.~Q.}\ \bibnamefont {Yao}}, \bibinfo {author} {\bibfnamefont {Y.~Z.}\ \bibnamefont {Xu}}, \bibinfo {author} {\bibfnamefont {D.}~\bibnamefont {Binosi}}, \bibinfo {author} {\bibfnamefont {Z.~F.}\ \bibnamefont {Cui}}, \bibinfo {author} {\bibfnamefont {M.}~\bibnamefont {Ding}}, \bibinfo {author} {\bibfnamefont {K.}~\bibnamefont {Raya}}, \bibinfo {author} {\bibfnamefont {C.~D.}\ \bibnamefont {Roberts}}, \bibinfo {author} {\bibfnamefont {J.}~\bibnamefont {Rodr\'\i{}guez-Quintero}},\ and\ \bibinfo {author} {\bibfnamefont {S.~M.}\ \bibnamefont {Schmidt}},\ }\bibfield  {title} {\bibinfo {title} {{Nucleon Gravitational Form Factors}},\ }\href@noop {} {\  (\bibinfo {year} {2024})},\ \Eprint {https://arxiv.org/abs/2409.15547} {arXiv:2409.15547 [hep-ph]} \BibitemShut {NoStop}%
\bibitem [{\citenamefont {Goharipour}\ \emph {et~al.}(2025{\natexlab{a}})\citenamefont {Goharipour}, \citenamefont {Hashamipour}, \citenamefont {Fatehi}, \citenamefont {Irani}, \citenamefont {Azizi},\ and\ \citenamefont {Goloskokov}}]{Goharipour:2025lep}%
  \BibitemOpen
  \bibfield  {author} {\bibinfo {author} {\bibfnamefont {M.}~\bibnamefont {Goharipour}}, \bibinfo {author} {\bibfnamefont {H.}~\bibnamefont {Hashamipour}}, \bibinfo {author} {\bibfnamefont {H.}~\bibnamefont {Fatehi}}, \bibinfo {author} {\bibfnamefont {F.}~\bibnamefont {Irani}}, \bibinfo {author} {\bibfnamefont {K.}~\bibnamefont {Azizi}},\ and\ \bibinfo {author} {\bibfnamefont {S.~V.}\ \bibnamefont {Goloskokov}} (\bibinfo {collaboration} {MMGPDs}),\ }\bibfield  {title} {\bibinfo {title} {{Mechanical properties of the nucleon from the generalized parton distributions}},\ }\href@noop {} {\  (\bibinfo {year} {2025}{\natexlab{a}})},\ \Eprint {https://arxiv.org/abs/2501.16257} {arXiv:2501.16257 [hep-ph]} \BibitemShut {NoStop}%
\bibitem [{\citenamefont {Dehghan}\ \emph {et~al.}(2025)\citenamefont {Dehghan}, \citenamefont {Almaksusi},\ and\ \citenamefont {Azizi}}]{Dehghan:2025ncw}%
  \BibitemOpen
  \bibfield  {author} {\bibinfo {author} {\bibfnamefont {Z.}~\bibnamefont {Dehghan}}, \bibinfo {author} {\bibfnamefont {F.}~\bibnamefont {Almaksusi}},\ and\ \bibinfo {author} {\bibfnamefont {K.}~\bibnamefont {Azizi}},\ }\bibfield  {title} {\bibinfo {title} {{Mechanical properties of proton using flavor-decomposed gravitational form factors}},\ }\href@noop {} {\  (\bibinfo {year} {2025})},\ \Eprint {https://arxiv.org/abs/2502.16689} {arXiv:2502.16689 [hep-ph]} \BibitemShut {NoStop}%
\bibitem [{\citenamefont {Goharipour}\ \emph {et~al.}(2025{\natexlab{b}})\citenamefont {Goharipour}, \citenamefont {Irani}, \citenamefont {Amiri}, \citenamefont {Fatehi}, \citenamefont {Falahi}, \citenamefont {Moradi},\ and\ \citenamefont {Azizi}}]{Goharipour:2025yxm}%
  \BibitemOpen
  \bibfield  {author} {\bibinfo {author} {\bibfnamefont {M.}~\bibnamefont {Goharipour}}, \bibinfo {author} {\bibfnamefont {F.}~\bibnamefont {Irani}}, \bibinfo {author} {\bibfnamefont {M.~H.}\ \bibnamefont {Amiri}}, \bibinfo {author} {\bibfnamefont {H.}~\bibnamefont {Fatehi}}, \bibinfo {author} {\bibfnamefont {B.}~\bibnamefont {Falahi}}, \bibinfo {author} {\bibfnamefont {A.}~\bibnamefont {Moradi}},\ and\ \bibinfo {author} {\bibfnamefont {K.}~\bibnamefont {Azizi}} (\bibinfo {collaboration} {MMGPDs}),\ }\bibfield  {title} {\bibinfo {title} {{Can we determine the exact size of the nucleon?: A comprehensive study of different radii}},\ }\href@noop {} {\  (\bibinfo {year} {2025}{\natexlab{b}})},\ \Eprint {https://arxiv.org/abs/2503.08847} {arXiv:2503.08847 [hep-ph]} \BibitemShut {NoStop}%
\bibitem [{\citenamefont {Broniowski}\ and\ \citenamefont {Ruiz~Arriola}(2025)}]{Broniowski:2025ctl}%
  \BibitemOpen
  \bibfield  {author} {\bibinfo {author} {\bibfnamefont {W.}~\bibnamefont {Broniowski}}\ and\ \bibinfo {author} {\bibfnamefont {E.}~\bibnamefont {Ruiz~Arriola}},\ }\bibfield  {title} {\bibinfo {title} {{Gravitational form factors and mechanical properties of the nucleon in a meson dominance approach}},\ }\href@noop {} {\  (\bibinfo {year} {2025})},\ \Eprint {https://arxiv.org/abs/2503.09297} {arXiv:2503.09297 [hep-ph]} \BibitemShut {NoStop}%
\bibitem [{\citenamefont {Ghim}\ \emph {et~al.}(2025)\citenamefont {Ghim}, \citenamefont {Won}, \citenamefont {Kim},\ and\ \citenamefont {Kim}}]{Ghim:2025gqo}%
  \BibitemOpen
  \bibfield  {author} {\bibinfo {author} {\bibfnamefont {N.-Y.}\ \bibnamefont {Ghim}}, \bibinfo {author} {\bibfnamefont {H.-Y.}\ \bibnamefont {Won}}, \bibinfo {author} {\bibfnamefont {J.-Y.}\ \bibnamefont {Kim}},\ and\ \bibinfo {author} {\bibfnamefont {H.-C.}\ \bibnamefont {Kim}},\ }\bibfield  {title} {\bibinfo {title} {{Nucleon tensor form factors at large $N_{c}$}},\ }\href@noop {} {\  (\bibinfo {year} {2025})},\ \Eprint {https://arxiv.org/abs/2501.12241} {arXiv:2501.12241 [hep-ph]} \BibitemShut {NoStop}%
\bibitem [{\citenamefont {Hatta}\ \emph {et~al.}(2025)\citenamefont {Hatta}, \citenamefont {Klest}, \citenamefont {Passek-K.},\ and\ \citenamefont {Schoenleber}}]{Hatta:2025vhs}%
  \BibitemOpen
  \bibfield  {author} {\bibinfo {author} {\bibfnamefont {Y.}~\bibnamefont {Hatta}}, \bibinfo {author} {\bibfnamefont {H.~T.}\ \bibnamefont {Klest}}, \bibinfo {author} {\bibfnamefont {K.}~\bibnamefont {Passek-K.}},\ and\ \bibinfo {author} {\bibfnamefont {J.}~\bibnamefont {Schoenleber}},\ }\bibfield  {title} {\bibinfo {title} {{Deeply virtual $\phi$-meson production near threshold}},\ }\href@noop {} {\  (\bibinfo {year} {2025})},\ \Eprint {https://arxiv.org/abs/2501.12343} {arXiv:2501.12343 [hep-ph]} \BibitemShut {NoStop}%
\bibitem [{\citenamefont {Hatta}\ and\ \citenamefont {Schoenleber}(2025)}]{Hatta:2025ryj}%
  \BibitemOpen
  \bibfield  {author} {\bibinfo {author} {\bibfnamefont {Y.}~\bibnamefont {Hatta}}\ and\ \bibinfo {author} {\bibfnamefont {J.}~\bibnamefont {Schoenleber}},\ }\bibfield  {title} {\bibinfo {title} {{Sullivan process near threshold and the pion gravitational form factors}},\ }\href@noop {} {\  (\bibinfo {year} {2025})},\ \Eprint {https://arxiv.org/abs/2502.12061} {arXiv:2502.12061 [hep-ph]} \BibitemShut {NoStop}%
\bibitem [{\citenamefont {Dehghan}\ and\ \citenamefont {Azizi}(2025)}]{Dehghan:2025eov}%
  \BibitemOpen
  \bibfield  {author} {\bibinfo {author} {\bibfnamefont {Z.}~\bibnamefont {Dehghan}}\ and\ \bibinfo {author} {\bibfnamefont {K.}~\bibnamefont {Azizi}},\ }\bibfield  {title} {\bibinfo {title} {{Mechanical properties of the $\Omega^-$ baryon from gravitational form factors}},\ }\href@noop {} {\  (\bibinfo {year} {2025})},\ \Eprint {https://arxiv.org/abs/2507.14840} {arXiv:2507.14840 [hep-ph]} \BibitemShut {NoStop}%
\bibitem [{\citenamefont {Guo}\ \emph {et~al.}(2025)\citenamefont {Guo}, \citenamefont {Yuan},\ and\ \citenamefont {Zhao}}]{Guo:2025jiz}%
  \BibitemOpen
  \bibfield  {author} {\bibinfo {author} {\bibfnamefont {Y.}~\bibnamefont {Guo}}, \bibinfo {author} {\bibfnamefont {F.}~\bibnamefont {Yuan}},\ and\ \bibinfo {author} {\bibfnamefont {W.}~\bibnamefont {Zhao}},\ }\bibfield  {title} {\bibinfo {title} {{Bayesian Inferring Nucleon's Gravitation Form Factors via Near-threshold $J/\psi$ Photoproduction}},\ }\href@noop {} {\  (\bibinfo {year} {2025})},\ \Eprint {https://arxiv.org/abs/2501.10532} {arXiv:2501.10532 [hep-ph]} \BibitemShut {NoStop}%
\bibitem [{\citenamefont {Liu}\ and\ \citenamefont {Watanabe}(2025)}]{Liu:2025vfe}%
  \BibitemOpen
  \bibfield  {author} {\bibinfo {author} {\bibfnamefont {Z.}~\bibnamefont {Liu}}\ and\ \bibinfo {author} {\bibfnamefont {A.}~\bibnamefont {Watanabe}},\ }\bibfield  {title} {\bibinfo {title} {{Gravitational form factor of the kaon in holographic QCD}},\ }\href@noop {} {\  (\bibinfo {year} {2025})},\ \Eprint {https://arxiv.org/abs/2503.18747} {arXiv:2503.18747 [hep-ph]} \BibitemShut {NoStop}%
\bibitem [{\citenamefont {Sugimoto}\ and\ \citenamefont {Tsukamoto}(2025)}]{Sugimoto:2025btn}%
  \BibitemOpen
  \bibfield  {author} {\bibinfo {author} {\bibfnamefont {S.}~\bibnamefont {Sugimoto}}\ and\ \bibinfo {author} {\bibfnamefont {T.}~\bibnamefont {Tsukamoto}},\ }\bibfield  {title} {\bibinfo {title} {{Energy-Momentum Tensor and D-term of Baryons in Top-down Holographic QCD}},\ }\href@noop {} {\  (\bibinfo {year} {2025})},\ \Eprint {https://arxiv.org/abs/2503.19492} {arXiv:2503.19492 [hep-th]} \BibitemShut {NoStop}%
\bibitem [{\citenamefont {Nair}\ \emph {et~al.}(2025)\citenamefont {Nair}, \citenamefont {Mondal}, \citenamefont {Xu}, \citenamefont {Zhao},\ and\ \citenamefont {Vary}}]{Nair:2025sfr}%
  \BibitemOpen
  \bibfield  {author} {\bibinfo {author} {\bibfnamefont {S.}~\bibnamefont {Nair}}, \bibinfo {author} {\bibfnamefont {C.}~\bibnamefont {Mondal}}, \bibinfo {author} {\bibfnamefont {S.}~\bibnamefont {Xu}}, \bibinfo {author} {\bibfnamefont {X.}~\bibnamefont {Zhao}},\ and\ \bibinfo {author} {\bibfnamefont {J.~P.}\ \bibnamefont {Vary}},\ }\bibfield  {title} {\bibinfo {title} {{Proton Gravitational Structure and Mass Decomposition on the Light Front}},\ }\href@noop {} {\  (\bibinfo {year} {2025})},\ \Eprint {https://arxiv.org/abs/2506.07554} {arXiv:2506.07554 [hep-ph]} \BibitemShut {NoStop}%
\bibitem [{\citenamefont {Cao}\ \emph {et~al.}(2025)\citenamefont {Cao}, \citenamefont {Guo}, \citenamefont {Li}, \citenamefont {Wu},\ and\ \citenamefont {Yao}}]{Cao:2025dkv}%
  \BibitemOpen
  \bibfield  {author} {\bibinfo {author} {\bibfnamefont {X.-H.}\ \bibnamefont {Cao}}, \bibinfo {author} {\bibfnamefont {F.-K.}\ \bibnamefont {Guo}}, \bibinfo {author} {\bibfnamefont {Q.-Z.}\ \bibnamefont {Li}}, \bibinfo {author} {\bibfnamefont {B.-W.}\ \bibnamefont {Wu}},\ and\ \bibinfo {author} {\bibfnamefont {D.-L.}\ \bibnamefont {Yao}},\ }\bibfield  {title} {\bibinfo {title} {{Gravitational form factors of pions, kaons and nucleons from dispersion relations}},\ }\href@noop {} {\  (\bibinfo {year} {2025})},\ \Eprint {https://arxiv.org/abs/2507.05375} {arXiv:2507.05375 [hep-ph]} \BibitemShut {NoStop}%
\bibitem [{\citenamefont {Corian{\`o}}\ \emph {et~al.}(2024)\citenamefont {Corian{\`o}}, \citenamefont {Lionetti}, \citenamefont {Melle},\ and\ \citenamefont {Tommasi}}]{Coriano:2024wrz}%
  \BibitemOpen
  \bibfield  {author} {\bibinfo {author} {\bibfnamefont {C.}~\bibnamefont {Corian{\`o}}}, \bibinfo {author} {\bibfnamefont {S.}~\bibnamefont {Lionetti}}, \bibinfo {author} {\bibfnamefont {D.}~\bibnamefont {Melle}},\ and\ \bibinfo {author} {\bibfnamefont {R.}~\bibnamefont {Tommasi}},\ }\bibfield  {title} {\bibinfo {title} {{The gravitational form factor of the pion and proton and the conformal anomaly}},\ }\href {https://doi.org/10.1051/epjconf/202431400030} {\bibfield  {journal} {\bibinfo  {journal} {EPJ Web Conf.}\ }\textbf {\bibinfo {volume} {314}},\ \bibinfo {pages} {00030} (\bibinfo {year} {2024})},\ \Eprint {https://arxiv.org/abs/2409.19586} {arXiv:2409.19586 [hep-ph]} \BibitemShut {NoStop}%
\bibitem [{\citenamefont {Corian{\`o}}\ \emph {et~al.}(2025)\citenamefont {Corian{\`o}}, \citenamefont {Lionetti}, \citenamefont {Melle}, \citenamefont {Tommasi},\ and\ \citenamefont {Torcellini}}]{Coriano:2025lge}%
  \BibitemOpen
  \bibfield  {author} {\bibinfo {author} {\bibfnamefont {C.}~\bibnamefont {Corian{\`o}}}, \bibinfo {author} {\bibfnamefont {S.}~\bibnamefont {Lionetti}}, \bibinfo {author} {\bibfnamefont {D.}~\bibnamefont {Melle}}, \bibinfo {author} {\bibfnamefont {R.}~\bibnamefont {Tommasi}},\ and\ \bibinfo {author} {\bibfnamefont {L.}~\bibnamefont {Torcellini}},\ }\bibfield  {title} {\bibinfo {title} {{Gravitational Form Factors and the QCD Dilaton at Large Momentum Transfer}},\ }in\ \href@noop {} {\emph {\bibinfo {booktitle} {{24th Hellenic School and Workshops on Elementary Particle Physics and Gravity}}}}\ (\bibinfo {year} {2025})\ \Eprint {https://arxiv.org/abs/2504.20884} {arXiv:2504.20884 [hep-ph]} \BibitemShut {NoStop}%
\bibitem [{\citenamefont {Stegeman}\ and\ \citenamefont {Zwicky}(2025)}]{Stegeman:2025sca}%
  \BibitemOpen
  \bibfield  {author} {\bibinfo {author} {\bibfnamefont {R.}~\bibnamefont {Stegeman}}\ and\ \bibinfo {author} {\bibfnamefont {R.}~\bibnamefont {Zwicky}},\ }\bibfield  {title} {\bibinfo {title} {{Gravitational $ D$-Form Factor: The $\sigma$-Meson as a Dilaton confronted with Lattice Data}},\ }\href@noop {} {\  (\bibinfo {year} {2025})},\ \Eprint {https://arxiv.org/abs/2508.18537} {arXiv:2508.18537 [hep-ph]} \BibitemShut {NoStop}%
\bibitem [{\citenamefont {Polyakov}\ and\ \citenamefont {Schweitzer}(2018)}]{Polyakov:2018zvc}%
  \BibitemOpen
  \bibfield  {author} {\bibinfo {author} {\bibfnamefont {M.~V.}\ \bibnamefont {Polyakov}}\ and\ \bibinfo {author} {\bibfnamefont {P.}~\bibnamefont {Schweitzer}},\ }\bibfield  {title} {\bibinfo {title} {{Forces inside hadrons: pressure, surface tension, mechanical radius, and all that}},\ }\href {https://doi.org/10.1142/S0217751X18300259} {\bibfield  {journal} {\bibinfo  {journal} {Int. J. Mod. Phys. A}\ }\textbf {\bibinfo {volume} {33}},\ \bibinfo {pages} {1830025} (\bibinfo {year} {2018})},\ \Eprint {https://arxiv.org/abs/1805.06596} {arXiv:1805.06596 [hep-ph]} \BibitemShut {NoStop}%
\bibitem [{\citenamefont {Burkert}\ \emph {et~al.}(2023)\citenamefont {Burkert}, \citenamefont {Elouadrhiri}, \citenamefont {Girod}, \citenamefont {Lorc\'e}, \citenamefont {Schweitzer},\ and\ \citenamefont {Shanahan}}]{Burkert:2023wzr}%
  \BibitemOpen
  \bibfield  {author} {\bibinfo {author} {\bibfnamefont {V.~D.}\ \bibnamefont {Burkert}}, \bibinfo {author} {\bibfnamefont {L.}~\bibnamefont {Elouadrhiri}}, \bibinfo {author} {\bibfnamefont {F.~X.}\ \bibnamefont {Girod}}, \bibinfo {author} {\bibfnamefont {C.}~\bibnamefont {Lorc\'e}}, \bibinfo {author} {\bibfnamefont {P.}~\bibnamefont {Schweitzer}},\ and\ \bibinfo {author} {\bibfnamefont {P.~E.}\ \bibnamefont {Shanahan}},\ }\bibfield  {title} {\bibinfo {title} {{Colloquium: Gravitational form factors of the proton}},\ }\href {https://doi.org/10.1103/RevModPhys.95.041002} {\bibfield  {journal} {\bibinfo  {journal} {Rev. Mod. Phys.}\ }\textbf {\bibinfo {volume} {95}},\ \bibinfo {pages} {041002} (\bibinfo {year} {2023})},\ \Eprint {https://arxiv.org/abs/2303.08347} {arXiv:2303.08347 [hep-ph]} \BibitemShut {NoStop}%
\bibitem [{\citenamefont {Ji}(1995)}]{Ji:1994av}%
  \BibitemOpen
  \bibfield  {author} {\bibinfo {author} {\bibfnamefont {X.-D.}\ \bibnamefont {Ji}},\ }\bibfield  {title} {\bibinfo {title} {{A QCD analysis of the mass structure of the nucleon}},\ }\href {https://doi.org/10.1103/PhysRevLett.74.1071} {\bibfield  {journal} {\bibinfo  {journal} {Phys. Rev. Lett.}\ }\textbf {\bibinfo {volume} {74}},\ \bibinfo {pages} {1071} (\bibinfo {year} {1995})},\ \Eprint {https://arxiv.org/abs/hep-ph/9410274} {arXiv:hep-ph/9410274} \BibitemShut {NoStop}%
\bibitem [{\citenamefont {Lanik}(1984)}]{Lanik:1984fc}%
  \BibitemOpen
  \bibfield  {author} {\bibinfo {author} {\bibfnamefont {J.}~\bibnamefont {Lanik}},\ }\bibfield  {title} {\bibinfo {title} {{A Possible Coupling of a Scalar Glueball to Pseudoscalar Goldstone Mesons}},\ }\href {https://doi.org/10.1016/0370-2693(84)91295-4} {\bibfield  {journal} {\bibinfo  {journal} {Phys. Lett. B}\ }\textbf {\bibinfo {volume} {144}},\ \bibinfo {pages} {439} (\bibinfo {year} {1984})}\BibitemShut {NoStop}%
\bibitem [{\citenamefont {Ellis}\ and\ \citenamefont {Lanik}(1985)}]{Ellis:1984jv}%
  \BibitemOpen
  \bibfield  {author} {\bibinfo {author} {\bibfnamefont {J.~R.}\ \bibnamefont {Ellis}}\ and\ \bibinfo {author} {\bibfnamefont {J.}~\bibnamefont {Lanik}},\ }\bibfield  {title} {\bibinfo {title} {{IS SCALAR GLUONIUM OBSERVABLE?}},\ }\href {https://doi.org/10.1016/0370-2693(85)91013-5} {\bibfield  {journal} {\bibinfo  {journal} {Phys. Lett. B}\ }\textbf {\bibinfo {volume} {150}},\ \bibinfo {pages} {289} (\bibinfo {year} {1985})}\BibitemShut {NoStop}%
\bibitem [{\citenamefont {Leung}\ \emph {et~al.}(1989)\citenamefont {Leung}, \citenamefont {Love},\ and\ \citenamefont {Bardeen}}]{Leung:1989hw}%
  \BibitemOpen
  \bibfield  {author} {\bibinfo {author} {\bibfnamefont {C.~N.}\ \bibnamefont {Leung}}, \bibinfo {author} {\bibfnamefont {S.~T.}\ \bibnamefont {Love}},\ and\ \bibinfo {author} {\bibfnamefont {W.~A.}\ \bibnamefont {Bardeen}},\ }\bibfield  {title} {\bibinfo {title} {{Aspects of Dynamical Symmetry Breaking in Gauge Field Theories}},\ }\href {https://doi.org/10.1016/0550-3213(89)90121-1} {\bibfield  {journal} {\bibinfo  {journal} {Nucl. Phys. B}\ }\textbf {\bibinfo {volume} {323}},\ \bibinfo {pages} {493} (\bibinfo {year} {1989})}\BibitemShut {NoStop}%
\bibitem [{\citenamefont {Campbell}\ \emph {et~al.}(1990)\citenamefont {Campbell}, \citenamefont {Ellis},\ and\ \citenamefont {Olive}}]{Campbell:1990ak}%
  \BibitemOpen
  \bibfield  {author} {\bibinfo {author} {\bibfnamefont {B.~A.}\ \bibnamefont {Campbell}}, \bibinfo {author} {\bibfnamefont {J.~R.}\ \bibnamefont {Ellis}},\ and\ \bibinfo {author} {\bibfnamefont {K.~A.}\ \bibnamefont {Olive}},\ }\bibfield  {title} {\bibinfo {title} {{{QCD} Phase Transitions in an Effective Field Theory}},\ }\href {https://doi.org/10.1016/0550-3213(90)90608-G} {\bibfield  {journal} {\bibinfo  {journal} {Nucl. Phys. B}\ }\textbf {\bibinfo {volume} {345}},\ \bibinfo {pages} {57} (\bibinfo {year} {1990})}\BibitemShut {NoStop}%
\bibitem [{\citenamefont {Donoghue}\ and\ \citenamefont {Leutwyler}(1991)}]{Donoghue:1991qv}%
  \BibitemOpen
  \bibfield  {author} {\bibinfo {author} {\bibfnamefont {J.~F.}\ \bibnamefont {Donoghue}}\ and\ \bibinfo {author} {\bibfnamefont {H.}~\bibnamefont {Leutwyler}},\ }\bibfield  {title} {\bibinfo {title} {{Energy and momentum in chiral theories}},\ }\href {https://doi.org/10.1007/BF01560453} {\bibfield  {journal} {\bibinfo  {journal} {Z. Phys. C}\ }\textbf {\bibinfo {volume} {52}},\ \bibinfo {pages} {343} (\bibinfo {year} {1991})}\BibitemShut {NoStop}%
\bibitem [{\citenamefont {Brown}\ and\ \citenamefont {Rho}(1991)}]{Brown:1991kk}%
  \BibitemOpen
  \bibfield  {author} {\bibinfo {author} {\bibfnamefont {G.~E.}\ \bibnamefont {Brown}}\ and\ \bibinfo {author} {\bibfnamefont {M.}~\bibnamefont {Rho}},\ }\bibfield  {title} {\bibinfo {title} {{Scaling effective Lagrangians in a dense medium}},\ }\href {https://doi.org/10.1103/PhysRevLett.66.2720} {\bibfield  {journal} {\bibinfo  {journal} {Phys. Rev. Lett.}\ }\textbf {\bibinfo {volume} {66}},\ \bibinfo {pages} {2720} (\bibinfo {year} {1991})}\BibitemShut {NoStop}%
\bibitem [{\citenamefont {Song}\ \emph {et~al.}(1997)\citenamefont {Song}, \citenamefont {Brown}, \citenamefont {Min},\ and\ \citenamefont {Rho}}]{Song:1997kx}%
  \BibitemOpen
  \bibfield  {author} {\bibinfo {author} {\bibfnamefont {C.}~\bibnamefont {Song}}, \bibinfo {author} {\bibfnamefont {G.~E.}\ \bibnamefont {Brown}}, \bibinfo {author} {\bibfnamefont {D.-P.}\ \bibnamefont {Min}},\ and\ \bibinfo {author} {\bibfnamefont {M.}~\bibnamefont {Rho}},\ }\bibfield  {title} {\bibinfo {title} {{Fluctuations in 'BR scaled' chiral Lagrangians}},\ }\href {https://doi.org/10.1103/PhysRevC.56.2244} {\bibfield  {journal} {\bibinfo  {journal} {Phys. Rev. C}\ }\textbf {\bibinfo {volume} {56}},\ \bibinfo {pages} {2244} (\bibinfo {year} {1997})},\ \Eprint {https://arxiv.org/abs/hep-ph/9705255} {arXiv:hep-ph/9705255} \BibitemShut {NoStop}%
\bibitem [{\citenamefont {Lee}\ \emph {et~al.}(2003)\citenamefont {Lee}, \citenamefont {Park}, \citenamefont {Rho},\ and\ \citenamefont {Vento}}]{Lee:2003eg}%
  \BibitemOpen
  \bibfield  {author} {\bibinfo {author} {\bibfnamefont {H.-J.}\ \bibnamefont {Lee}}, \bibinfo {author} {\bibfnamefont {B.-Y.}\ \bibnamefont {Park}}, \bibinfo {author} {\bibfnamefont {M.}~\bibnamefont {Rho}},\ and\ \bibinfo {author} {\bibfnamefont {V.}~\bibnamefont {Vento}},\ }\bibfield  {title} {\bibinfo {title} {{Sliding vacua in dense skyrmion matter}},\ }\href {https://doi.org/10.1016/S0375-9474(03)01626-9} {\bibfield  {journal} {\bibinfo  {journal} {Nucl. Phys. A}\ }\textbf {\bibinfo {volume} {726}},\ \bibinfo {pages} {69} (\bibinfo {year} {2003})},\ \Eprint {https://arxiv.org/abs/hep-ph/0304066} {arXiv:hep-ph/0304066} \BibitemShut {NoStop}%
\bibitem [{\citenamefont {Park}\ \emph {et~al.}(2004)\citenamefont {Park}, \citenamefont {Rho},\ and\ \citenamefont {Vento}}]{Park:2003sd}%
  \BibitemOpen
  \bibfield  {author} {\bibinfo {author} {\bibfnamefont {B.-Y.}\ \bibnamefont {Park}}, \bibinfo {author} {\bibfnamefont {M.}~\bibnamefont {Rho}},\ and\ \bibinfo {author} {\bibfnamefont {V.}~\bibnamefont {Vento}},\ }\bibfield  {title} {\bibinfo {title} {{Vector mesons and dense Skyrmion matter}},\ }\href {https://doi.org/10.1016/j.nuclphysa.2004.01.131} {\bibfield  {journal} {\bibinfo  {journal} {Nucl. Phys. A}\ }\textbf {\bibinfo {volume} {736}},\ \bibinfo {pages} {129} (\bibinfo {year} {2004})},\ \Eprint {https://arxiv.org/abs/hep-ph/0310087} {arXiv:hep-ph/0310087} \BibitemShut {NoStop}%
\bibitem [{\citenamefont {Park}\ \emph {et~al.}(2008)\citenamefont {Park}, \citenamefont {Rho},\ and\ \citenamefont {Vento}}]{Park:2008zg}%
  \BibitemOpen
  \bibfield  {author} {\bibinfo {author} {\bibfnamefont {B.-Y.}\ \bibnamefont {Park}}, \bibinfo {author} {\bibfnamefont {M.}~\bibnamefont {Rho}},\ and\ \bibinfo {author} {\bibfnamefont {V.}~\bibnamefont {Vento}},\ }\bibfield  {title} {\bibinfo {title} {{The Role of the Dilaton in Dense Skyrmion Matter}},\ }\href {https://doi.org/10.1016/j.nuclphysa.2008.03.015} {\bibfield  {journal} {\bibinfo  {journal} {Nucl. Phys. A}\ }\textbf {\bibinfo {volume} {807}},\ \bibinfo {pages} {28} (\bibinfo {year} {2008})},\ \Eprint {https://arxiv.org/abs/0801.1374} {arXiv:0801.1374 [hep-ph]} \BibitemShut {NoStop}%
\bibitem [{\citenamefont {Li}\ \emph {et~al.}(2018)\citenamefont {Li}, \citenamefont {Wen}, \citenamefont {Ma},\ and\ \citenamefont {Rho}}]{Li:2018gng}%
  \BibitemOpen
  \bibfield  {author} {\bibinfo {author} {\bibfnamefont {Y.-L.}\ \bibnamefont {Li}}, \bibinfo {author} {\bibfnamefont {P.-S.}\ \bibnamefont {Wen}}, \bibinfo {author} {\bibfnamefont {Y.-L.}\ \bibnamefont {Ma}},\ and\ \bibinfo {author} {\bibfnamefont {M.}~\bibnamefont {Rho}},\ }\bibfield  {title} {\bibinfo {title} {{Scale-Chiral Effective Field Theory for Nuclear Interactions in the Veneziano Limit}},\ }\href@noop {} {\  (\bibinfo {year} {2018})},\ \Eprint {https://arxiv.org/abs/1802.08140} {arXiv:1802.08140 [nucl-th]} \BibitemShut {NoStop}%
\bibitem [{\citenamefont {Crewther}\ and\ \citenamefont {Tunstall}(2015)}]{Crewther:2013vea}%
  \BibitemOpen
  \bibfield  {author} {\bibinfo {author} {\bibfnamefont {R.~J.}\ \bibnamefont {Crewther}}\ and\ \bibinfo {author} {\bibfnamefont {L.~C.}\ \bibnamefont {Tunstall}},\ }\bibfield  {title} {\bibinfo {title} {{$\Delta I=1/2$ rule for kaon decays derived from QCD infrared fixed point}},\ }\href {https://doi.org/10.1103/PhysRevD.91.034016} {\bibfield  {journal} {\bibinfo  {journal} {Phys. Rev. D}\ }\textbf {\bibinfo {volume} {91}},\ \bibinfo {pages} {034016} (\bibinfo {year} {2015})},\ \Eprint {https://arxiv.org/abs/1312.3319} {arXiv:1312.3319 [hep-ph]} \BibitemShut {NoStop}%
\bibitem [{\citenamefont {Li}\ \emph {et~al.}(2017)\citenamefont {Li}, \citenamefont {Ma},\ and\ \citenamefont {Rho}}]{Li:2016uzn}%
  \BibitemOpen
  \bibfield  {author} {\bibinfo {author} {\bibfnamefont {Y.-L.}\ \bibnamefont {Li}}, \bibinfo {author} {\bibfnamefont {Y.-L.}\ \bibnamefont {Ma}},\ and\ \bibinfo {author} {\bibfnamefont {M.}~\bibnamefont {Rho}},\ }\bibfield  {title} {\bibinfo {title} {{Chiral-scale effective theory including a dilatonic meson}},\ }\href {https://doi.org/10.1103/PhysRevD.95.114011} {\bibfield  {journal} {\bibinfo  {journal} {Phys. Rev. D}\ }\textbf {\bibinfo {volume} {95}},\ \bibinfo {pages} {114011} (\bibinfo {year} {2017})},\ \Eprint {https://arxiv.org/abs/1609.07014} {arXiv:1609.07014 [hep-ph]} \BibitemShut {NoStop}%
\bibitem [{\citenamefont {Kasai}\ \emph {et~al.}(2016)\citenamefont {Kasai}, \citenamefont {Okumura},\ and\ \citenamefont {Suzuki}}]{Kasai:2016ifi}%
  \BibitemOpen
  \bibfield  {author} {\bibinfo {author} {\bibfnamefont {A.}~\bibnamefont {Kasai}}, \bibinfo {author} {\bibfnamefont {K.-i.}\ \bibnamefont {Okumura}},\ and\ \bibinfo {author} {\bibfnamefont {H.}~\bibnamefont {Suzuki}},\ }\bibfield  {title} {\bibinfo {title} {{A dilaton-pion mass relation}},\ }\href@noop {} {\  (\bibinfo {year} {2016})},\ \Eprint {https://arxiv.org/abs/1609.02264} {arXiv:1609.02264 [hep-lat]} \BibitemShut {NoStop}%
\bibitem [{\citenamefont {Hansen}\ \emph {et~al.}(2017)\citenamefont {Hansen}, \citenamefont {Lang\ae{}ble},\ and\ \citenamefont {Sannino}}]{Hansen:2016fri}%
  \BibitemOpen
  \bibfield  {author} {\bibinfo {author} {\bibfnamefont {M.}~\bibnamefont {Hansen}}, \bibinfo {author} {\bibfnamefont {K.}~\bibnamefont {Lang\ae{}ble}},\ and\ \bibinfo {author} {\bibfnamefont {F.}~\bibnamefont {Sannino}},\ }\bibfield  {title} {\bibinfo {title} {{Extending Chiral Perturbation Theory with an Isosinglet Scalar}},\ }\href {https://doi.org/10.1103/PhysRevD.95.036005} {\bibfield  {journal} {\bibinfo  {journal} {Phys. Rev. D}\ }\textbf {\bibinfo {volume} {95}},\ \bibinfo {pages} {036005} (\bibinfo {year} {2017})},\ \Eprint {https://arxiv.org/abs/1610.02904} {arXiv:1610.02904 [hep-ph]} \BibitemShut {NoStop}%
\bibitem [{\citenamefont {Appelquist}\ \emph {et~al.}(2017)\citenamefont {Appelquist}, \citenamefont {Ingoldby},\ and\ \citenamefont {Piai}}]{Appelquist:2017wcg}%
  \BibitemOpen
  \bibfield  {author} {\bibinfo {author} {\bibfnamefont {T.}~\bibnamefont {Appelquist}}, \bibinfo {author} {\bibfnamefont {J.}~\bibnamefont {Ingoldby}},\ and\ \bibinfo {author} {\bibfnamefont {M.}~\bibnamefont {Piai}},\ }\bibfield  {title} {\bibinfo {title} {{Dilaton EFT Framework For Lattice Data}},\ }\href {https://doi.org/10.1007/JHEP07(2017)035} {\bibfield  {journal} {\bibinfo  {journal} {JHEP}\ }\textbf {\bibinfo {volume} {07}},\ \bibinfo {pages} {035}},\ \Eprint {https://arxiv.org/abs/1702.04410} {arXiv:1702.04410 [hep-ph]} \BibitemShut {NoStop}%
\bibitem [{\citenamefont {Appelquist}\ \emph {et~al.}(2018)\citenamefont {Appelquist}, \citenamefont {Ingoldby},\ and\ \citenamefont {Piai}}]{Appelquist:2017vyy}%
  \BibitemOpen
  \bibfield  {author} {\bibinfo {author} {\bibfnamefont {T.}~\bibnamefont {Appelquist}}, \bibinfo {author} {\bibfnamefont {J.}~\bibnamefont {Ingoldby}},\ and\ \bibinfo {author} {\bibfnamefont {M.}~\bibnamefont {Piai}},\ }\bibfield  {title} {\bibinfo {title} {{Analysis of a Dilaton EFT for Lattice Data}},\ }\href {https://doi.org/10.1007/JHEP03(2018)039} {\bibfield  {journal} {\bibinfo  {journal} {JHEP}\ }\textbf {\bibinfo {volume} {03}},\ \bibinfo {pages} {039}},\ \Eprint {https://arxiv.org/abs/1711.00067} {arXiv:1711.00067 [hep-ph]} \BibitemShut {NoStop}%
\bibitem [{\citenamefont {Cat\`a}\ and\ \citenamefont {M\"uller}(2020)}]{Cata:2019edh}%
  \BibitemOpen
  \bibfield  {author} {\bibinfo {author} {\bibfnamefont {O.}~\bibnamefont {Cat\`a}}\ and\ \bibinfo {author} {\bibfnamefont {C.}~\bibnamefont {M\"uller}},\ }\bibfield  {title} {\bibinfo {title} {{Chiral effective theories with a light scalar at one loop}},\ }\href {https://doi.org/10.1016/j.nuclphysb.2020.114938} {\bibfield  {journal} {\bibinfo  {journal} {Nucl. Phys. B}\ }\textbf {\bibinfo {volume} {952}},\ \bibinfo {pages} {114938} (\bibinfo {year} {2020})},\ \Eprint {https://arxiv.org/abs/1906.01879} {arXiv:1906.01879 [hep-ph]} \BibitemShut {NoStop}%
\bibitem [{\citenamefont {Appelquist}\ \emph {et~al.}(2020)\citenamefont {Appelquist}, \citenamefont {Ingoldby},\ and\ \citenamefont {Piai}}]{Appelquist:2019lgk}%
  \BibitemOpen
  \bibfield  {author} {\bibinfo {author} {\bibfnamefont {T.}~\bibnamefont {Appelquist}}, \bibinfo {author} {\bibfnamefont {J.}~\bibnamefont {Ingoldby}},\ and\ \bibinfo {author} {\bibfnamefont {M.}~\bibnamefont {Piai}},\ }\bibfield  {title} {\bibinfo {title} {{Dilaton potential and lattice data}},\ }\href {https://doi.org/10.1103/PhysRevD.101.075025} {\bibfield  {journal} {\bibinfo  {journal} {Phys. Rev. D}\ }\textbf {\bibinfo {volume} {101}},\ \bibinfo {pages} {075025} (\bibinfo {year} {2020})},\ \Eprint {https://arxiv.org/abs/1908.00895} {arXiv:1908.00895 [hep-ph]} \BibitemShut {NoStop}%
\bibitem [{\citenamefont {Brown}\ \emph {et~al.}(2019)\citenamefont {Brown}, \citenamefont {Golterman}, \citenamefont {Kr\o{}jer}, \citenamefont {Shamir},\ and\ \citenamefont {Splittorff}}]{Brown:2019ipr}%
  \BibitemOpen
  \bibfield  {author} {\bibinfo {author} {\bibfnamefont {T.~V.}\ \bibnamefont {Brown}}, \bibinfo {author} {\bibfnamefont {M.}~\bibnamefont {Golterman}}, \bibinfo {author} {\bibfnamefont {S.}~\bibnamefont {Kr\o{}jer}}, \bibinfo {author} {\bibfnamefont {Y.}~\bibnamefont {Shamir}},\ and\ \bibinfo {author} {\bibfnamefont {K.}~\bibnamefont {Splittorff}},\ }\bibfield  {title} {\bibinfo {title} {{The $\epsilon$-regime of dilaton chiral perturbation theory}},\ }\href {https://doi.org/10.1103/PhysRevD.100.114515} {\bibfield  {journal} {\bibinfo  {journal} {Phys. Rev. D}\ }\textbf {\bibinfo {volume} {100}},\ \bibinfo {pages} {114515} (\bibinfo {year} {2019})},\ \Eprint {https://arxiv.org/abs/1909.10796} {arXiv:1909.10796 [hep-lat]} \BibitemShut {NoStop}%
\bibitem [{\citenamefont {Matsuzaki}\ and\ \citenamefont {Yamawaki}(2014)}]{Matsuzaki:2013eva}%
  \BibitemOpen
  \bibfield  {author} {\bibinfo {author} {\bibfnamefont {S.}~\bibnamefont {Matsuzaki}}\ and\ \bibinfo {author} {\bibfnamefont {K.}~\bibnamefont {Yamawaki}},\ }\bibfield  {title} {\bibinfo {title} {{Dilaton Chiral Perturbation Theory: Determining the Mass and Decay Constant of the Technidilaton on the Lattice}},\ }\href {https://doi.org/10.1103/PhysRevLett.113.082002} {\bibfield  {journal} {\bibinfo  {journal} {Phys. Rev. Lett.}\ }\textbf {\bibinfo {volume} {113}},\ \bibinfo {pages} {082002} (\bibinfo {year} {2014})},\ \Eprint {https://arxiv.org/abs/1311.3784} {arXiv:1311.3784 [hep-lat]} \BibitemShut {NoStop}%
\bibitem [{\citenamefont {Zwicky}(2024{\natexlab{a}})}]{Zwicky:2023bzk}%
  \BibitemOpen
  \bibfield  {author} {\bibinfo {author} {\bibfnamefont {R.}~\bibnamefont {Zwicky}},\ }\bibfield  {title} {\bibinfo {title} {{QCD with an infrared fixed point: The pion sector}},\ }\href {https://doi.org/10.1103/PhysRevD.109.034009} {\bibfield  {journal} {\bibinfo  {journal} {Phys. Rev. D}\ }\textbf {\bibinfo {volume} {109}},\ \bibinfo {pages} {034009} (\bibinfo {year} {2024}{\natexlab{a}})},\ \Eprint {https://arxiv.org/abs/2306.06752} {arXiv:2306.06752 [hep-ph]} \BibitemShut {NoStop}%
\bibitem [{\citenamefont {Zwicky}(2024{\natexlab{b}})}]{Zwicky:2023krx}%
  \BibitemOpen
  \bibfield  {author} {\bibinfo {author} {\bibfnamefont {R.}~\bibnamefont {Zwicky}},\ }\bibfield  {title} {\bibinfo {title} {{QCD with an infrared fixed point and a dilaton}},\ }\href {https://doi.org/10.1103/PhysRevD.110.014048} {\bibfield  {journal} {\bibinfo  {journal} {Phys. Rev. D}\ }\textbf {\bibinfo {volume} {110}},\ \bibinfo {pages} {014048} (\bibinfo {year} {2024}{\natexlab{b}})},\ \Eprint {https://arxiv.org/abs/2312.13761} {arXiv:2312.13761 [hep-ph]} \BibitemShut {NoStop}%
\bibitem [{\citenamefont {Shifman}\ and\ \citenamefont {Zwicky}(2023)}]{Shifman:2023jqn}%
  \BibitemOpen
  \bibfield  {author} {\bibinfo {author} {\bibfnamefont {M.}~\bibnamefont {Shifman}}\ and\ \bibinfo {author} {\bibfnamefont {R.}~\bibnamefont {Zwicky}},\ }\bibfield  {title} {\bibinfo {title} {{Relating {\ensuremath{\beta}}*' and {\ensuremath{\gamma}}Q*' in the N=1 SQCD conformal window}},\ }\href {https://doi.org/10.1103/PhysRevD.108.114013} {\bibfield  {journal} {\bibinfo  {journal} {Phys. Rev. D}\ }\textbf {\bibinfo {volume} {108}},\ \bibinfo {pages} {114013} (\bibinfo {year} {2023})},\ \Eprint {https://arxiv.org/abs/2310.16449} {arXiv:2310.16449 [hep-th]} \BibitemShut {NoStop}%
\bibitem [{\citenamefont {Skyrme}(1962)}]{Skyrme:1962vh}%
  \BibitemOpen
  \bibfield  {author} {\bibinfo {author} {\bibfnamefont {T.~H.~R.}\ \bibnamefont {Skyrme}},\ }\bibfield  {title} {\bibinfo {title} {{A Unified Field Theory of Mesons and Baryons}},\ }\href {https://doi.org/10.1016/0029-5582(62)90775-7} {\bibfield  {journal} {\bibinfo  {journal} {Nucl. Phys.}\ }\textbf {\bibinfo {volume} {31}},\ \bibinfo {pages} {556} (\bibinfo {year} {1962})}\BibitemShut {NoStop}%
\bibitem [{\citenamefont {Adkins}\ \emph {et~al.}(1983)\citenamefont {Adkins}, \citenamefont {Nappi},\ and\ \citenamefont {Witten}}]{Adkins:1983ya}%
  \BibitemOpen
  \bibfield  {author} {\bibinfo {author} {\bibfnamefont {G.~S.}\ \bibnamefont {Adkins}}, \bibinfo {author} {\bibfnamefont {C.~R.}\ \bibnamefont {Nappi}},\ and\ \bibinfo {author} {\bibfnamefont {E.}~\bibnamefont {Witten}},\ }\bibfield  {title} {\bibinfo {title} {{Static Properties of Nucleons in the Skyrme Model}},\ }\href {https://doi.org/10.1016/0550-3213(83)90559-X} {\bibfield  {journal} {\bibinfo  {journal} {Nucl. Phys. B}\ }\textbf {\bibinfo {volume} {228}},\ \bibinfo {pages} {552} (\bibinfo {year} {1983})}\BibitemShut {NoStop}%
\bibitem [{\citenamefont {Polyakov}(2003)}]{Polyakov:2002yz}%
  \BibitemOpen
  \bibfield  {author} {\bibinfo {author} {\bibfnamefont {M.~V.}\ \bibnamefont {Polyakov}},\ }\bibfield  {title} {\bibinfo {title} {{Generalized parton distributions and strong forces inside nucleons and nuclei}},\ }\href {https://doi.org/10.1016/S0370-2693(03)00036-4} {\bibfield  {journal} {\bibinfo  {journal} {Phys. Lett. B}\ }\textbf {\bibinfo {volume} {555}},\ \bibinfo {pages} {57} (\bibinfo {year} {2003})},\ \Eprint {https://arxiv.org/abs/hep-ph/0210165} {arXiv:hep-ph/0210165} \BibitemShut {NoStop}%
\bibitem [{\citenamefont {Adler}\ \emph {et~al.}(1977)\citenamefont {Adler}, \citenamefont {Collins},\ and\ \citenamefont {Duncan}}]{Adler:1976zt}%
  \BibitemOpen
  \bibfield  {author} {\bibinfo {author} {\bibfnamefont {S.~L.}\ \bibnamefont {Adler}}, \bibinfo {author} {\bibfnamefont {J.~C.}\ \bibnamefont {Collins}},\ and\ \bibinfo {author} {\bibfnamefont {A.}~\bibnamefont {Duncan}},\ }\bibfield  {title} {\bibinfo {title} {{Energy-Momentum-Tensor Trace Anomaly in Spin 1/2 Quantum Electrodynamics}},\ }\href {https://doi.org/10.1103/PhysRevD.15.1712} {\bibfield  {journal} {\bibinfo  {journal} {Phys. Rev. D}\ }\textbf {\bibinfo {volume} {15}},\ \bibinfo {pages} {1712} (\bibinfo {year} {1977})}\BibitemShut {NoStop}%
\bibitem [{\citenamefont {Collins}\ \emph {et~al.}(1977)\citenamefont {Collins}, \citenamefont {Duncan},\ and\ \citenamefont {Joglekar}}]{Collins:1976yq}%
  \BibitemOpen
  \bibfield  {author} {\bibinfo {author} {\bibfnamefont {J.~C.}\ \bibnamefont {Collins}}, \bibinfo {author} {\bibfnamefont {A.}~\bibnamefont {Duncan}},\ and\ \bibinfo {author} {\bibfnamefont {S.~D.}\ \bibnamefont {Joglekar}},\ }\bibfield  {title} {\bibinfo {title} {{Trace and Dilatation Anomalies in Gauge Theories}},\ }\href {https://doi.org/10.1103/PhysRevD.16.438} {\bibfield  {journal} {\bibinfo  {journal} {Phys. Rev. D}\ }\textbf {\bibinfo {volume} {16}},\ \bibinfo {pages} {438} (\bibinfo {year} {1977})}\BibitemShut {NoStop}%
\bibitem [{\citenamefont {Laue}(1911)}]{Laue:1911lrk}%
  \BibitemOpen
  \bibfield  {author} {\bibinfo {author} {\bibfnamefont {M.}~\bibnamefont {Laue}},\ }\bibfield  {title} {\bibinfo {title} {{Zur Dynamik der Relativit{\"a}tstheorie}},\ }\href {https://doi.org/10.1002/andp.19113400808} {\bibfield  {journal} {\bibinfo  {journal} {Annalen Phys.}\ }\textbf {\bibinfo {volume} {340}},\ \bibinfo {pages} {524} (\bibinfo {year} {1911})}\BibitemShut {NoStop}%
\bibitem [{\citenamefont {Perevalova}\ \emph {et~al.}(2016)\citenamefont {Perevalova}, \citenamefont {Polyakov},\ and\ \citenamefont {Schweitzer}}]{Perevalova:2016dln}%
  \BibitemOpen
  \bibfield  {author} {\bibinfo {author} {\bibfnamefont {I.~A.}\ \bibnamefont {Perevalova}}, \bibinfo {author} {\bibfnamefont {M.~V.}\ \bibnamefont {Polyakov}},\ and\ \bibinfo {author} {\bibfnamefont {P.}~\bibnamefont {Schweitzer}},\ }\bibfield  {title} {\bibinfo {title} {{On LHCb pentaquarks as a baryon-$\psi$(2S) bound state: prediction of isospin-$\frac3{2}$ pentaquarks with hidden charm}},\ }\href {https://doi.org/10.1103/PhysRevD.94.054024} {\bibfield  {journal} {\bibinfo  {journal} {Phys. Rev. D}\ }\textbf {\bibinfo {volume} {94}},\ \bibinfo {pages} {054024} (\bibinfo {year} {2016})},\ \Eprint {https://arxiv.org/abs/1607.07008} {arXiv:1607.07008 [hep-ph]} \BibitemShut {NoStop}%
\bibitem [{\citenamefont {Lorc\'e}(2018)}]{Lorce:2017xzd}%
  \BibitemOpen
  \bibfield  {author} {\bibinfo {author} {\bibfnamefont {C.}~\bibnamefont {Lorc\'e}},\ }\bibfield  {title} {\bibinfo {title} {{On the hadron mass decomposition}},\ }\href {https://doi.org/10.1140/epjc/s10052-018-5561-2} {\bibfield  {journal} {\bibinfo  {journal} {Eur. Phys. J. C}\ }\textbf {\bibinfo {volume} {78}},\ \bibinfo {pages} {120} (\bibinfo {year} {2018})},\ \Eprint {https://arxiv.org/abs/1706.05853} {arXiv:1706.05853 [hep-ph]} \BibitemShut {NoStop}%
\bibitem [{\citenamefont {Ponciano}\ and\ \citenamefont {Garcia~Canal}(2005)}]{Ponciano:2004cs}%
  \BibitemOpen
  \bibfield  {author} {\bibinfo {author} {\bibfnamefont {J.~A.}\ \bibnamefont {Ponciano}}\ and\ \bibinfo {author} {\bibfnamefont {C.~A.}\ \bibnamefont {Garcia~Canal}},\ }\bibfield  {title} {\bibinfo {title} {{Approximate solutions for the single soliton in a Skyrmion-type model with a dilaton scalar field}},\ }\href {https://doi.org/10.1103/PhysRevC.72.065206} {\bibfield  {journal} {\bibinfo  {journal} {Phys. Rev. C}\ }\textbf {\bibinfo {volume} {72}},\ \bibinfo {pages} {065206} (\bibinfo {year} {2005})},\ \Eprint {https://arxiv.org/abs/hep-ph/0405258} {arXiv:hep-ph/0405258} \BibitemShut {NoStop}%
\bibitem [{\citenamefont {Carson}(1991)}]{Carson:1991fu}%
  \BibitemOpen
  \bibfield  {author} {\bibinfo {author} {\bibfnamefont {L.}~\bibnamefont {Carson}},\ }\bibfield  {title} {\bibinfo {title} {{Static properties of He-3 and H-3 in the Skyrme model}},\ }\href {https://doi.org/10.1016/0375-9474(91)90472-I} {\bibfield  {journal} {\bibinfo  {journal} {Nucl. Phys. A}\ }\textbf {\bibinfo {volume} {535}},\ \bibinfo {pages} {479} (\bibinfo {year} {1991})}\BibitemShut {NoStop}%
\bibitem [{\citenamefont {Garcia Martin-Caro}\ \emph {et~al.}(2023)\citenamefont {Garcia Martin-Caro}, \citenamefont {Huidobro},\ and\ \citenamefont {Hatta}}]{GarciaMartin-Caro:2023klo}%
  \BibitemOpen
  \bibfield  {author} {\bibinfo {author} {\bibfnamefont {A.}~\bibnamefont {Garcia Martin-Caro}}, \bibinfo {author} {\bibfnamefont {M.}~\bibnamefont {Huidobro}},\ and\ \bibinfo {author} {\bibfnamefont {Y.}~\bibnamefont {Hatta}},\ }\bibfield  {title} {\bibinfo {title} {{Gravitational form factors of nuclei in the Skyrme model}},\ }\href {https://doi.org/10.1103/PhysRevD.108.034014} {\bibfield  {journal} {\bibinfo  {journal} {Phys. Rev. D}\ }\textbf {\bibinfo {volume} {108}},\ \bibinfo {pages} {034014} (\bibinfo {year} {2023})},\ \Eprint {https://arxiv.org/abs/2304.05994} {arXiv:2304.05994 [nucl-th]} \BibitemShut {NoStop}%
\bibitem [{\citenamefont {Brown}\ \emph {et~al.}(1984)\citenamefont {Brown}, \citenamefont {Jackson}, \citenamefont {Rho},\ and\ \citenamefont {Vento}}]{Brown:1984sx}%
  \BibitemOpen
  \bibfield  {author} {\bibinfo {author} {\bibfnamefont {G.~E.}\ \bibnamefont {Brown}}, \bibinfo {author} {\bibfnamefont {A.~D.}\ \bibnamefont {Jackson}}, \bibinfo {author} {\bibfnamefont {M.}~\bibnamefont {Rho}},\ and\ \bibinfo {author} {\bibfnamefont {V.}~\bibnamefont {Vento}},\ }\bibfield  {title} {\bibinfo {title} {{THE NUCLEON AS A TOPOLOGICAL CHIRAL SOLITON}},\ }\href {https://doi.org/10.1016/0370-2693(84)90754-8} {\bibfield  {journal} {\bibinfo  {journal} {Phys. Lett. B}\ }\textbf {\bibinfo {volume} {140}},\ \bibinfo {pages} {285} (\bibinfo {year} {1984})}\BibitemShut {NoStop}%
\bibitem [{\citenamefont {Cebulla}\ \emph {et~al.}(2007)\citenamefont {Cebulla}, \citenamefont {Goeke}, \citenamefont {Ossmann},\ and\ \citenamefont {Schweitzer}}]{Cebulla:2007ei}%
  \BibitemOpen
  \bibfield  {author} {\bibinfo {author} {\bibfnamefont {C.}~\bibnamefont {Cebulla}}, \bibinfo {author} {\bibfnamefont {K.}~\bibnamefont {Goeke}}, \bibinfo {author} {\bibfnamefont {J.}~\bibnamefont {Ossmann}},\ and\ \bibinfo {author} {\bibfnamefont {P.}~\bibnamefont {Schweitzer}},\ }\bibfield  {title} {\bibinfo {title} {{The Nucleon form-factors of the energy momentum tensor in the Skyrme model}},\ }\href {https://doi.org/10.1016/j.nuclphysa.2007.08.004} {\bibfield  {journal} {\bibinfo  {journal} {Nucl. Phys. A}\ }\textbf {\bibinfo {volume} {794}},\ \bibinfo {pages} {87} (\bibinfo {year} {2007})},\ \Eprint {https://arxiv.org/abs/hep-ph/0703025} {arXiv:hep-ph/0703025} \BibitemShut {NoStop}%
\bibitem [{\citenamefont {Ji}\ and\ \citenamefont {Yang}(2025{\natexlab{a}})}]{Ji:2025gsq}%
  \BibitemOpen
  \bibfield  {author} {\bibinfo {author} {\bibfnamefont {X.}~\bibnamefont {Ji}}\ and\ \bibinfo {author} {\bibfnamefont {C.}~\bibnamefont {Yang}},\ }\bibfield  {title} {\bibinfo {title} {{Momentum Flow and Forces on Quarks in the Nucleon}},\ }\href@noop {} {\  (\bibinfo {year} {2025}{\natexlab{a}})},\ \Eprint {https://arxiv.org/abs/2503.01991} {arXiv:2503.01991 [hep-ph]} \BibitemShut {NoStop}%
\bibitem [{\citenamefont {Diakonov}\ and\ \citenamefont {Petrov}(1986)}]{Diakonov:1986yh}%
  \BibitemOpen
  \bibfield  {author} {\bibinfo {author} {\bibfnamefont {D.}~\bibnamefont {Diakonov}}\ and\ \bibinfo {author} {\bibfnamefont {V.~Y.}\ \bibnamefont {Petrov}},\ }\bibfield  {title} {\bibinfo {title} {{Chiral Theory of Nucleons}},\ }\href@noop {} {\bibfield  {journal} {\bibinfo  {journal} {JETP Lett.}\ }\textbf {\bibinfo {volume} {43}},\ \bibinfo {pages} {75} (\bibinfo {year} {1986})}\BibitemShut {NoStop}%
\bibitem [{\citenamefont {Diakonov}\ \emph {et~al.}(1988)\citenamefont {Diakonov}, \citenamefont {Petrov},\ and\ \citenamefont {Pobylitsa}}]{Diakonov:1987ty}%
  \BibitemOpen
  \bibfield  {author} {\bibinfo {author} {\bibfnamefont {D.}~\bibnamefont {Diakonov}}, \bibinfo {author} {\bibfnamefont {V.~Y.}\ \bibnamefont {Petrov}},\ and\ \bibinfo {author} {\bibfnamefont {P.~V.}\ \bibnamefont {Pobylitsa}},\ }\bibfield  {title} {\bibinfo {title} {{A Chiral Theory of Nucleons}},\ }\href {https://doi.org/10.1016/0550-3213(88)90443-9} {\bibfield  {journal} {\bibinfo  {journal} {Nucl. Phys. B}\ }\textbf {\bibinfo {volume} {306}},\ \bibinfo {pages} {809} (\bibinfo {year} {1988})}\BibitemShut {NoStop}%
\bibitem [{\citenamefont {Goeke}\ \emph {et~al.}(2007)\citenamefont {Goeke}, \citenamefont {Grabis}, \citenamefont {Ossmann}, \citenamefont {Polyakov}, \citenamefont {Schweitzer}, \citenamefont {Silva},\ and\ \citenamefont {Urbano}}]{Goeke:2007fp}%
  \BibitemOpen
  \bibfield  {author} {\bibinfo {author} {\bibfnamefont {K.}~\bibnamefont {Goeke}}, \bibinfo {author} {\bibfnamefont {J.}~\bibnamefont {Grabis}}, \bibinfo {author} {\bibfnamefont {J.}~\bibnamefont {Ossmann}}, \bibinfo {author} {\bibfnamefont {M.~V.}\ \bibnamefont {Polyakov}}, \bibinfo {author} {\bibfnamefont {P.}~\bibnamefont {Schweitzer}}, \bibinfo {author} {\bibfnamefont {A.}~\bibnamefont {Silva}},\ and\ \bibinfo {author} {\bibfnamefont {D.}~\bibnamefont {Urbano}},\ }\bibfield  {title} {\bibinfo {title} {{Nucleon form-factors of the energy momentum tensor in the chiral quark-soliton model}},\ }\href {https://doi.org/10.1103/PhysRevD.75.094021} {\bibfield  {journal} {\bibinfo  {journal} {Phys. Rev. D}\ }\textbf {\bibinfo {volume} {75}},\ \bibinfo {pages} {094021} (\bibinfo {year} {2007})},\ \Eprint {https://arxiv.org/abs/hep-ph/0702030} {arXiv:hep-ph/0702030} \BibitemShut {NoStop}%
\bibitem [{\citenamefont {Guichon}(1988)}]{Guichon:1987jp}%
  \BibitemOpen
  \bibfield  {author} {\bibinfo {author} {\bibfnamefont {P.~A.~M.}\ \bibnamefont {Guichon}},\ }\bibfield  {title} {\bibinfo {title} {{A Possible Quark Mechanism for the Saturation of Nuclear Matter}},\ }\href {https://doi.org/10.1016/0370-2693(88)90762-9} {\bibfield  {journal} {\bibinfo  {journal} {Phys. Lett. B}\ }\textbf {\bibinfo {volume} {200}},\ \bibinfo {pages} {235} (\bibinfo {year} {1988})}\BibitemShut {NoStop}%
\bibitem [{\citenamefont {Saito}\ and\ \citenamefont {Thomas}(1994)}]{Saito:1994ki}%
  \BibitemOpen
  \bibfield  {author} {\bibinfo {author} {\bibfnamefont {K.}~\bibnamefont {Saito}}\ and\ \bibinfo {author} {\bibfnamefont {A.~W.}\ \bibnamefont {Thomas}},\ }\bibfield  {title} {\bibinfo {title} {{A Quark - meson coupling model for nuclear and neutron matter}},\ }\href {https://doi.org/10.1016/0370-2693(94)91520-2} {\bibfield  {journal} {\bibinfo  {journal} {Phys. Lett. B}\ }\textbf {\bibinfo {volume} {327}},\ \bibinfo {pages} {9} (\bibinfo {year} {1994})},\ \Eprint {https://arxiv.org/abs/nucl-th/9403015} {arXiv:nucl-th/9403015} \BibitemShut {NoStop}%
\bibitem [{\citenamefont {Miller}(2019)}]{Miller:2018ybm}%
  \BibitemOpen
  \bibfield  {author} {\bibinfo {author} {\bibfnamefont {G.~A.}\ \bibnamefont {Miller}},\ }\bibfield  {title} {\bibinfo {title} {{Defining the proton radius: A unified treatment}},\ }\href {https://doi.org/10.1103/PhysRevC.99.035202} {\bibfield  {journal} {\bibinfo  {journal} {Phys. Rev. C}\ }\textbf {\bibinfo {volume} {99}},\ \bibinfo {pages} {035202} (\bibinfo {year} {2019})},\ \Eprint {https://arxiv.org/abs/1812.02714} {arXiv:1812.02714 [nucl-th]} \BibitemShut {NoStop}%
\bibitem [{\citenamefont {Jaffe}(2021)}]{Jaffe:2020ebz}%
  \BibitemOpen
  \bibfield  {author} {\bibinfo {author} {\bibfnamefont {R.~L.}\ \bibnamefont {Jaffe}},\ }\bibfield  {title} {\bibinfo {title} {{Ambiguities in the definition of local spatial densities in light hadrons}},\ }\href {https://doi.org/10.1103/PhysRevD.103.016017} {\bibfield  {journal} {\bibinfo  {journal} {Phys. Rev. D}\ }\textbf {\bibinfo {volume} {103}},\ \bibinfo {pages} {016017} (\bibinfo {year} {2021})},\ \Eprint {https://arxiv.org/abs/2010.15887} {arXiv:2010.15887 [hep-ph]} \BibitemShut {NoStop}%
\bibitem [{\citenamefont {Fujii}\ and\ \citenamefont {Tanaka}(2025)}]{Fujii:2025paw}%
  \BibitemOpen
  \bibfield  {author} {\bibinfo {author} {\bibfnamefont {D.}~\bibnamefont {Fujii}}\ and\ \bibinfo {author} {\bibfnamefont {M.}~\bibnamefont {Tanaka}},\ }\bibfield  {title} {\bibinfo {title} {{Scale-anomaly-induced confining pressure within hadrons}},\ }\href@noop {} {\  (\bibinfo {year} {2025})},\ \Eprint {https://arxiv.org/abs/2507.23786} {arXiv:2507.23786 [hep-ph]} \BibitemShut {NoStop}%
\bibitem [{\citenamefont {Ji}\ and\ \citenamefont {Yang}(2025{\natexlab{b}})}]{Ji:2025qax}%
  \BibitemOpen
  \bibfield  {author} {\bibinfo {author} {\bibfnamefont {X.}~\bibnamefont {Ji}}\ and\ \bibinfo {author} {\bibfnamefont {C.}~\bibnamefont {Yang}},\ }\bibfield  {title} {\bibinfo {title} {{A Journey of Seeking Pressures and Forces in the Nucleon}},\ }\href@noop {} {\  (\bibinfo {year} {2025}{\natexlab{b}})},\ \Eprint {https://arxiv.org/abs/2508.16727} {arXiv:2508.16727 [hep-ph]} \BibitemShut {NoStop}%
\end{thebibliography}%

\end{document}